\documentclass[amstex,epsf,amssymb,usenatbib]{mnras}
\usepackage{epsf,graphics,subfig}
\usepackage{amsmath,amssymb,natbib}
\usepackage{lmodern}
\usepackage{cite}

\usepackage{rotating,times,graphicx,latexsym,color,longtable,lscape} 

\usepackage{nicefrac}
\def\aj{AJ}                   
\def\araa{ARA\&A}             
\def\apj{ApJ}                 
\def\apjl{ApJ}                
\def\apjs{ApJS}               
\def\aap{A\&A}                
\def\mnras{MNRAS}             
\def\nat{Nature}              

\makeatletter

\newcommand{\Rmnum}[1]{\expandafter\@slowromancap\romannumeral #1@}
\makeatother

\providecommand{\numberTblEq}[1]{\refstepcounter{tblEqCounter}\label{#1}\thetag{\thetblEqCounter}}

\title[Simulations \& signatures of protoplanetary nebulae]{Numerical simulations of wind-driven protoplanetary nebulae. \Rmnum{1}. near-infrared emission}
    
\author[I. Novikov \& M.D. Smith]
{Igor Novikov$^{1}$\thanks{E-mail: i-novikov@hotmail.com}
\& Michael Smith$^{1}$\thanks{E-mail: M.D.Smith@kent.ac.uk}\\
$^{1}$Centre for Astrophysics \& Planetary Science, The University of Kent, Canterbury, Kent CT2 7NH, U.K. }                                                                                                                                                             
\date{Accepted .....
      Received ..... ;
      in original form .....}

\pagerange{\pageref{firstpage}--\pageref{lastpage}}
\pubyear{2018}

\begin{document}
                                                                                                                                                             
\maketitle
                                                                                                                                                             
\label{firstpage}
                                                                                                                                                             
\begin{abstract}
To understand how the circumstellar environments of post-AGB stars develop into planetary nebulae, we initiate a systematic study of 2D axisymmetric hydrodynamic simulations 
of protoplanetary nebula (pPN) with a modified ZEUS code. The aim of this first work  is to compare the structure of prolate ellipsoidal  winds into a stationary ambient medium where both 
media can be either atomic or molecular.  We specifically model the early twin-shock phase which generates a decelerating shell. A thick deformed and turbulent shell grows when
 an atomic wind expands into an atomic medium. In all other cases, the interaction shell region fragments into radial protrusions due to molecular cooling and chemistry.
The resulting fingers eliminate any global slip parallel to the shell surface. This rough surface implies that weak shocks are prominent in the excitation of the gas despite the fast speed of advance. This may explain why low excitation molecular hydrogen is found  towards the front of elliptical pPN.
We constrain molecular dissociative fractions and timescales of fast $\mathrm H_2$ winds and the pPN lifetime with wind densities $\mathrm{\sim10^{5}cm^{-3}}$ and shock speeds of $\mathrm{80\sim200\,km\,s^{-1}}$.  
We identify a variety of stages associated with thermal excitation of H$_2$ near-infrared emission. 
Generated line emission maps and position-velocity diagrams enable a comparison and distinction with post-AGB survey results. The $\mathrm{1\to0 \, S(1)}$ $\&$ $\mathrm{2\to1 \, S(1)}$ lines are 
lobe-dominated bows rather than bipolar shells.  
\noindent 
\end{abstract} 
  
\section{Introduction} 
\label{intro}

Images of planetary nebula (PN) have long been admired. Their  diverse shapes and intricate structure in visible light raises questions of how they form and develop. The debate has recognised  several types of interacting winds, including slow, fast and super.
The generation of these winds, however,  is the basis for the model uncertainties especially with the influence of binarity, accretion and magnetic fields.
Winds of high speed are clearly blown off from stars in a number of circumstances associated with planetary nebula \citep{2002ARA&A..40..439B}. These winds display a range of speeds (30 - 300 ~km~s$^{-1}$), asymmetry (from spherical to collimated) and shape (from elliptical to bipolar). The interaction with the immediate environment can be observed on  large scales of 1000\,AU to a few parsecs  and at high spatial resolution from the infrared through to the optical  \citep{1987AJ.....94..671B,1996ApJ...462..777K,1997MNRAS.284...32P}.
We may thus gain detailed information on the initial ramming of global winds into the interstellar medium  through the interpretation of the shock excitation of atomic and molecular species 
\citep{1978ApJ...219L.125K}. 

The programme of simulations begun here is intended to reveal the physical properties of the wind interactions and to display the molecular and atomic emission lines as
diagnostic tools. In particular, we wish to systematically investigate the early ramming of fast winds in the context of protoplanetary nebula (pPN) for which near-infrared observations provide detailed information through integral field spectroscopy \citep{2015MNRAS.447.1080G}. The objective is to perform simulations to learn how molecular and atomic gas behaves under many diverse conditions. By comparison to sets of observations, this will inform us how the star's mass loss and environment evolve. Knowing this may then aid our understanding of how much mass and energy is returned into the interstellar medium, 
how  interstellar dust is replenished and how the driving star evolves for which there remains considerable discrepancies  \citep{2016A&A...586A...5M,2016A&A...588A..25M}.

The circumstellar envelope, however, may not itself be static but also a part of a previously ejected slower wind or outflow with a radial density gradient and clumpy gas distribution.
To some extent, we know how the initial winds build up a circumstellar environment, based on theory and observation of the Active Giant Branch (AGB) phase \citep{1978ApJ...219L.125K}. Thermal pulses during this stage drive bursts of dense winds which appear as molecular shells \citep[][e.g.]{1988A&A...196L...1O,2016A&A...586A...5M}. This slow wind expands at just 10\,-\,20~km~s$^{-1}$ \citep{1990A&A...227L..29L}.  Subsequently, the star evolves through the last phases of the AGB,  corresponding to a super-wind phase of lower rate of mass-loss but higher speed at 50\,-\,60~km~s$^{-1}$ \citep{2009IAUS..259...35B}.

There is indeed evidence that the subsequent evolutionary stages from AGB to PN correspond to a sequence of winds which can be described as 
slow, super and fast \citep{2001A&A...378..958M,2005ARA&A..43..435H,2006ApJ...650..237H}. The final  fast post-AGB wind evolves with velocities of several $\mathrm{100 \,km\,s^{-1}}$ \citep{2001A&A...377..868B}.
The post-AGB stage is, however, a very short period in stellar evolution, just $\sim 10^3 $ years \citep{2000ApJ...528..861U}, making these objects scarce. Only about 100 are identified, designated as protoplanetary nebula \citep{2011MNRAS.417...32L}.

Furthermore, as it evolves off the Main Sequence, the stages in the life journey of low and intermediate-mass stars, in the broad mass range of 1\,$\sim$\,8  M$_\odot$, may be contrasting.   The fast winds from the more massive stars, which drive into the previously ejected  `superwind', are more likely driven by radiation pressure.  In addition, each wind potentially possesses a different mass loss rate and geometry \citep{2012MNRAS.424.2055H}. For these various reasons, it is still not clear which configuration sequences are actually followed \citep{2012MNRAS.424.2055H}. 

The well-defined phases which are observationally identified are Active Giant Branch (AGB), post-AGB, protoplanetary nebula (pPN), planetary nebula (PN)  and, eventually, the exposed white dwarf stage. Not all stars necessarily proceed through all stages. These transitions are marked by a changing environment in which not only winds and  shells but jets and/or bullets \citep{2016ApJ...820..134H} of diverse speed and density also appear.  As a result, pPN and PN  display a myriad of morphologies \citep{2002ARA&A..40..439B} that  challenge our innate desire  to classify.

 We are foremost interested in finding the configurations which can plausibly explain the expanding pPN as seen in H$_2$ emission lines. \citet{2008ApJ...688..327H} found the emission to be from the walls of the nebula producing an ellipsoidal velocity structure. \citet{2011MNRAS.411.1453G} identified a fast wind impinging on the cavity walls and tips. 
\citet{2008A&A...480..775V} detected shock-excited molecular hydrogen emission as well as collisionally-excited emission lines from $\mathrm{[O I]}$, $\mathrm{[C I]}$, and $\mathrm{[Fe II]}$. Finally, \citet{2015MNRAS.447.1080G} used molecular hydrogen emission lines together with hydrogen and helium recombination lines to explore the distribution of molecular and atomic gas and distinguish a range of evolutionary stages. 

Wind interactions have, nevertheless, already been well studied \citep[][e.g.]{1982ApJ...258..280K,2005MNRAS.360..963M,2012MNRAS.424.2055H}. The present work improves on these studies by including a full molecular and atomic cooling function. In doing so, we are able to 
predict the infrared line properties in some detail. This is particularly important  to the  interpretation of the initial ramming, where a shell structure is formed between the two shock waves: the reverse shock which decelerates the wind and the advancing shock which sweeps up the ambient medium.

As our backdrop, we first intend to model winds through a set of axisymmetric simulations  with an initial uniform stationary ambient medium. Axisymmetry is indeed observed to be relevant from the earliest expansion  stages  with elongated shells of typical ellipticity 0.44
\citep{2000ApJ...528..861U} .

We have introduced an elliptical wind profile to be consistent with the  findings of \citet{2011MNRAS.417...32L} that pPN with a detached shell or a visible central star appear elliptical in nature.   There are 36 simulations discussed in detail here: (1) the wind ellipticity is varied through an anisotropic wind speed with plane-axis aspect ratios of 1:1, 1:2 and 1:4, (2) the axial speed of the wind is taken as slow, medium or fast and (3) the wind and ambient medium can both either be molecular or atomic. After this fundamental study, we then intend to perform three dimensional simulations of winds into winds.  
 
 \begin{figure*}
\includegraphics[width=0.89\textwidth,keepaspectratio]{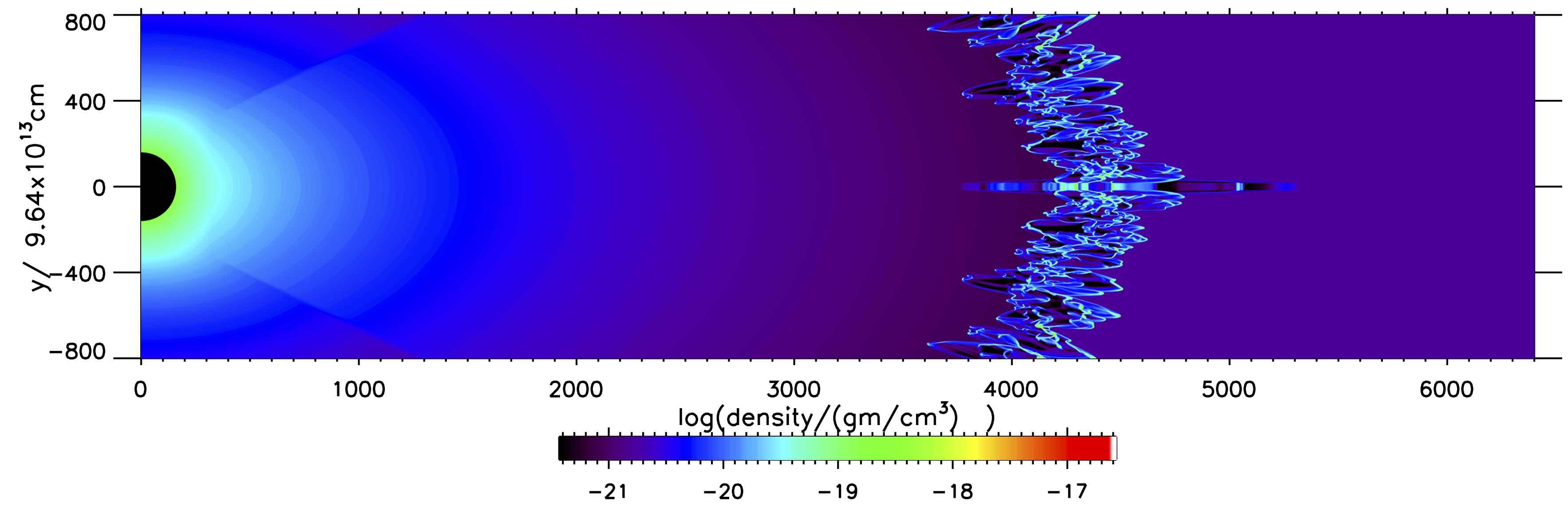} 
\includegraphics[width=0.89\textwidth,keepaspectratio]{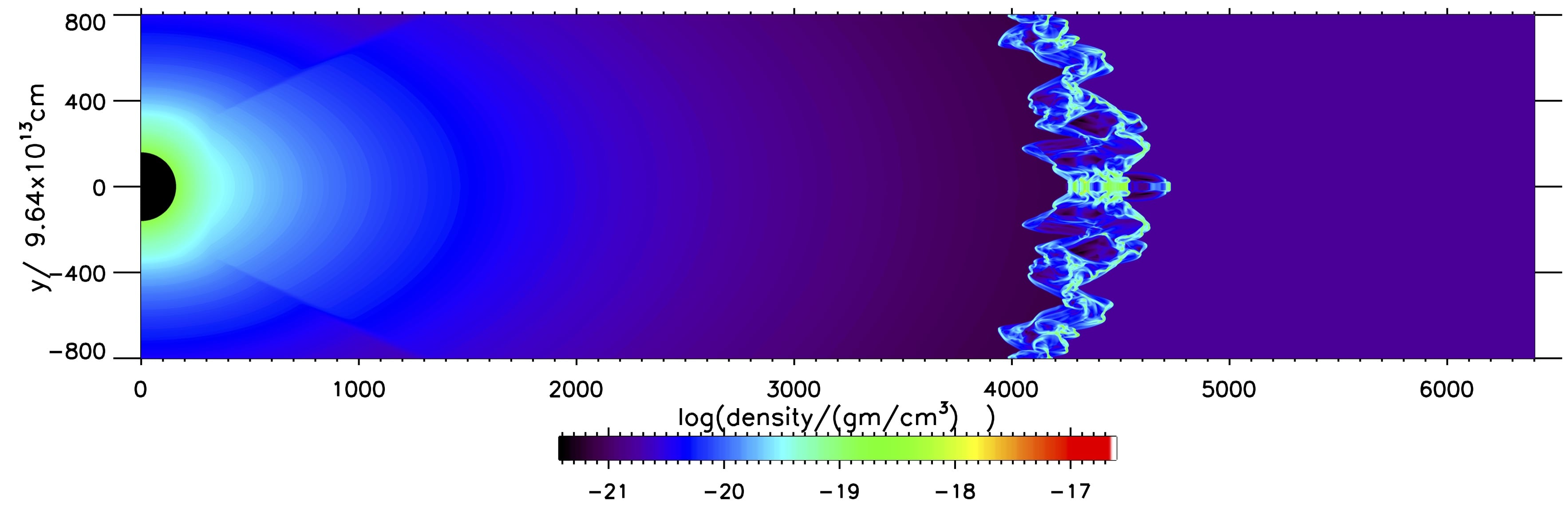} 
\caption{Molecular wind/ambient (upper panel) and atomic wind/ambient (lower panel)  density distributions for 2:1 elliptical winds of speed
$V_w =140$\,km\,s$^{-1}$ along the symmetry axis. Only the displayed portion has been simulated here in order to confirm the detailed features at high resolution of the global winds modelled  below. The wind is driven from a shell at 0.01\,pc for a time of 1,000 years, onto a grid of length 0.2\,pc. An inner mask is superimposed to improve the colour range. }  
\label{comparison}
\end{figure*}

 A feature of the molecular simulations, as exemplified  in Fig.\,\ref{comparison}, are the growth of fingers along the shock surface. The fingers grow through hydrodynamic shell instabilities and we find that the sound speed of the
cold shell determines the maximum coherent wavelength.  Observationally, there is considerable evidence that shells may turn into shrapnel, after breaking up into filaments and fingers as well as cometary-shaped globules \citep{1984Natur.311..236T,2009ApJ...700.1067M, 2005AJ....129.1625U,2015A&A...573A..56S}. 

Such protrusions can arise  through the clumpy nature of the ambient medium
\citep{2013MNRAS.436..470S}, through variable winds or through hydrodynamical instabilities \citep{1995Natur.377..315S}. 
Such hydrodynamic instability also dominates jet-driven molecular interactions in which the advancing bow shock splits up into many mini-bows \citep{2004MNRAS.347.1097R}.   However, these are enhanced in wind-driven flows through the Rayleigh-Taylor instability as the heavy shell rapidly decelerates.
In addition, strong cooling in the shocked zone can very easily result in growing protrusions which promote thermal \citep{2003MNRAS.339..133S}  as well as hydrodynamical instability. 

\section{Method} 
\label{Method}

\subsection{The code}
 
  The Eulerian ZEUS code \citep{1992ApJS...80..753S}  is employed as our basis  to carry out the hydrodynamical calculations on a two dimension cylindrical grid with axial symmetry.  The code uses Van Leer advection \citep{1977JCoPh..23..276V}, consistent transport of the magnetic field, von Neumann and Richtmyer artificial viscosity \citep{1950JAP....21..232V} and an upwinded scheme, efficient for solving supersonic flow configurations. It is versatile, robust and well-tested \citep{2010ApJS..187..119C}. Although higher order codes are potentially more accurate, the high speed of the algorithms means that parameter space can be explored without sacrificing resolution.
We modify version 3.5 of ZEUS-3D (dzeus35) which is freely available for use by the scientific community and can be downloaded from the Institute of Computational Astrophysics (ICA) at St.Mary's University, Nova Scotia, Canada\,\footnote{\url{http://www.ica.smu.ca/zeus3d/}}. 
 
 The models are set up to generate winds with speeds along the axis ranging from $v_w \mathrm{\sim 80 - 200\, km\, s^{-1}}$ into a uniform density medium as specified in Table~\ref{partable}.   The steady wind of  hydrogen nucleon density $n_w$ (plus 10\% helium)  is injected from a simulated spherical surface of radius $R_{w}= 0.01$\,parsec.   
  This relatively large injection radius was chosen so as to allow a good numerically approximate to the surface of a sphere on the grid. The majority of the 2D axially symmetric winds were set up on a cylindrical grid of $3,200\times1,600$ zones, corresponding to $0.2$ pc on the jet axis and $0.1$ pc of radial extent. 
   
 \begin{figure} 
\includegraphics[width=0.24\textwidth,keepaspectratio]{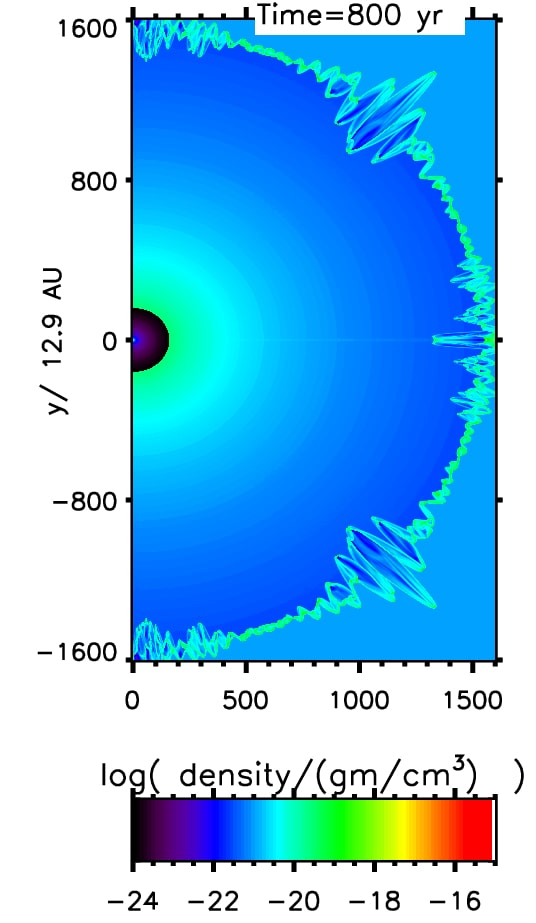}
\includegraphics[width=0.24\textwidth,keepaspectratio]{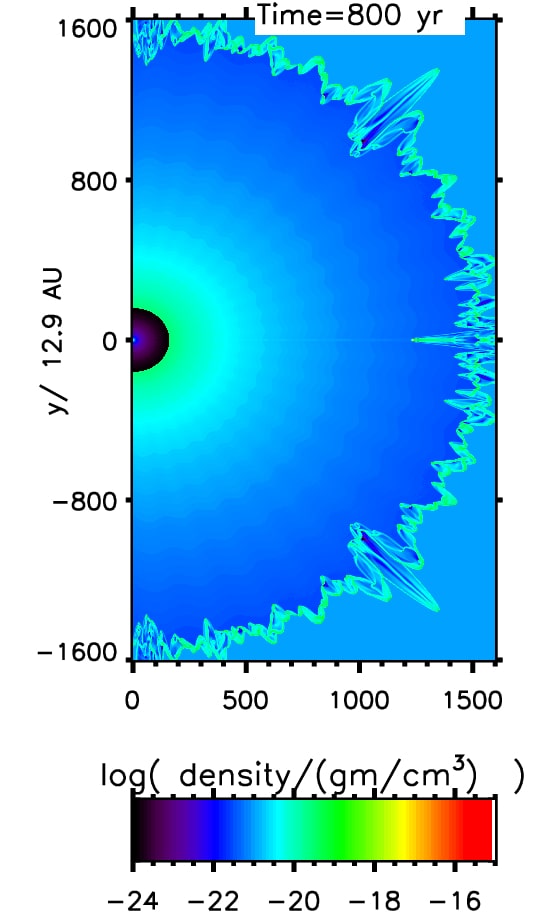}
\caption{Comparison of simulations for a uniform radial wind (left panel) and a density-corrugated wind (right panel). Cross-sectional distributions of the density after 750 years for a spherical molecular wind of radial speed 200~km\,s$^{-1}$ into  a uniform ambient molecular medium. The imposed density corrugations in the wind are of ten sinusoidal 
waves of amplitude 2.5\% between the axis and 90$^\circ$.  The cross-section displayed is 0.1\,pc~$\times$~0.2\,pc and the input sphere is of radius 0.01\,pc corresponding to 160 zones.}  \label{sphere}
\end{figure}
 
\begin{table*}
\centering
\caption[Atomic \& Molecular wind parameters]
 {Initial wind/ambient parameters. Top \& middle: Macroscopic wind/ambient parameters for 2:1, 140~km\,s$^{-1}$, $\mathrm{10^{-4} M_{\odot}yr^{-1}}$ outflows at injection radius. Bottom: wind mach numbers.  \label{partable} }
 \begin{tabular}{ | l  c  r|}
 \hline
 Property   &     Atomic & Molecular \\
\hline
Wind: \\
wind density, $\mathrm{\rm{\rho_w (gm/cm^3)}}$    & $3.76\times 10^{-20}$ & $3.76\times 10^{-20}$ \\
wind internal energy, $\mathrm{\rm{e_{w} (erg/cm^{3})}}$ & $1.83 \times 10^{-09}$ & $1.55 \times 10^{-09}$ \\
molecular fraction, $\mathrm{n(H_2)/n}$                      & $0$ & $0.5$ \\
specific heat ratio, $\mathrm{\gamma}$        & $1.66667$ & $1.42857$ \\
temperature, $\mathrm{T_{w}(K)}$      & $500$ & $500$ \\
\hline
Ambient medium: \\
 density, $\mathrm{\rho_{a} (gm/cm^3)}$          & $1.51\times 10^{-21}$  & $1.51\times 10^{-21}$ \\ 
 int. energy, $\mathrm{e_{a} (erg/cm^{3})}$ & $2.92 \times 10^{-12}$ & $2.48 \times 10^{-12}$ \\ 
 temperature, $\mathrm{T_{a}(K)}$       & $20$ & $20$ \\
 \hline 
Mach number, $\rm{M_{w}}$            &        &        \\
 $\mathrm{80 \, km/s}$                    & $34.40$ & $50.32$ \\
 $\mathrm{100 \, km/s}$                  & $43.01$ & $62.91$ \\
 $ \mathrm{140 \, km/s}$                   & $60.22$ & $88.07$\\
 $ \mathrm{200 \, km/s}$                   & $86.02$ & $125.82$\\
\hline
\end{tabular}

\end{table*}

 \subsection{Parameters and wind velocity profile}

 We first model a pure radial  outflow emanating from a spherical surface at constant speed. We take  a mass outflow rate of $10^{-4}$~M$_\odot$~yr$^{-1}$ and use this to determine the initial 
parameter set which is listed in Table~\ref{partable}.

We then introduce an ellipsoidal wind profile, consistent with findings by \citet{2011MNRAS.417...32L}: pPN with a detached shell or a visible central star appear elliptical in nature and have two spectral energy distribution (SED) peaks. The short and long wavelength components are  photospheric and dust emission, respectively.  We explore two other wind speed distributions with axial-to-plane speed ratios of 2:1 and 4:1. The axial and plane speed components at $\mathrm{(z,r)}$ on the inner spherical surface are then adjusted accordingly, and given by $\mathrm{v_z =  (z/R_{w}) v_{w} }$ and $\mathrm{v_r =  \epsilon  (r/R_{w}) v_{w} }$ with $\mathrm{\epsilon}$ equal to 1, 0.5 or 0.25. Thus the speed on the surface is  fixed by $\mathrm{v_{z} =  v_{w} cos\theta}$ and $\mathrm{v_{r} = \epsilon v_{w} sin\theta}$ where $\mathrm{\theta}$ is the polar angle. The wind input radius is $\mathrm{R_{w} = 0.01 \, pc}$.
 
It is noted that by taking cylindrical symmetry, we achieve the high resolution necessary to study radiative shocks. However, this often introduces well-known numerical artefacts. In particular for this study, the imposed numerical approximation to a sphere is over a finite number of zones. Therefore,  in addition, we manually break the spherical symmetry by implementing small wavelength corrugations on to the input density. In so doing, we can verify how far  the results depend on the numerical approximation. As shown in Fig. \ref{sphere}, imposing  perturbations (right panel) does modify the stability of the shell, promoting growth much more uniformly around the shell. The shell structure is not directly related to the imposed oscillations. We thus suggest below that the sound speed in the thin interaction layer determines the wavelength of the perturbations which grow. We also find that the evolving complex shell structure is not significantly influenced when elliptical winds are present. The symmetry of the underlying physical problem is automatically broken. 

The axial symmetry also creates a protrusion along the axis, a phenomenon eliminated in full 3D simulations and minimised in 2D by taking sufficiently high resolution. This numerical artefact is found directly along the axis where a narrow nose cone does tend to develop in cylindrical symmetry \citep{2006MNRAS.371.1448M}. We also find a fast growth at about 45$^\circ$ for the spherical wind caused by the directional splitting computations combined with the flow symmetry.

The ambient medium is initially at rest in all the simulations of this first work. The density is chosen in order to create the conditions with propagating forward and backwards shocks which decrease and increase in strength, respectively, as the wind expands over the defined domain. This set-up enables the reverse  shock to become dominant after passing through a distance $\mathrm{R_{ppn}}$. Hence the introduced winds with particle densities of above $\mathrm{\sim 10^{5}\,cm^{-3}}$ are very overdense with respect to the ambient medium with $\mathrm{\sim 10^{3}\,cm^{-3}}$ resulting in an initial
mass density ratio of $\mathrm{\eta_o=25}$  and an initial ballistic  expansion. The temperature of the ambient medium is set to $\mathrm{20 \, K}$, and the input temperature of the wind is $\mathrm{500 \, K}$.
 In all cases the wind is over-pressured with respect to the ambient medium by the factor $\mathrm{\kappa\sim10^{3}}$.
   
Toward image generation, the ZEUS fortran  code delivers the relevant mass density, molecular fraction, velocity and temperature  for each cell. Separate IDL codes then determine the flux contribution to a zone in a 3-D image-velocity data cube for any given molecular or atomic line assuming the flux depends only on the above parameters.. The data is adjusted according to the angle to the line of sight $\mathrm{\theta}$ to produce the predicted 2--D image.   


\newcounter{tblEqCounter} 

\setcounter{tblEqCounter}{\theequation} 
\begin{table*}
\caption[$\mathrm H_2$ collisional de-excitation rates]
 {$\mathrm H_2$ collisional de-excitation rates. Parameters: $T_{v}=1635 \, \rm{K}$, inelastic rotational cross section $\sigma_{0}= 10^{-16} \, \rm{cm^{2}}$ and $\overline{v}=({8kT/ \pi m_{H}})^{1/2}$ is the average thermal speed of an incident hydrogen molecule with $\rm{\Delta{E}=E(v^{\prime},J^{\prime})-E(v,J)}$ denoting decrease in de-excitation rates. Tabulated collisional de-excitation rate coefficients are presented graphically in Fig.~\ref{emission}. References: \citep{1983ApJ...270..578L}$\mathrm{^1}$, \citep{1982ARA&A..20..163S}$\mathrm{^2}$, \citep{1979ApJS...41..555H}$\mathrm{^3}$. \label{de-excitation rates} }
\centering
\def\arraystretch{1.5}
\begin{tabular}{ | l  l  r|}
 \hline
 Collision type   & \,\,\,\,\,\,\,\,\,\,\,\,\,\,\,\,\,\,  Rate coefficients ${\rm(cm^{3}\,s^{-1})}$ & Reference \\
\hline
Vibrational H-H$_2$ & $k_{H}^{(1,0)}= \begin{cases}
 1.4\times 10^{-13} \exp[(T/125)-(T/577)^2], &  T \textless T_{v} \\
 1.0 \times 10^{-12}T^{1/2}\exp(-1000/T)   &  T \textgreater T_{v}.
\end{cases} $ \numberTblEq{eq1} & 1,2 \\
 & $k_{H}^{(2,1)}=4.5 \times 10^{-12}T^{1/2}\exp(-(500/T)^2)$ \,\,\,\,\,\,\,\,\,\,\,\,\,\,\,\,\,\,\,\,\,\,\,\,\,\,\,\,\,\,\,\,\,\,\,\,\,\,\,\,\,\,\ \numberTblEq{eq2} & 3 \\ 
 & $k_{H}^{(2,0)}=1.6 \times 10^{-12}T^{1/2}\exp(-(400/T)^2)$ \,\,\,\,\,\,\,\,\,\,\,\,\,\,\,\,\,\,\,\,\,\,\,\,\,\,\,\,\,\,\,\,\,\,\,\,\,\,\,\,\,\,\ \numberTblEq{eq3} & 3 \\
Vibrational H$_2$-H$_2$ & $k_{H_2}^{(1,0)}= \sigma_{0} \overline{v}\exp(-4.2 \Delta E/k)/(T+1200))$ \,\,\,\,\,\,\,\,\,\,\,\,\,\,\,\,\,\,\,\,\,\,\,\,\,\,\,\,\,\,\,\,\,\,\,\,\,\,\,\,\,\,\ \numberTblEq{eq4} &  1,2\\ 
 & $k_{H_2}^{(2,1)}=k_{H_2}^{(1,0)}$ \,\,\,\,\,\,\,\,\,\,\,\,\,\,\,\,\,\,\,\,\,\,\,\,\,\,\,\,\,\,\,\,\,\,\,\,\,\,\,\,\,\,\,\,\,\,\,\,\,\,\,\,\,\,\,\,\,\,\,\,\,\,\,\,\,\,\,\,\,\,\,\,\,\,\,\,\,\,\,\,\,\,\,\,\,\,\,\,\,\,\,\,\,\,\,\,\,\,\,\,\,\,\,\,\,\,\,\,\,\,\,\,\,\,\,\,\,\,\,\ \numberTblEq{eq5}& 3 \\ 
  & $k_{H_2}^{(2,0)}=0$ \,\,\,\,\,\,\,\,\,\,\,\,\,\,\,\,\,\,\,\,\,\,\,\,\,\,\,\,\,\,\,\,\,\,\,\,\,\,\,\,\,\,\,\,\,\,\,\,\,\,\,\,\,\,\,\,\,\,\,\,\,\,\,\,\,\,\,\,\,\,\,\,\,\,\,\,\,\,\,\,\,\,\,\,\,\,\,\,\,\,\,\,\,\,\,\,\,\,\,\,\,\,\,\,\,\,\,\,\,\,\,\,\,\,\,\,\,\,\,\,\,\,\,\,\,\,\,\,\,\,\,\ \numberTblEq{eq5} & 3 \\
\hline
\end{tabular}
\end{table*}

\setcounter{equation}{\thetblEqCounter} 

\begin{figure}
\includegraphics[width=\columnwidth,height=0.26\textheight]{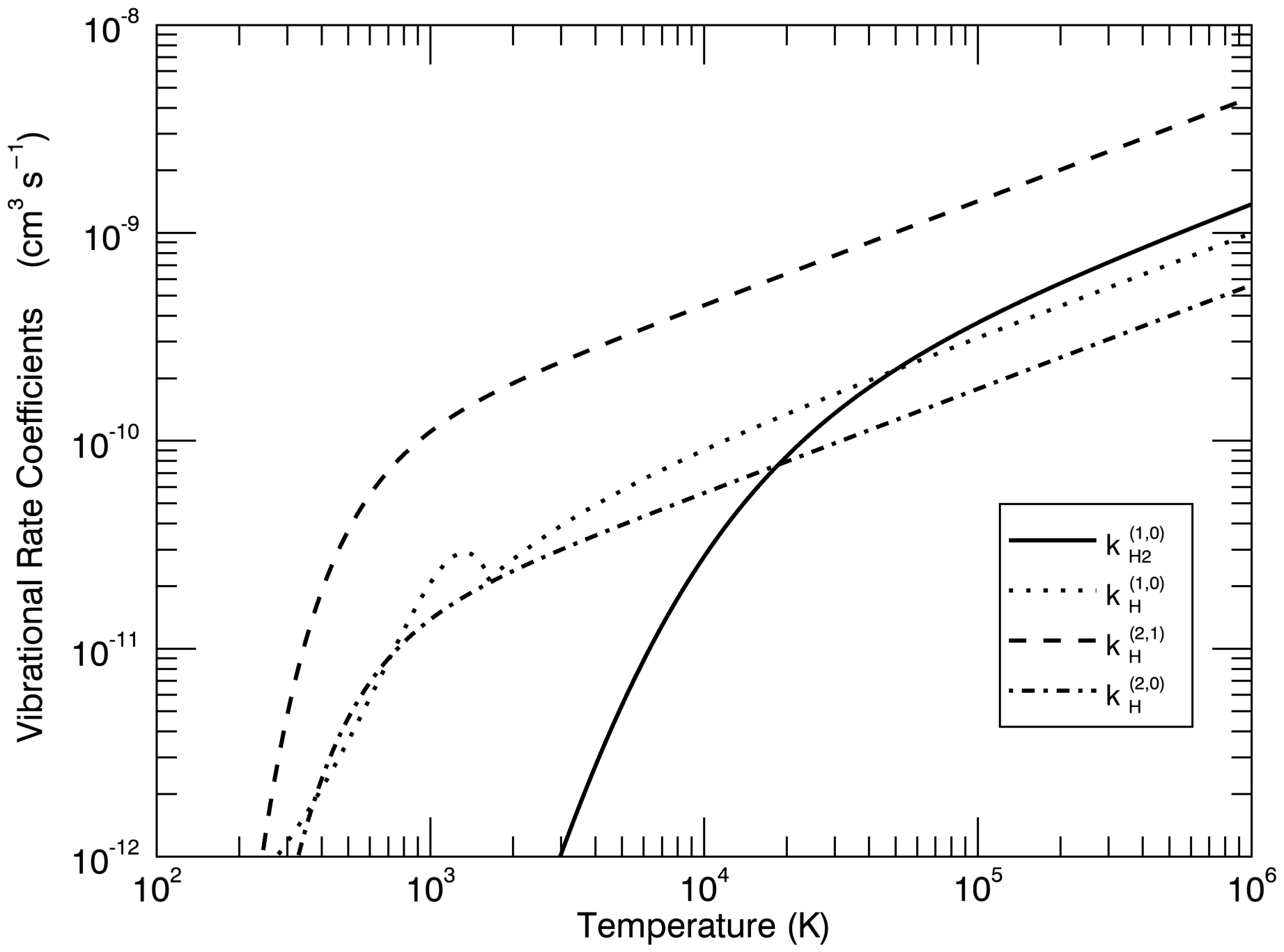}
\caption[Emission rates]
{Collisional de-excitation rate coefficients for H-H$_2$ and H$_2$-H$_2$ from Table~\ref{de-excitation rates} indicating higher rates for vibrational H-H$_2$ collisions in the range of $10^{2}-10^{4} \, \rm{K}$ and dominant cooling by collisional dissosiation for $T\,\textgreater\, 4000 \,\rm{K}$. }  
\label{emission}
\end{figure}

\subsection{Numerical and physical limitations}
 
\begin{figure}
\subfloat[Subfigure 1 list of figures text][ $1,300\times650$ grid zones.]{
\includegraphics[width=0.45\columnwidth,height=0.20\textheight]{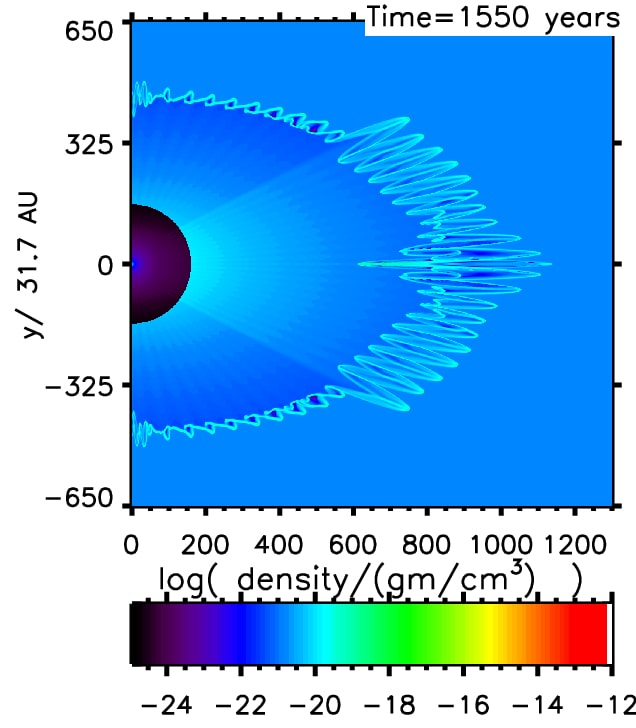}
\label{fig:12x65}}
\subfloat[Subfigure 2 list of figures text][ $2,600\times1,300$ grid zones.]{
\includegraphics[width=0.45\columnwidth,height=0.20\textheight]{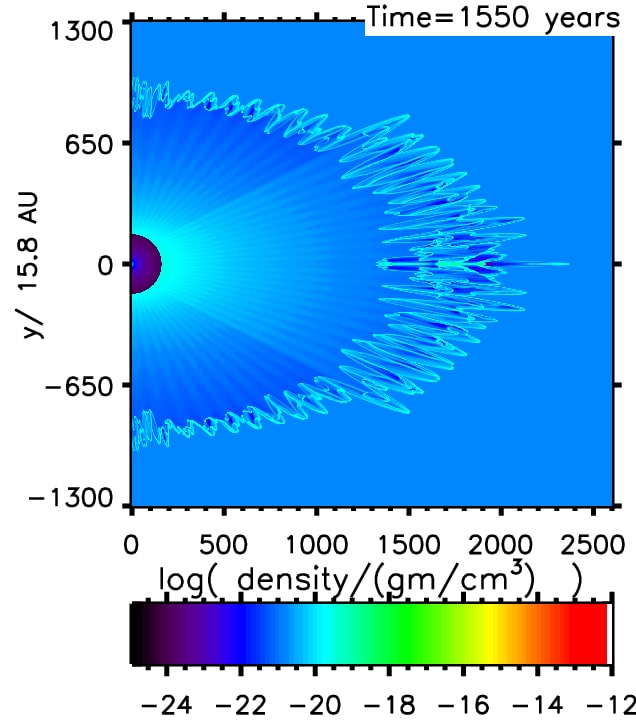}
\label{fig:26x13}}
\qquad
\subfloat[Subfigure 3 list of figures text][$3,200\times1,600$ grid zones.]{
\includegraphics[width=0.45\columnwidth,height=0.20\textheight]{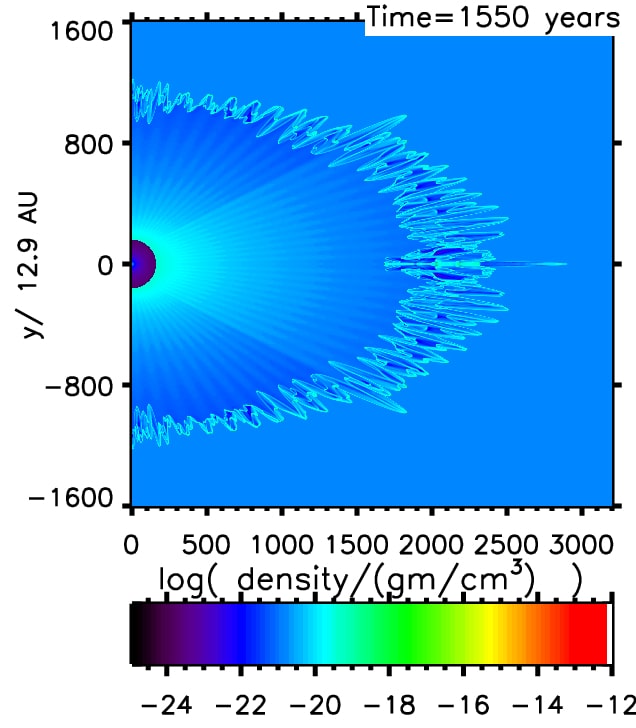}
\label{fig:32x16}}
\subfloat[Subfigure 4 list of figures text][$3,800\times1,900$ grid zones.]{
\includegraphics[width=0.45\columnwidth,height=0.20\textheight]{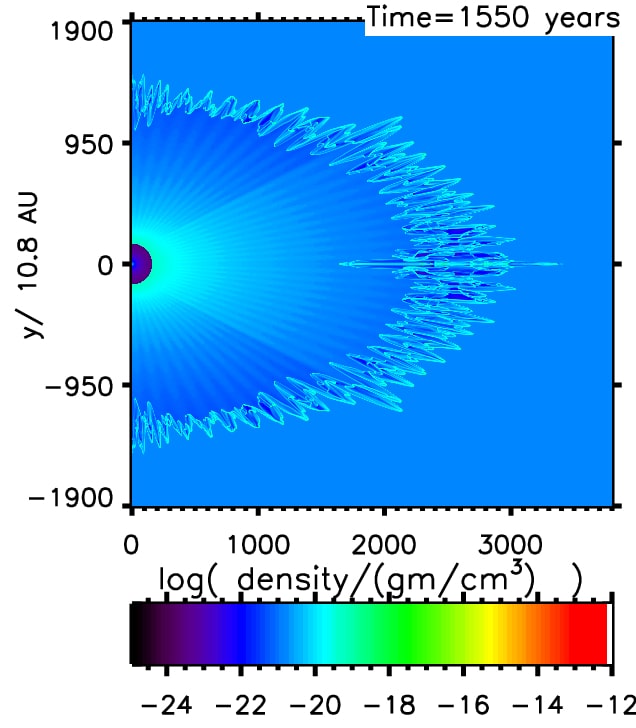}
\label{fig:42x21}}
\qquad
\caption{Cross-sectional distributions of mass density  for a 2:1 $\rm{H_2}$ 140~km~s$^{-1}$ wind into $\rm{H_2}$ ambient medium at low/high grid resolutions.}
\label{resolutions}
\end{figure}

 A clear distinction needs to be made between physical structure and structure induced by numerical effects. We have therefore performed  resolution studies. This is particularly difficult in three dimensions since shock resolution is sacrificed for the spatial integrity. However, in two dimensional simulations we can better resolve the major section of the radiative layers where the infrared emission is generated. Several molecular runs were set up at the grid resolutions indicated in Fig.~\ref{resolutions}. The executed simulations included corrugation effects with 20 sinusoidal density oscillation in the wind across the quadrant in which the wind enters. Attempts at higher resolutions than those displayed resulted in extreme parameter ranges and eventual code crashes.
The simulations show a remarkable similarity, consistent with numerical convergence. This provides confidence that the predicted global emission patterns predicted here are representative of the physics included. 

Photo-dissociation of molecular hydrogen is not included in these calculations. Molecule formation on grains and dissociation through collisions with atoms and molecules are incorporated.
This means that we assume that the central UV source is still ramping up but not significant enough to destroy the developing molecular layer. This clearly does not hold once the pPN transitions 
into a fully-fledged planetary nebula and will depend on the mass and evolution of the emerging star.

Self-irradiation from high-speed shocks is strong and will introduce extra photo-dissociation and photo-ionisation from the emitted UV
  photons. Here, the roughness of the shock surface implies that the ultraviolet can be absorbed locally, adjacent to the strong shocks,  and so can be effective before the gas has been advected though the front. However, this will be somewhat counter-balanced by the  roughness of the shock surface which implies that vast regions of the surface are highly oblique to the approaching flow.
  Overall, these radiation effects may well be crucial for the high speed wind of 200\,km\,s$^{-1}$ where also an ionisation precursor would propagate ahead of the front and
  alter the atomic emission. Such effects are not easy to accomodate beyond one dimensional simulations.  
  
  In addition, the post-processing of the simulations assumes an LTE approximation for the rotational levels of H$_2$ but non-LTE between vibrational levels. This approximation can only be used for 
  warm H$_2$ in which the low-J levels are likely to be thermalised and the high-J levels are sparsely populated. Although often applied as a better approach than assuming full LTE, its accuracy has not been tested against a full calculation.

\subsection{Cooling and chemistry}  

We summarise the most important modifications made to the generic ZEUS code due to the added  molecular and atomic cooling functions as previously formulated in routines written by \citet{1997A&A...318..595S} and translated into a similar ZEUS code by \citet{2003MNRAS.339..133S}.  We have also taken atomic cooling parameters as summarised by \citet{1989ApJ...342..306H} in order to  generate atomic emission maps.  
 Modelling actual electron transitions in atoms and molecules during a simulation would be computationally intensive. Therefore, we use routines to calculate the emission flux per unit volume from fixed electron and atomic abundances.  Hence, it should be noted that these formulae only generate indicative values for the high excitation lines. 
  In fact, the strong post-shock atomic cooling immediately behind the discontinuity of a fast atomic shock is not resolved in these simulations.
 This is also relevant at a hot molecular shock front, where both heating and dissociative cooling can  take place within the same zones. Therefore, the peak temperature as given in the simulations is significantly lower than in a discontinuous shock jump. Nevertheless, the subsequent density, molecular fraction and and temperature in the cooling layer is accurately simulated
  \citep{1997A&A...318..595S}.

 The internal energy equation is altered by including an additional cooling term $\Lambda$,
\begin{equation}
\rho\frac{D}{Dt}\frac{e}{\rho}=-\rho\nabla\cdot v - \Lambda(T,n,f).
\end{equation}
The cooling term is taken to be  a function of three variables: temperature $\mathrm{T}$, hydrogen nuclei density $\mathrm{n}$, and the abundance of molecular hydrogen, $\mathrm{f}$. Cooling from the dissociation of molecular hydrogen takes the form
\begin{equation}
    \Lambda_{dis}=7.18\times10^{-12}(n^2_{H}k_{D,H_2}+n_{H}n_{H_2}k_{D,H}). \label{eq:8}
\end{equation}
The first term is the $\mathrm{4.48\,eV}$ of dissociation energy of $\mathrm H_2$ multiplied by reaction rate coefficients provided by \citet{1987ApJ...318...32S}.Collisional heating in shock waves triggers the rotational and vibrational excitation  of constituent $\mathrm H_2$ molecules. The separate vibrational and rotational cooling functions were listed by \citet{2003MNRAS.339..133S}.
\begin{equation}
	\Lambda_{col}=n_{H_2} \left[  \frac{L_v^H}{1 + (L_v^H)/(L_v^L)} + \frac{L_r^H}{1 + (L_r^H)/(L_r^H)} \right]. \label{eq:9}
\end{equation}

Molecular hydrogen possesses a permanent quadrupole moment  enabling rotational transitions in $\mathrm O$, $\mathrm Q$ and $\mathrm S$ branches, 
corresponding to $\mathrm{ \Delta J =2,\, 0, \, -2} $, respectively. Of these, the most common observed ro-vibrational transitions which are useful as diagnostics are the $\mathrm{1\to0 \, S(1)}$ line at $\mathrm{2.12 \, \mu m}$  and the higher energy $\mathrm{2\to1 \, S(1)}$ line at $\mathrm{2.24 \,\mu m}$. Both lines are ortho transitions with equal upper degeneracies  and fall within the 
near-infrared K-band window.

The H$_2$ line emission is determined from the population density of the upper level $\mathrm{N_j}$ of the transition. A non-equilibrium chemistry is assumed with a non-LTE state, in which the vibrational levels are not distributed according to the Boltzmann distribution. Unlike the vibrational levels, the rotational levels within each vibrational level are assumed to be in LTE  due to much lower radiative transition rates. The time in between vibrationally-exciting collisions in NLTE is long enough to allow for radiative decays to lower levels via spontaneous radiative transitions. This reduces the upper level population to a fraction which we call the deviation factor $\mathrm{\eta_{j}}$. 
 The deviation factor is approximately calculated in a stepwise fashion for subsequent vibrational populations:   

\begin{equation}
    \eta_{1} = \frac{{{\rm V}_{1}}/{{\rm V}_{0}}}{1+{{\rm V}_{1}}/{{\rm V}_{0}}},
\end{equation}
where V$_{j}$ is the populations of the $\mathrm{j^{th}}$ vibrational level and is given by
\begin{equation}
   {\rm V}_{1}/{\rm V}_{0} = \frac{\Gamma_{1,0}\,{\rm exp}\,(-T_{\rm v}/T)}{\Gamma_{1,0}\,+{\rm A}_{1,0}},
\end{equation}
where $\mathrm{\Gamma_{j\prime \, j\prime\prime}}$ is the collisional de-excitation rate
and ${\mathrm A}_{j\prime,j\prime\prime}$ is the Einstein radiation coefficient or  spontaneous transition probability. The latter are given by \citet{1977ApJS...35..281T} as suitably averaged for $\mathrm{J}$ levels in the given $\mathrm{v}$ state: ${\mathrm A}_{1,0}=8.3 \times 10^{-7} $\, s$^{-1}, \, {\mathrm A}_{2,0}=4.1 \times 10^{-7}$\, s$^{-1}$ and ${\mathrm A}_{2,1}=1.1 \times 10^{-6} \, s^{-1}$; degeneracies are $\mathrm{g_{0}=g_{1}=g_{2}=1}$ and transition energies $\mathrm{E_{1,0}/k=E_{2,1}/k=0.5E_{2,0}/k=5860\,K}$ given by \citet{1979ApJS...41..555H}.

The second vibrational level is evaluated in this approximation through
\begin{equation}
    {{\rm V}_{2}/{{\rm V}_{1}}}= \frac{{{\rm V}_{\rm{2up}}}} {{{\rm V}_{\rm{2down}}}}\,\,{\rm exp}\,(-5648/T),
\end{equation}
where
\begin{gather}
     \rm{V_{2up}}=\rm{\Gamma_{2,1}}\,+(1+\rm{A_{1,0}}/\rm{\Gamma_{1,0}}) \, \Gamma_{2,0}, \\
\rm{V_{2down}}=\Gamma_{2,1}+\Gamma_{2,0}+\rm{A_{2,0}}+\rm{A_{2,1}}, \\
\Gamma_{j\prime \, j\prime\prime}=k_{H_2}^{j\prime \, j\prime\prime}\,n({\rm H_2})+k_{H}^{j\prime \, j\prime\prime}\,n({\rm H}).
\end{gather}
We take $\mathrm{T}_{\rm v}=5987$\,K as the temperature difference between the upper and lower vibrational levels. The rate coefficients are $k_{H_2}^{j\prime,j\prime\prime}$ and $k_{H}^{j\prime,j\prime\prime}$ for de-excitation on collision with H$_2$ and H as tabulated in Table~\ref{de-excitation rates} and shown in Fig.~\ref{emission}. These parameters form a three-level cooling system for $\mathrm{v=0,1,2}$ vibrational levels with condensed rotational levels of a given vibrational state. The cooling function takes into account all the vibrational and rotational transitions. Observational data including images up to $\mathrm{v=2}$ are available. Higher vibrational level images are very rare or faint. Since the excitation energy of the first vibrational state is $\rm{T} \sim 6000 \, \rm{K}$ but the emission from these vibrational levels is from gas well below that at $T\,\textless \,3000 \, \rm{K}$, we can employ the above approximation without the need for considerable computational expense from the treatment of each individual excited level. The level density $\mathrm{N_j}$ is then given by
\begin{equation}
N_j=\frac{{\rm g_j} \,n({\rm H_2}) \,\eta_{j}\, {\rm exp}\,(-T_r/T)}{{\cal Q}\,(T)}\,\,\,\,\,\,{\rm \bigl[cm^{-3}\bigr]},
\end{equation}
where ${\rm g_j}=(2I+1)(2J+1)$ is the degeneracy of the level satisfying the exclusion principle  with $\mathrm{I=0}$ `para' states paired with even $\mathrm{J}$, while $\mathrm{I=1}$ 
`ortho' states are paired with odd $\mathrm{J}$. Hence the line intensity is proportional to the statistical weight of $21$ for both modelled transitions and $T_r=1015$\,K is the temperature difference between the upper and lower rotational levels within $\mathrm{V_j}$. ${\cal Q}\,(T)$ is the molecule partition function which is taken as ${\cal Q}\,(T)=(T/40.75)\,\bigl[ 1-{\rm exp}\,(-5\,987/T) \bigr]^{-1}$ for H$_2$. The radiation contribution for each shock element can then be determined from
\begin{gather}
\dot{E}_{1,0} = \frac{N_j{\rm Z_j}\,{\rm h}\,{\rm c}}{\lambda_j}\,\,\,\,\,\,\,\,\,\,\,\,\,\,{\rm \bigl[erg\,s^{-1}\,cm^{-3}\bigr]}\, ,\\
\dot{E}_{2,1} = 1.35\, \frac{{{\rm V}_{2}}}{{{\rm V}_{1}}} \, \dot{E}_{1,0}\,\,\,\, {\rm \bigl[erg\,s^{-1}\,cm^{-3}\bigr]}\, ,
\end{gather}
where h and c are the usual constants and Z$_{\rm j}$ is the spontaneous radiative decay tabulated by \citet{1977ApJS...35..281T} i.e. 
$\rm{Z_{1,0}} = 3.47\times \, 10^{-7} \, s^{-1}$ and $\rm{Z_{2,1}} = 4.98\times \, 10^{-7} \, s^{-1}$ . The $\mathrm{\dot{E}}$ K-band spectra is then integrated over the $\nicefrac{\pi}{2}$ degrees due to cylindrical symmetry, thus maintaining a constant volume for each element of gas that is projected on to the 2--D CCD image in Section 4. The total volumetric emission is summed up through the whole grid and multiplied by the zone size cubed resulting in the dissipation of energy from the entire flow. A 3--D data cube is constructed along with the velocity information to produce position-velocity (PV) diagrams described in Section 5.

\subsection{Specific heat ratio}

The simultaneous determination of the specific heat ratio, $\mathrm{\gamma}$,  and temperature is a critical part of  the ZEUS chemistry. 
 To begin, we fix the wind hydrogen nucleon density as $\mathrm{n_{w} = \rho_{w}/1.4 m_p}$ (with 10\% helium $\mathrm{n(He)=0.1n_{w}}$).

The total number of hydrogen nuclei present per unit volume is
\begin{equation}
n(H_{nuclei})=2 \times n(H_{mol})+n(H_{atoms}).
\end{equation} 
 Hence, the molecular abundance $\mathrm{n(H_{mol}) = f \,n(H_{nuclei})}$, where $\mathrm{f}$ ranges between $\mathrm{0}$ for full atomic and  $\mathrm{0.5}$  for fully molecular. Therefore
\begin{equation}
n(H_{atoms})=(1-2f) \times n(H_{nuclei}),
\end{equation}
and on simplifying the notation:
\begin{equation}
n(H)=(1-2f)n.
\end{equation} 
 Summing all three chemical constituents  brings the particle number density $\mathrm{<n>}$
 \begin{equation}
<n>=n(H)+n(H_2)+n(He)=(1.1-f)n.
\end{equation}

The total specific heat at constant volume of the system $\mathrm{c_v}$, is given by the sum of the individual specific heats of the components that are, in turn, determined by the number of degrees of freedom. For the temperatures attained in the shocks, we can take the molecular hydrogen to possess three translational and two rotational degrees of freedom making five degrees of freedom in total. Two additional vibrational degrees are ignored due to the high activation temperatures of  greater than 6,000\,K, under which molecular hydrogen would be fully dissociated. Therefore
\begin{equation}
 c_v=\frac{3}{2}k_B\frac{n(H)}{<n>}+\frac{5}{2}k_B\frac{n(H_2)}{<n>}+\frac{3}{2}k_B\frac{n(He)}{<n>}
\end{equation}
with $\mathrm{k_B}$ representing the Boltzmann constant. Substituting into the above equation and solving for $\mathrm{c_v}$ yields
\begin{equation}
\label{parameters}
c_v=\frac{3.3-f}{2.2-2f}k_B.
\end{equation}
The specific heat ratio $\mathrm{\gamma=c_p/c_v}$ together with the specific heat at constant pressure $\mathrm{c_p=c_v+k_B}$ gives
\begin{equation}
\gamma=\frac{c_p}{c_v}
      =\frac{c_v+k_B}{c_v}
      =\frac{5.5-3f}{3.3-f}.
\end{equation}
Hence $\mathrm{\gamma=\frac{5}{3}}$ and $\mathrm{\gamma=\frac{10}{7}}$ represent atomic and fully molecular media, respectively. Heavier elements such as carbon and oxygen are considered in the code through the molecular cooling functions.

\subsection{Scaling}

The properties of the wind and shell expansion are described in a similar manner to a freely expanding jet \citep{2004orst.book.....S}. 
For a constant spherical wind, the mass loss rate is
\begin{equation}
{\dot M}= 4\pi R_{w}^2\rho_{w}v_{w}=const.
\end{equation}
In general, for molecular outflows at these moderate densities, the electron dissociation becomes dominant at speeds above $\mathrm{\sim50\,km\,s^{-1}}$ \citep{1994MNRAS.266..238S}. 
The wind driver is a hydrodynamic jump  or J-type shock, followed by a radiative cooling layer. This layer is a combination of  a hot zone in which $\mathrm H_2$ is dissociated, producing dissociative cooling, and generating atomic line emission, followed by the radiative cooling zone, where the remaining $\mathrm H_2$ is excited without dissociation. We model an optically thin ambient medium, hence no absorption occurs. The excited molecules then decay via electric quadrupule transitions on time scales of order $10^{6}$\,s \citep{1976ApJ...203..132B}.

\begin{figure}
\subfloat[Subfigure 1 list of figures text][M1: $\rm{H_2}$ wind into $\rm{H_2}$ amb.]{
\includegraphics[width=0.49\columnwidth]{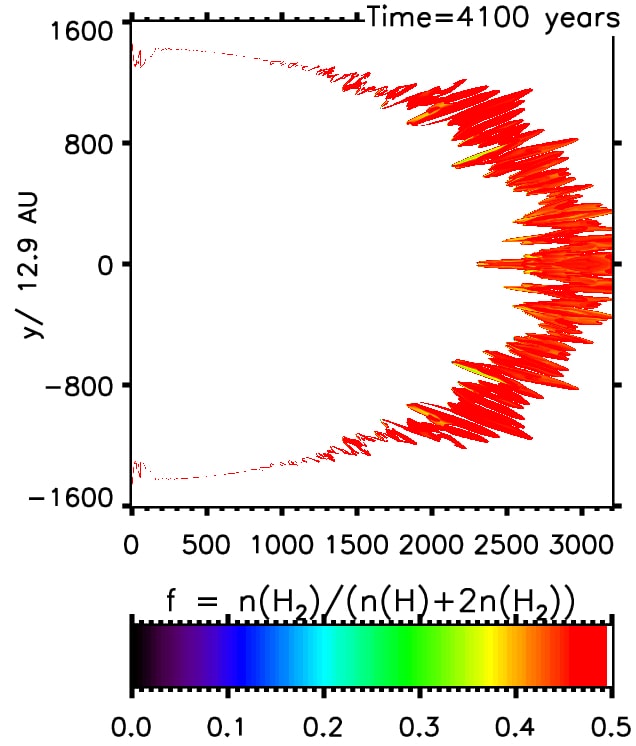}
\label{fig:subfig1a}}
\subfloat[Subfigure 2 list of figures text][M2: $\rm{H_2}$ wind into $\rm{H}$ amb.]{
\includegraphics[width=0.49\columnwidth]{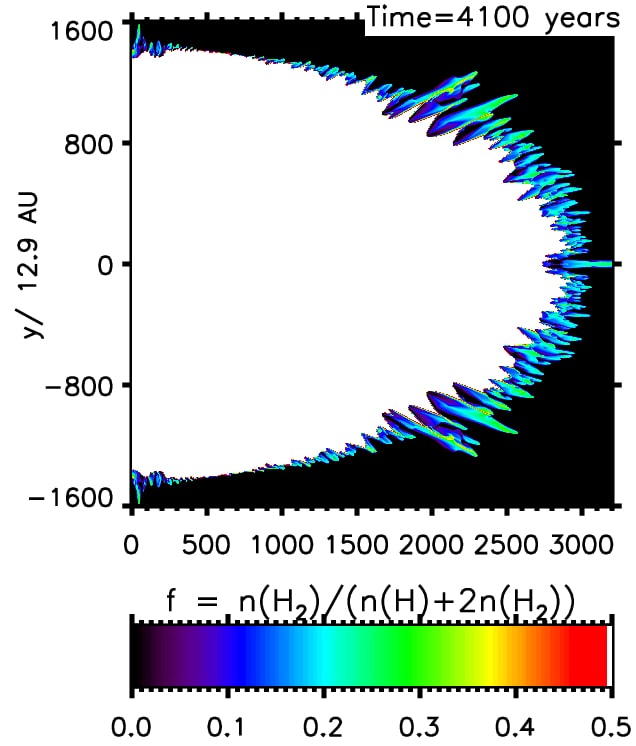}
\label{fig:subfig2a}}
\qquad
\subfloat[Subfigure 3 list of figures text][M3: $\rm{H}$ wind into $\rm{H_2}$ amb.]{
\includegraphics[width=0.49\columnwidth]{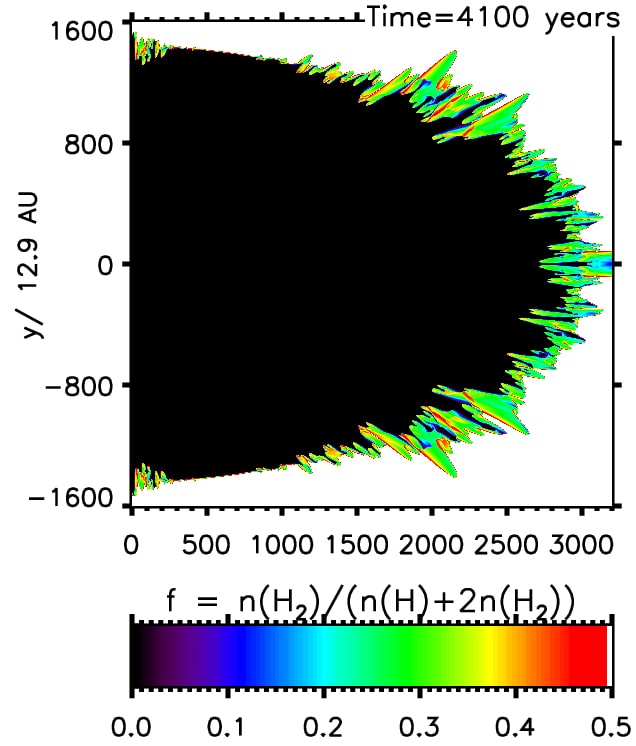}
\label{fig:subfig3a}}
\subfloat[Subfigure 4 list of figures text][H$_2$ mass fractions v. time]{
\includegraphics[width=3.6cm, height=4.8cm]{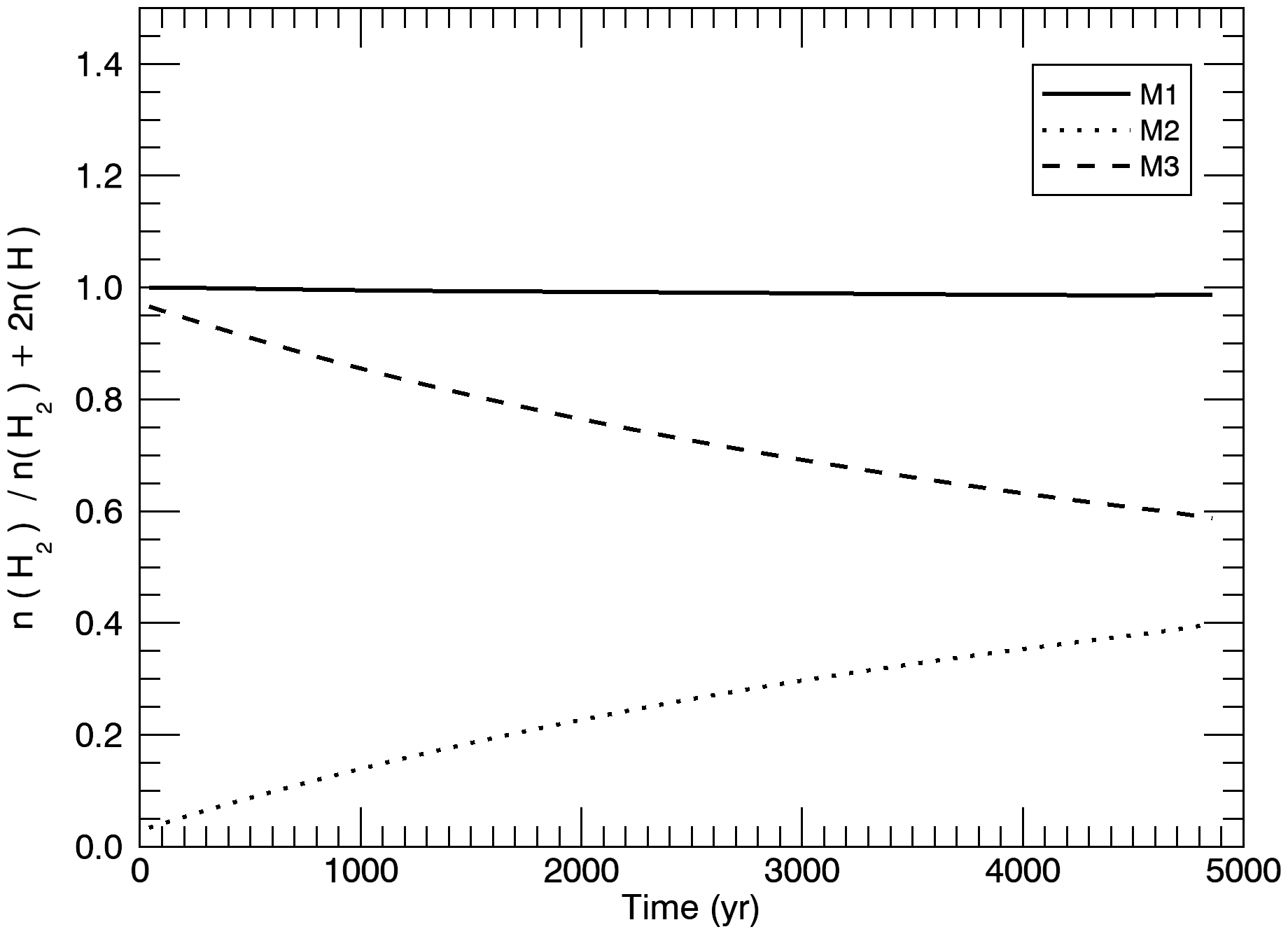} 
\label{fig:subfig4a}}
\qquad
\caption{Cross-sectional distributions of the molecular fraction of 2:1 outflows from a wind with axial speed of 80~km~s$^{-1}$. The displayed region is 0.2\,pc$^2$. The lower-right panel displays the ratio pf total molecular hydrogen mass to the total hydrogen mass as a function of time for the three runs.}
\label{molfrac_2.1_80}
\end{figure}

\begin{figure}
\subfloat[Subfigure 1 list of figures text][M1: $\rm{H_2}$ wind into $\rm{H_2}$ amb.]{
\includegraphics[width=0.49\columnwidth]{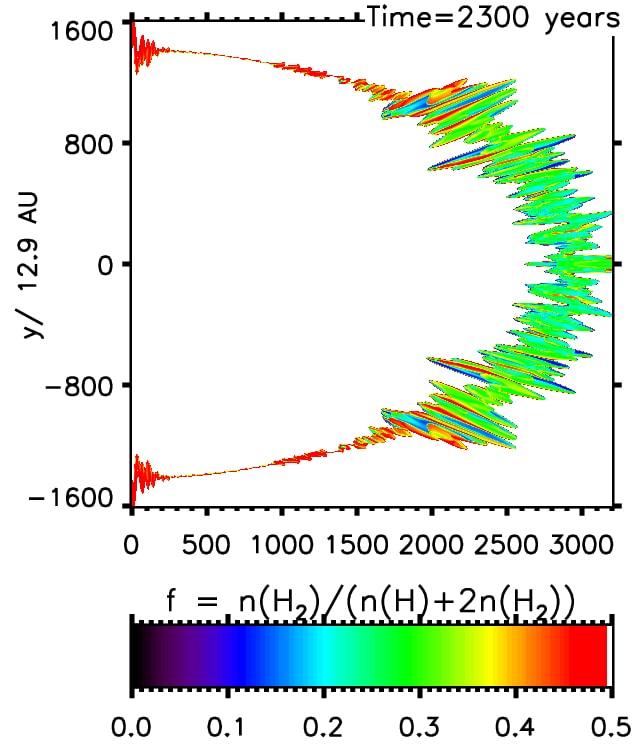}
\label{fig:subfig1a}}
\subfloat[Subfigure 2 list of figures text][M2: $\rm{H_2}$ wind into $\rm{H}$ amb.]{
\includegraphics[width=0.49\columnwidth]{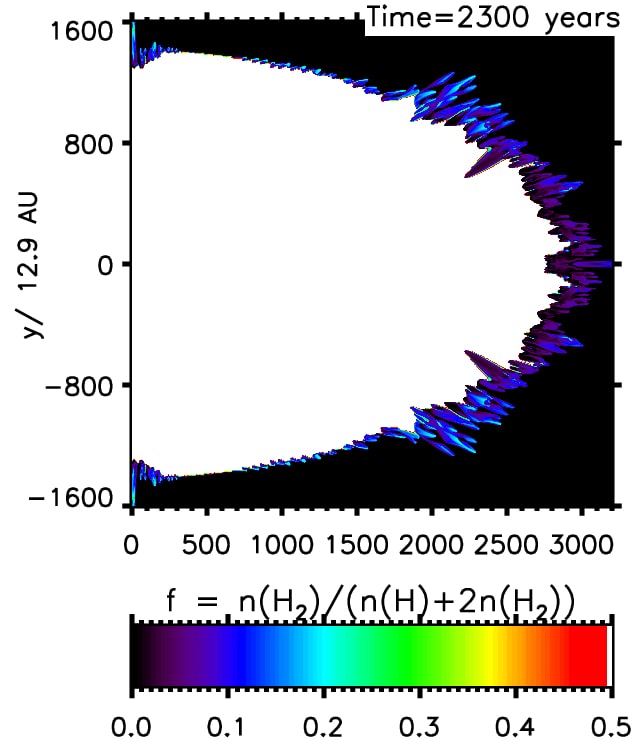}
\label{fig:subfig2a}}
\qquad
\subfloat[Subfigure 3 list of figures text][M3: $\rm{H}$ wind into $\rm{H_2}$ amb.]{
\includegraphics[width=0.49\columnwidth]{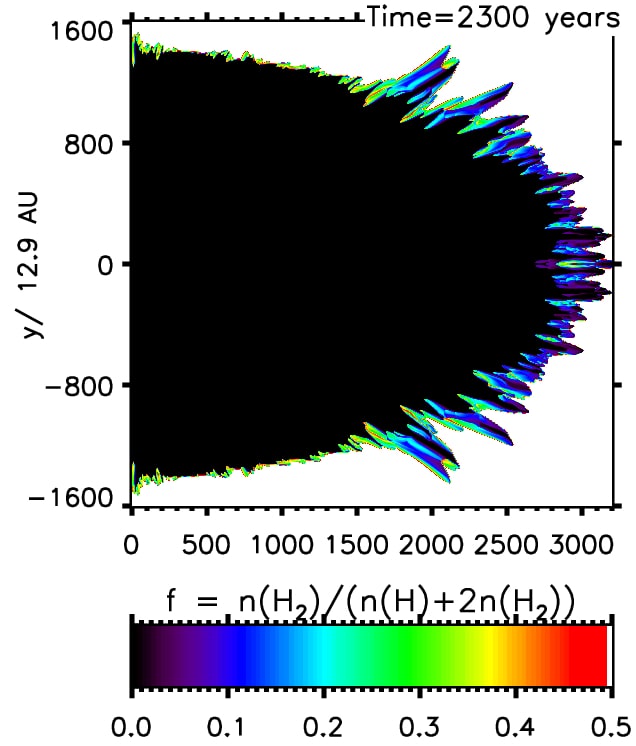}
\label{fig:subfig3a}}
\subfloat[Subfigure 4 list of figures text][H$_2$ mass fractions v. time]{
\includegraphics[width=3.6cm, height=4.8cm]{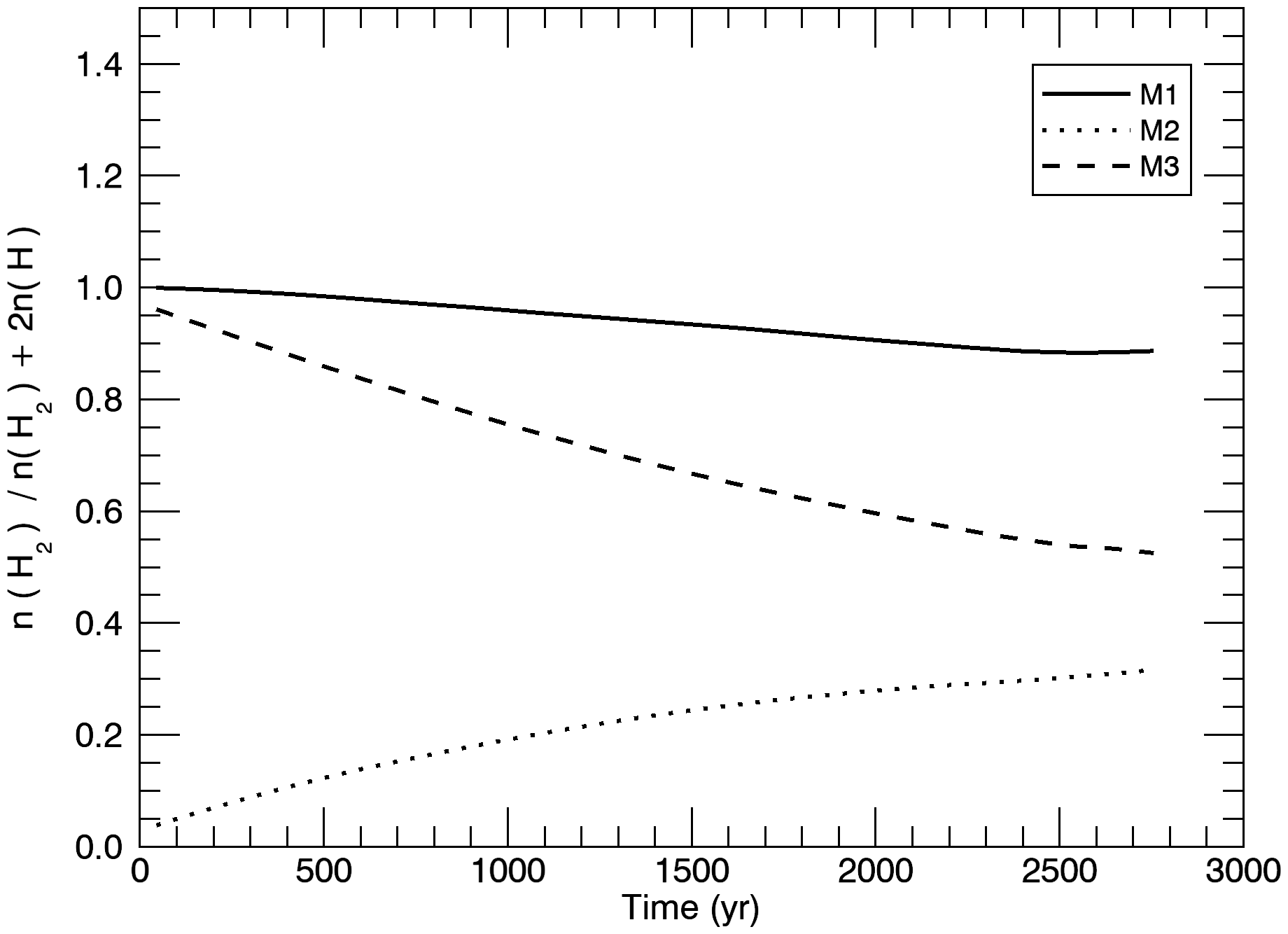} 
\label{fig:subfig4a}}
\qquad
\caption{Cross-sectional distributions of the molecular fraction of 2:1 outflows from a wind with axial speed of 140~km~s$^{-1}$. The displayed region is 0.2\,pc$^2$. The lower-right panel displays the ratio pf total molecular hydrogen mass to the total hydrogen mass as a function of time for the three runs.}\label{molfrac_2.1_140}
\end{figure}

\begin{figure}
\subfloat[Subfigure 1 list of figures text][M1: $\rm{H_2}$ wind into $\rm{H_2}$ amb.]{
\includegraphics[width=0.49\columnwidth]{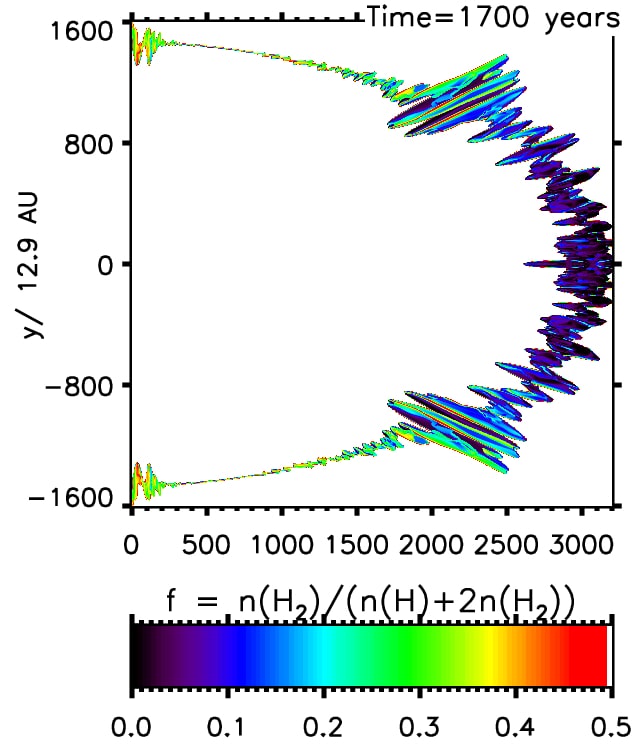}
\label{fig:subfig1a}}
\subfloat[Subfigure 2 list of figures text][M2: $\rm{H_2}$ wind into $\rm{H}$ amb.]{
\includegraphics[width=0.49\columnwidth]{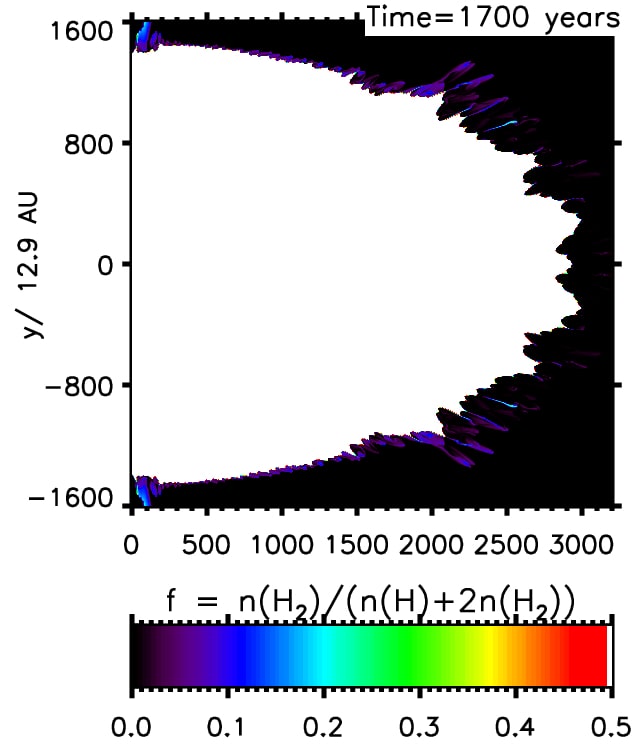}
\label{fig:subfig2a}}
\qquad
\subfloat[Subfigure 3 list of figures text][M3: $\rm{H}$ wind into $\rm{H_2}$ amb.]{
\includegraphics[width=0.49\columnwidth]{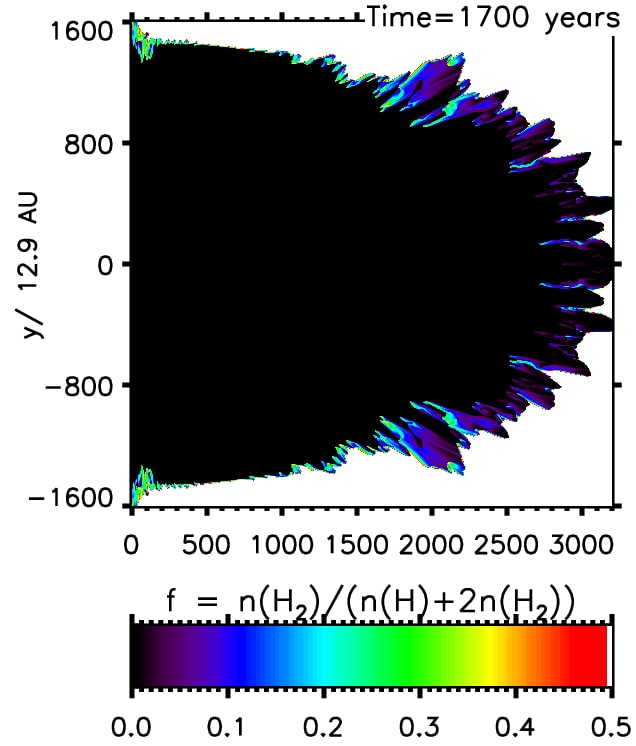}
\label{fig:subfig3a}}
\subfloat[Subfigure 4 list of figures text][H$_2$ mass fractions v. time]{
\includegraphics[width=3.6cm, height=4.8cm]{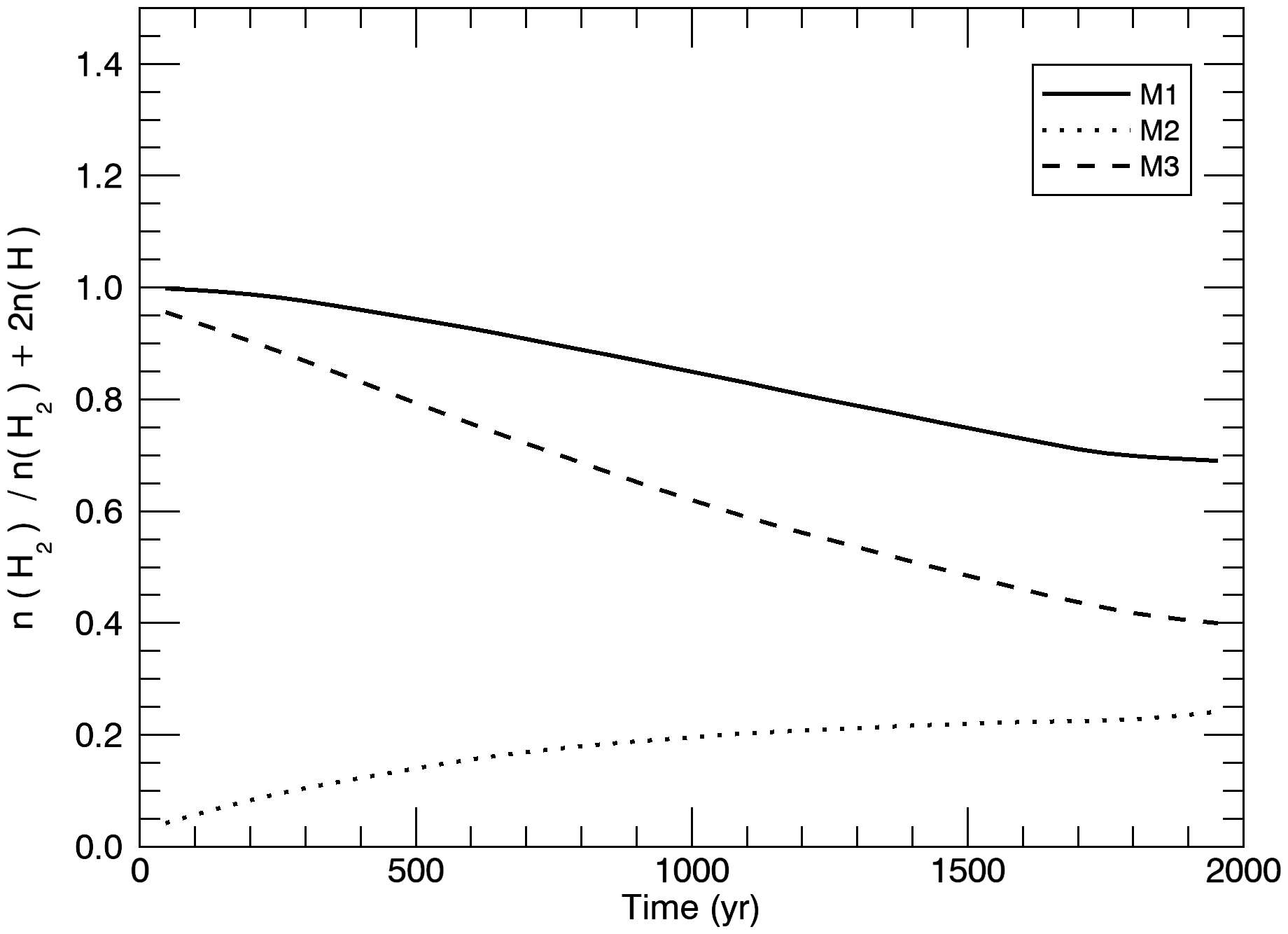} 
\label{fig:subfig4a}}
\qquad
\captionsetup{justification=raggedright,singlelinecheck=false}
\caption{Cross-sectional distributions of the molecular fraction of 2:1 outflows from a wind with axial speed of 200~km~s$^{-1}$. The displayed region is 0.2\,pc$^2$. The lower-right panel displays the ratio pf total molecular hydrogen mass to the total hydrogen mass as a function of time for the three runs.} 
\label{molfrac_2.1_200}
\end{figure}

\begin{figure}
\subfloat[Subfigure 1 list of figures text][M1: $\rm{H_2}$ wind into $\rm{H_2}$ amb.]{
\includegraphics[width=0.45\columnwidth,height=0.20\textheight]{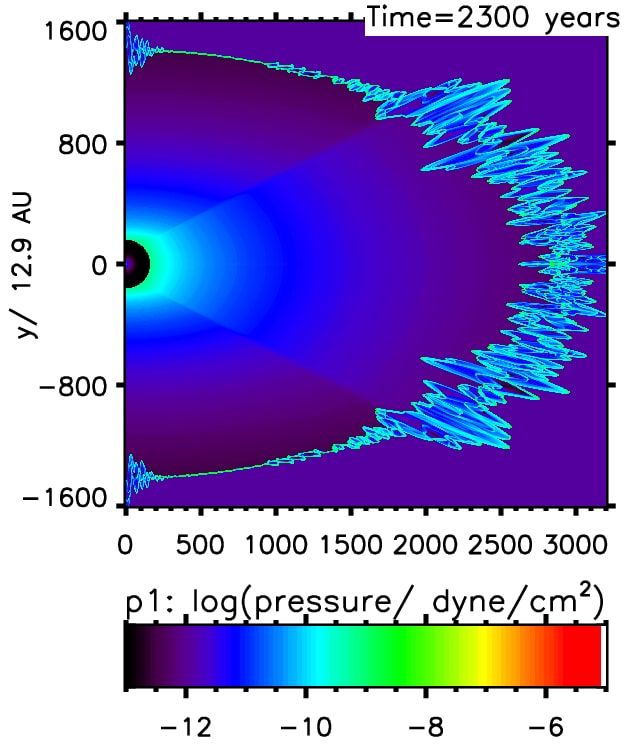}
\label{fig:subfig1a}}
\subfloat[Subfigure 2 list of figures text][M2: $\rm{H_2}$ wind into $\rm{H}$ amb.]{
\includegraphics[width=0.45\columnwidth,height=0.20\textheight]{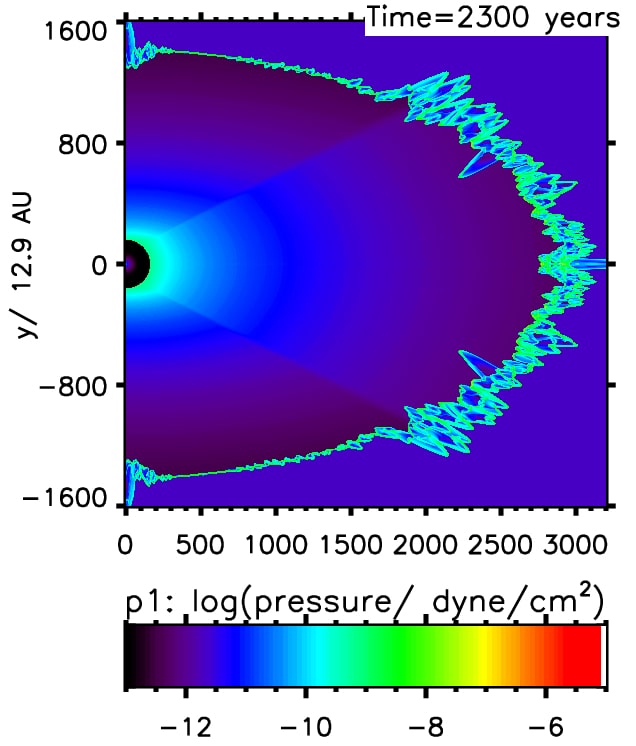}
\label{fig:subfig2a}}
\qquad
\subfloat[Subfigure 3 list of figures text][M3: $\rm{H}$ wind into $\rm{H_2}$ amb.]{
\includegraphics[width=0.45\columnwidth,height=0.20\textheight]{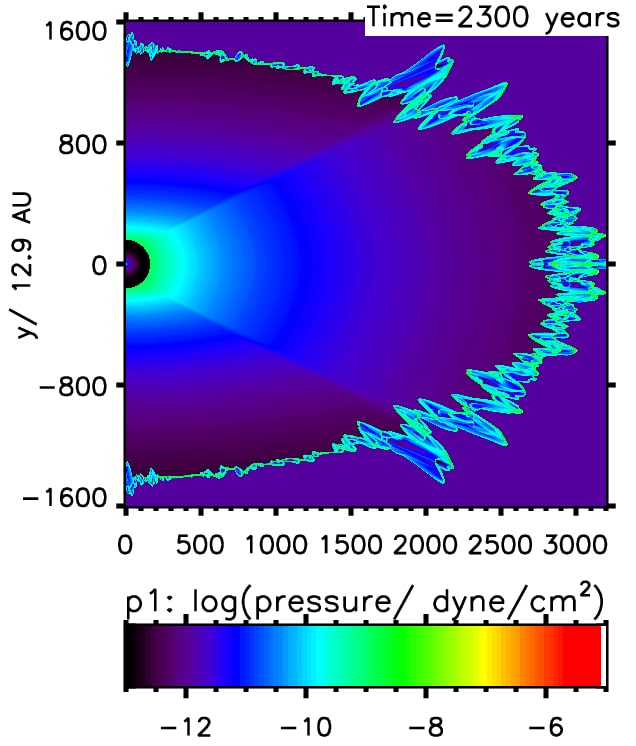}
\label{fig:subfig3a}}
\subfloat[Subfigure 4 list of figures text][M4: $\rm{H}$ wind into $\rm{H}$ amb.]{
\includegraphics[width=0.45\columnwidth,height=0.20\textheight]{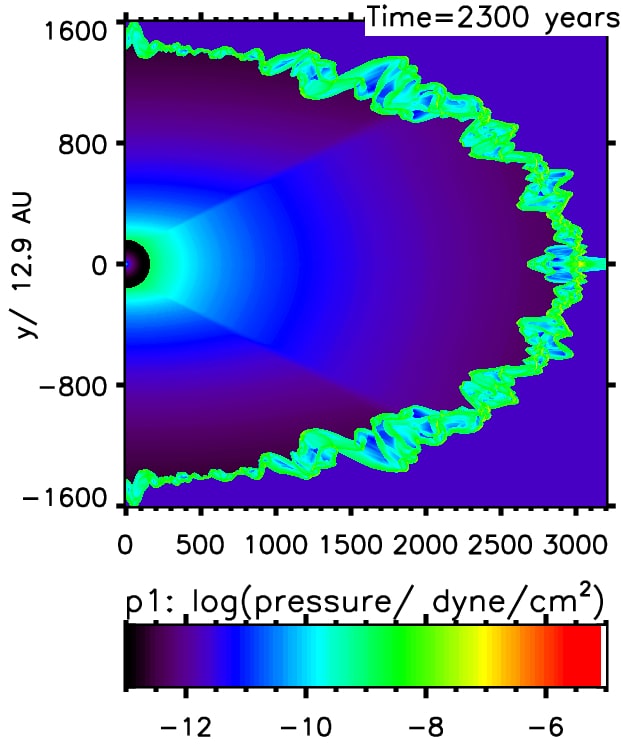}
\label{fig:subfig4a}}
\qquad
\caption{Cross-sectional distributions of pressure produced by 2:1 outflows from a and with axial speed of 140~km~s$^{-1}$. The lower half of each distribution is the mirror image of the top half. A mask of $160$ grid zones is placed over the central wind region to cover up the radius around a bright star.}
\label{pressure21}
\end{figure}

Assuming a stationary ambient medium, the speed of the advancing
shock front in terms of the density ratio $\mathrm{\eta = \rho_{w}/\rho_{a}}$, is worked out through ram-pressure balance:
\begin{equation}
U=\frac{\sqrt{\eta}}{1+\sqrt{\eta}} \,\, v_{w}.
\end{equation} 
For a `heavy' wind with $\mathrm{\eta \gg 1}$,\,and $\mathrm{v_w \gg U}$, the wind ploughs ahead. With the initial value of $\mathrm{\eta_o=25}$, the leading shock is initially strong, moving with $\mathrm{\sim \nicefrac{1}{3}\,{v_{w}}}$. We have thus chosen 
\begin{equation}
n_{a}=4 \times n_{w} \left( \frac{R_{w}} {R_{ppn}}\right) ^{2}.
\end{equation}
So having expanded to $\mathrm{R_{ppn}}$,  $\mathrm{\eta \sim 0.25}$ and $\mathrm{U \sim \frac{1}{3} \,v_{w}}$.Therefore, these simulations focus on the transition phase in which the shell experiences a strong deceleration. 
\begin{equation}
t_{lim}=\int_{R_{w}}^{R_{ppn}} \frac{{d}r }{U} \sim \frac{R^2_{ppn}}{2v_{w}\sqrt{\eta_o}R_{w}}.
\end{equation}
This yields a timescale of order  1,000 years for $\mathrm{R_{ppn} = 0.1}$\,pc and the parameters in Table\,\ref{partable}.

\begin{figure}
\subfloat[Subfigure 1 list of figures text][M1: H$_2$ wind into H$_2$ amb.]{
\includegraphics[width=0.45\columnwidth,height=0.20\textheight]{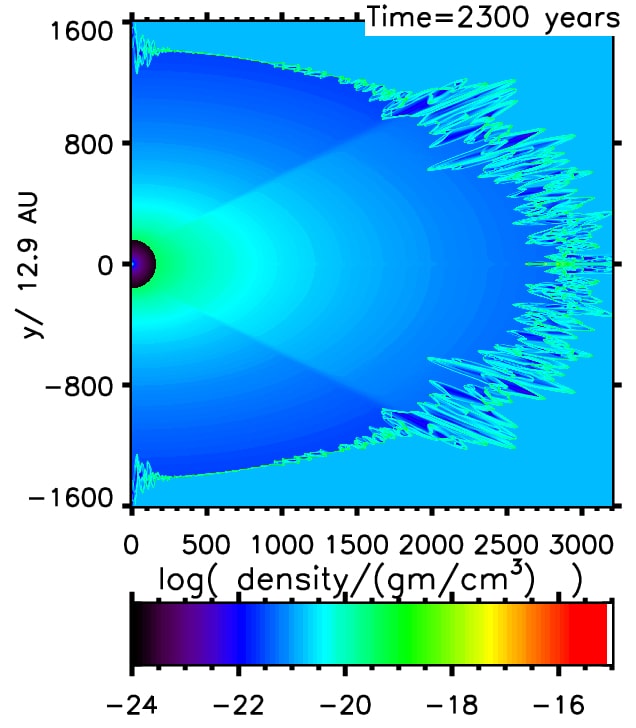}
\label{fig:subfig1b}}
\subfloat[Subfigure 2 list of figures text][M2: $\rm{H_2}$ wind into $\rm{H}$ amb.]{
\includegraphics[width=0.45\columnwidth,height=0.20\textheight]{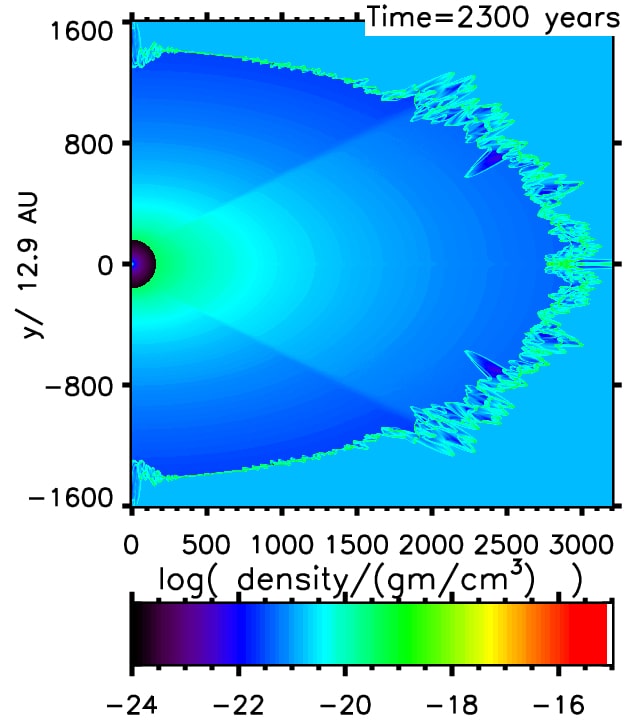}
\label{fig:subfig2b}}
\qquad
\subfloat[Subfigure 3 list of figures text][M3: $\rm{H}$ wind into $\rm{H_2}$ amb.]{
\includegraphics[width=0.45\columnwidth,height=0.20\textheight]{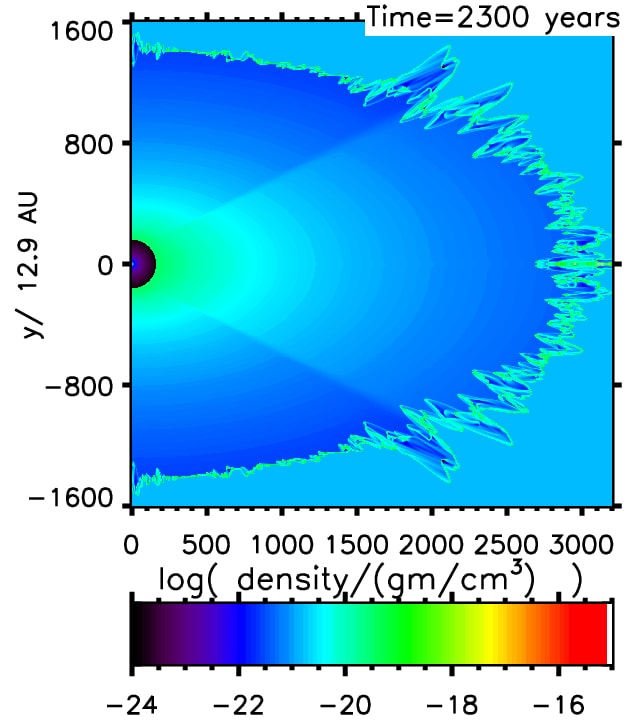}
\label{fig:subfig3b}}
\subfloat[Subfigure 4 list of figures text][M4: $\rm{H}$ wind into $\rm{H}$ amb.]{
\includegraphics[width=0.45\columnwidth,height=0.20\textheight]{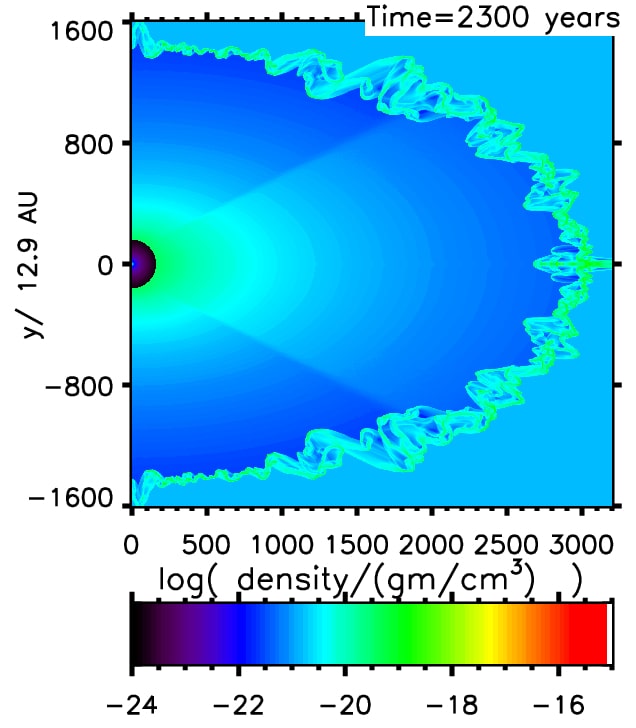}
\label{fig:subfig4b}} 
\qquad
\caption{Cross-sectional distributions of density for $V_w = 140$\,km~s$^{-1}$ winds.} 
\label{density21}
\end{figure}

\begin{figure}
\subfloat[Subfigure 1 list of figures text][M1: $\rm{H_2}$ wind into $\rm{H_2}$ amb.]{
\includegraphics[width=0.45\columnwidth,height=0.20\textheight]{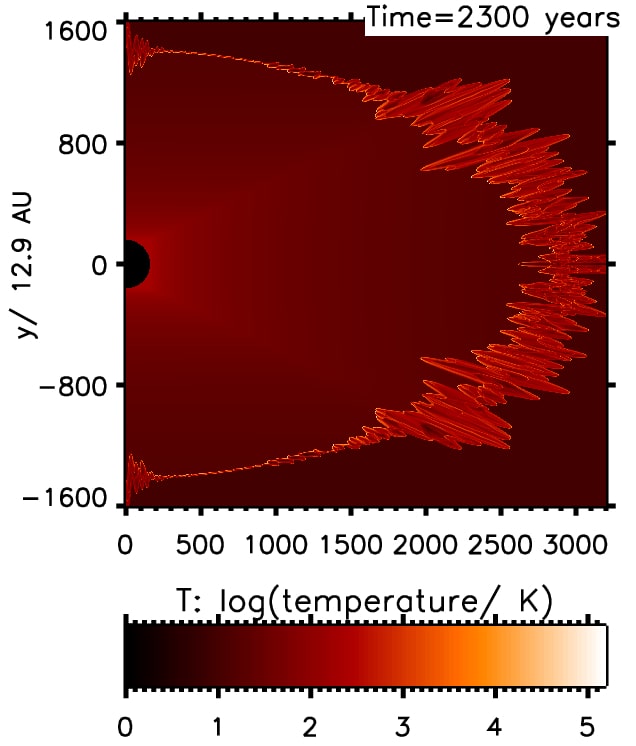}
\label{fig:subfig1c}}
\subfloat[Subfigure 2 list of figures text][M2: $\rm{H_2}$ wind into $\rm{H}$ amb.]{
\includegraphics[width=0.45\columnwidth,height=0.20\textheight]{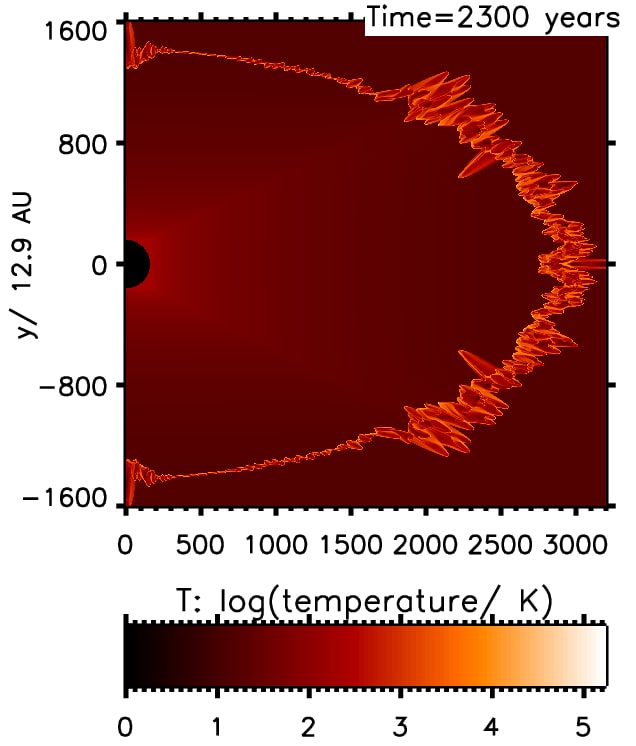}
\label{fig:subfig2c}}
\qquad
\subfloat[Subfigure 3 list of figures text][M3: $\rm{H}$ wind into $\rm{H_2}$ amb.]{
\includegraphics[width=0.45\columnwidth,height=0.20\textheight]{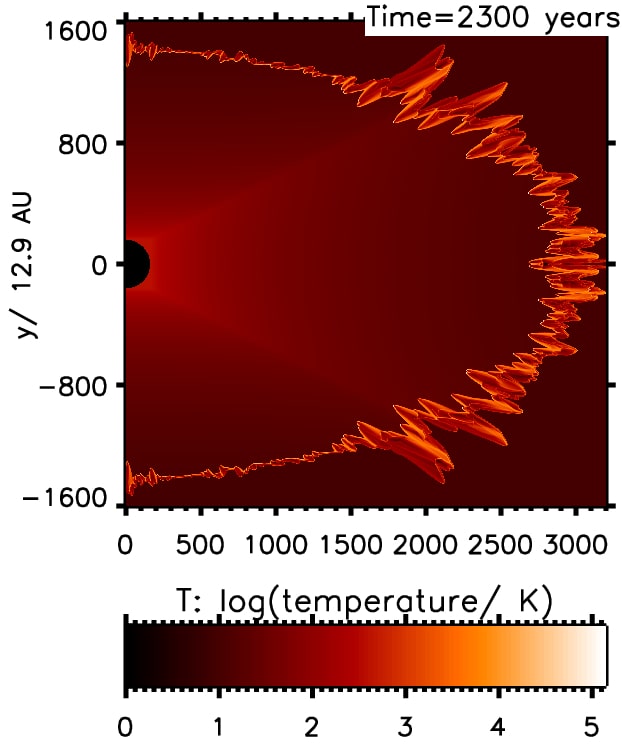}
\label{fig:subfig3c}}
\subfloat[Subfigure 4 list of figures text][M4: $\rm{H}$ wind into $\rm{H}$ amb.]{
\includegraphics[width=0.45\columnwidth,height=0.20\textheight]{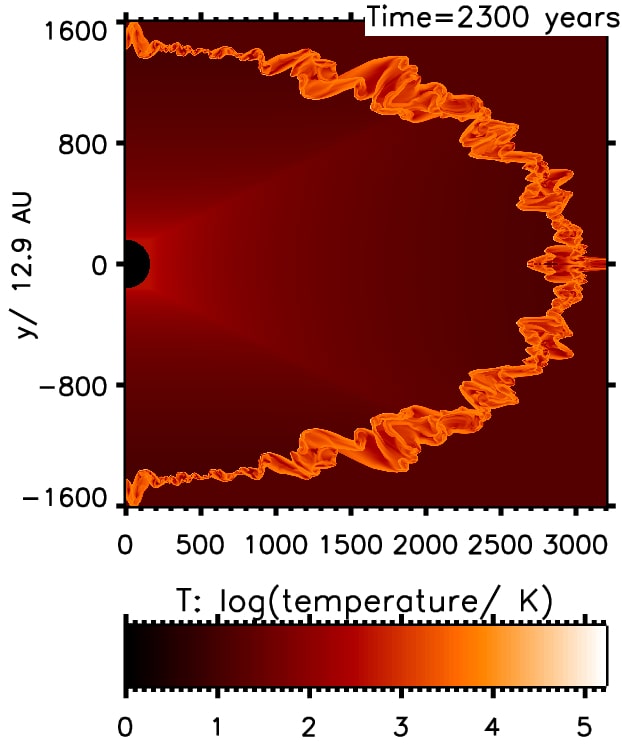}
\label{fig:subfig4c}}
\qquad
\caption{Logarithmic temperature maps of  140~km~s$^{-1}$ 2:1 wind, all scaled to the atomic model (d). Images indicate maximum temperature for molecular run (a) at $\rm{T} \sim 3,000 \, \rm{K}$, in case of atomic run (d), temperature reaches $\rm{T} \sim 20,000 \, \rm{K}$.}
\label{temperature21}
\end{figure}

\begin{figure*}
\includegraphics[width=0.45\textwidth]{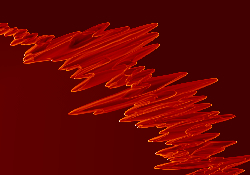} 
\includegraphics[width=0.45\textwidth]{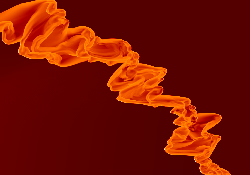}
\qquad
\caption{Detail from the scaled logarithmic temperature maps of 140~km~s$^{-1}$ 2:1 wind, for the pure molecular run (MWMA, left panel) and pure atomic run (AWAA, right panel).}
\label{temperature21zoom}
\end{figure*}

\begin{figure}
\subfloat[ $\mathrm{V_w=80\, kms^{-1}}$]{
        \label{subfig:correct}
        \includegraphics[width=0.99\columnwidth]{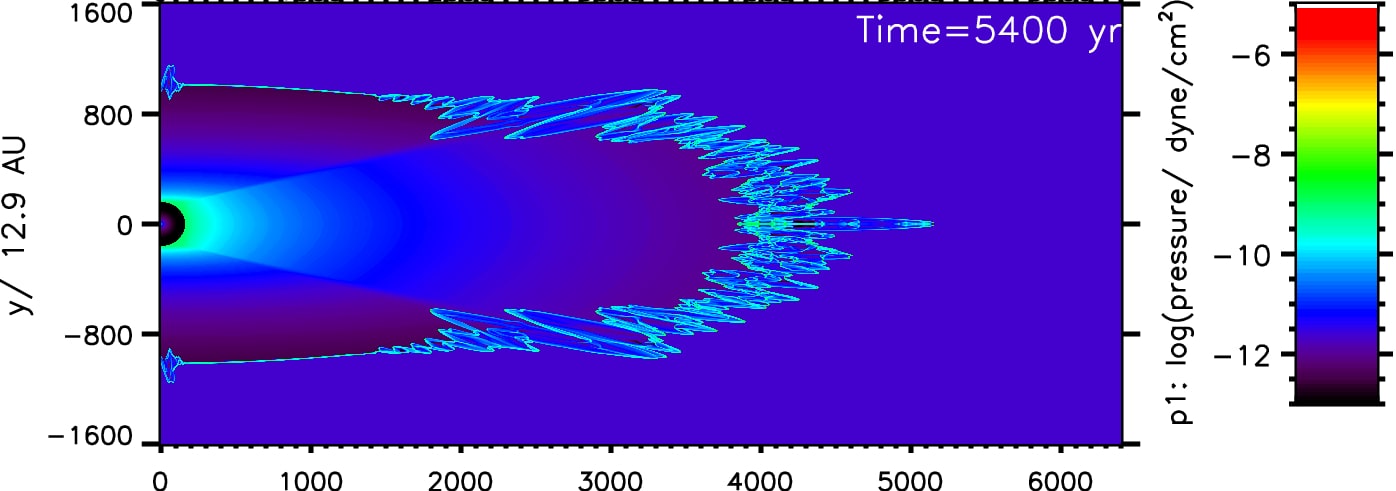} } \\
\subfloat[$\mathrm{V_w=140\, kms^{-1}}$]{
        \label{subfig:notwhitelight}
        \includegraphics[width=0.99\columnwidth]{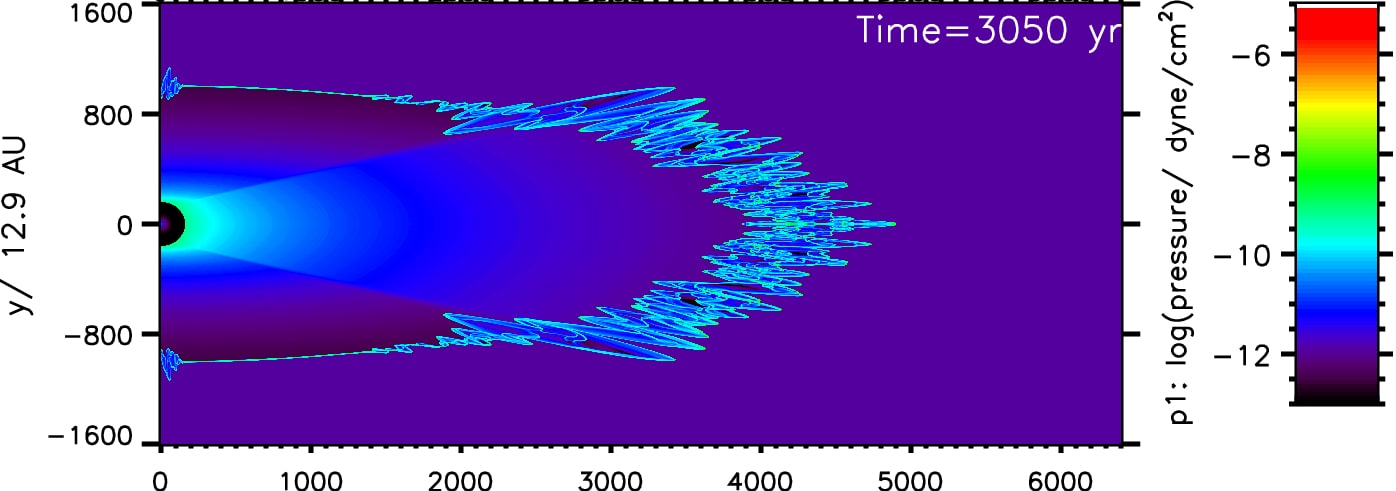} }\\
\subfloat[$\mathrm{V_w=200\, kms^{-1}}$]{
        \label{subfig:nonkohler}
        \includegraphics[width=0.99\columnwidth]{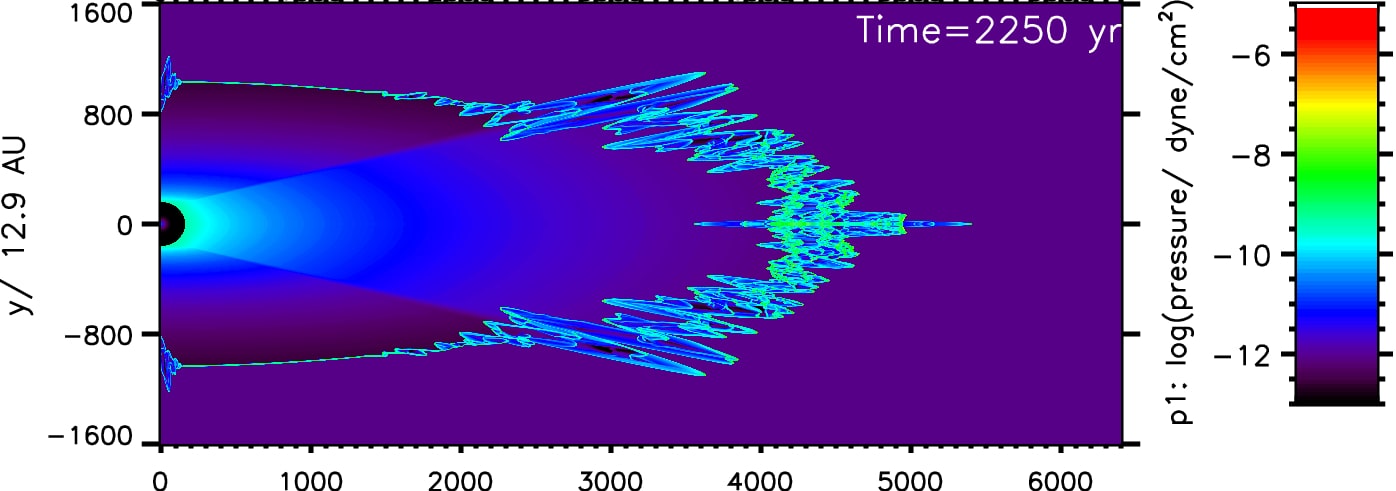}} 
        \caption{Variation of pressure distributions for three axial wind speeds in a collimated 4:1 $\rm{H}$2 wind model interacting with an $\rm{H}$2 ambient medium.}
\label{wind3mwma}
\end{figure}

\begin{figure}
\subfloat[ $\mathrm{V_w=80\, kms^{-1}}$]{
        \label{subfig:correct}
        \includegraphics[width=0.95\columnwidth,keepaspectratio]{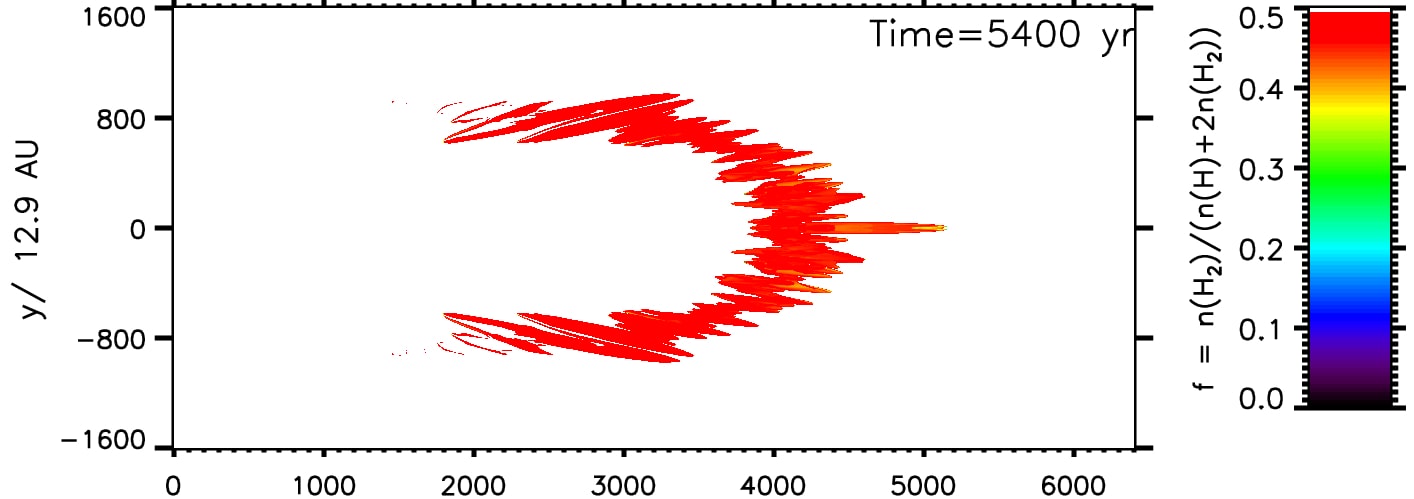} } \\
\subfloat[$\mathrm{V_w=140\, kms^{-1}}$]{
        \label{subfig:notwhitelight}
        \includegraphics[width=0.95\columnwidth,keepaspectratio]{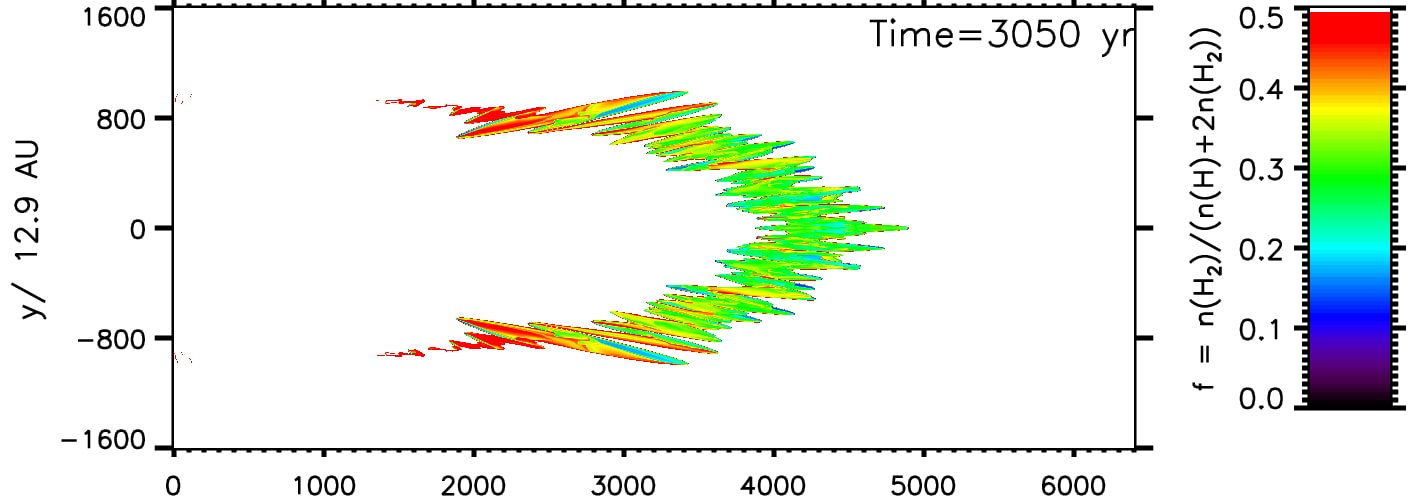} } \\
\subfloat[$\mathrm{V_w=200\, kms^{-1}}$]{
        \label{subfig:nonkohler}
        \includegraphics[width=0.95\columnwidth,keepaspectratio]{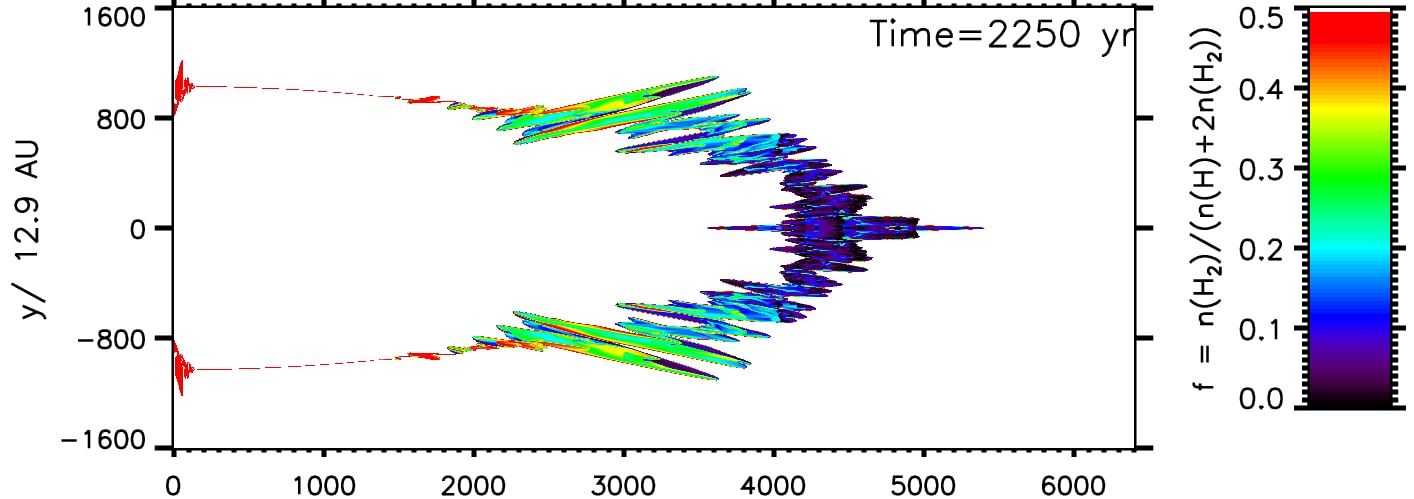}}         
\caption{M1: $\rm{H}$2 wind interacting with an $\rm{H}$2 ambient medium. Corresponding cross-sectional distributions of molecular fraction produced by 4.1 outflows.}
\label{wind4mwma}
\end{figure}

\begin{figure}
\subfloat[$\mathrm{V_w=80\, kms^{-1}}$]{
        \label{subfig:correct}
        \includegraphics[width=0.95\columnwidth,keepaspectratio]{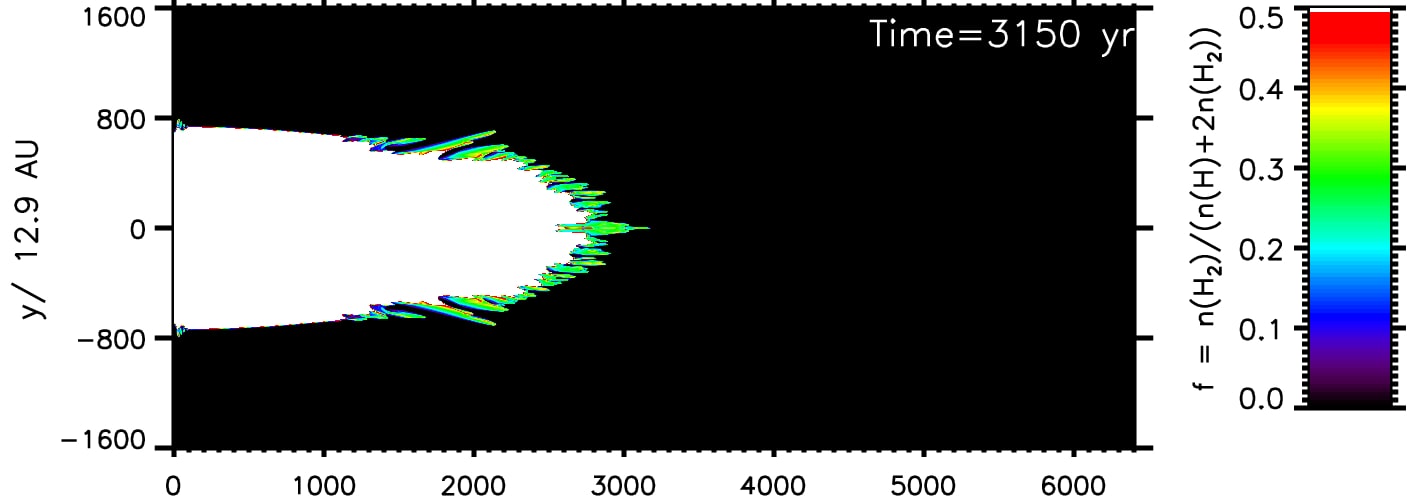} } \\
\subfloat[$\mathrm{V_w=140\, kms^{-1}}$]{
        \label{subfig:notwhitelight}
        \includegraphics[width=0.95\columnwidth,height=0.3\textheight,keepaspectratio]{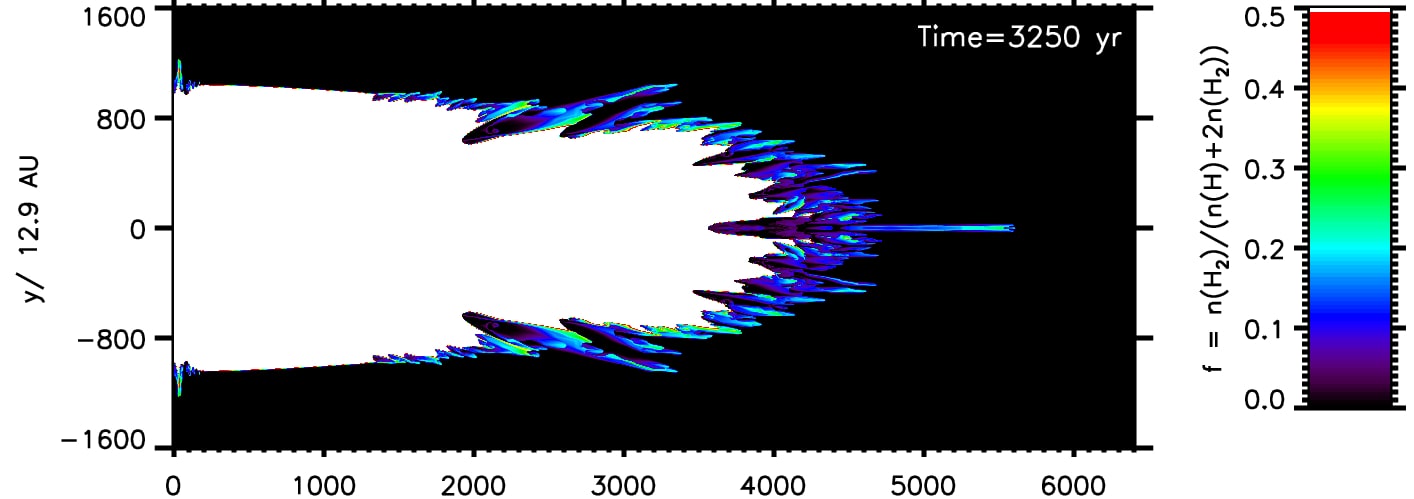} } \\
\subfloat[$\mathrm{V_w=200\, kms^{-1}}$]{
        \label{subfig:nonkohler}
        \includegraphics[width=0.95\columnwidth,height=0.3\textheight,keepaspectratio]{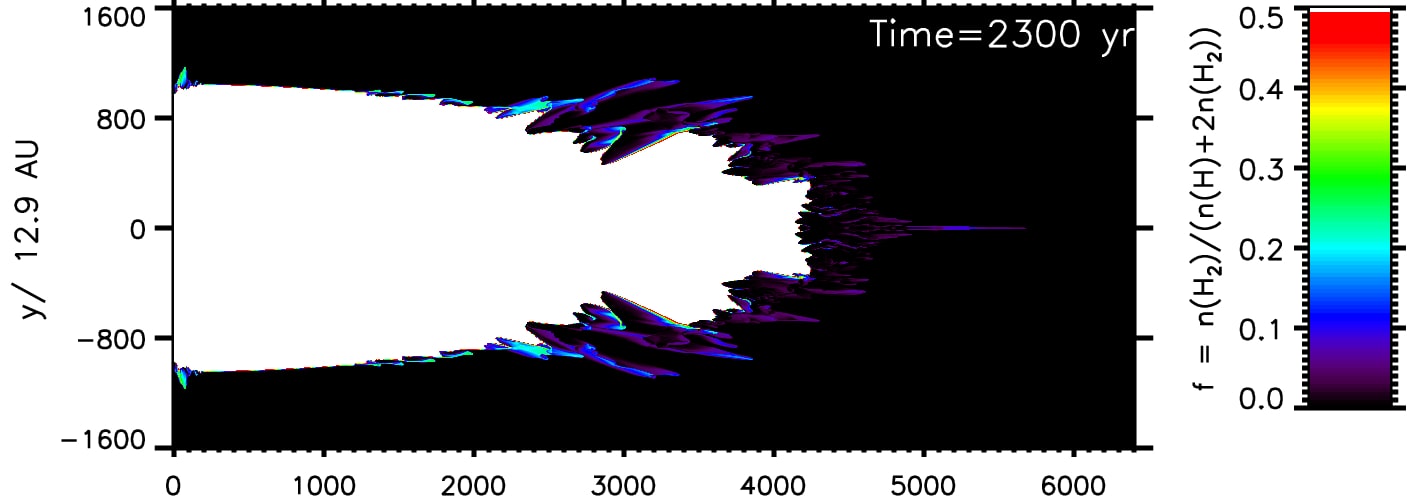}} 
  \caption{M2: $\rm{H}$2 wind interacting with an $\rm{H}$ ambient medium. Corresponding cross-sectional distributions of molecular fraction produced by 4.1 outflows.}      
\label{wind4mwaa}
\qquad
\end{figure}

\begin{figure}
 \subfloat[$\mathrm{V_w=80\, kms^{-1}}$]{
\label{subfig:correct}
\includegraphics[width=0.95\columnwidth]{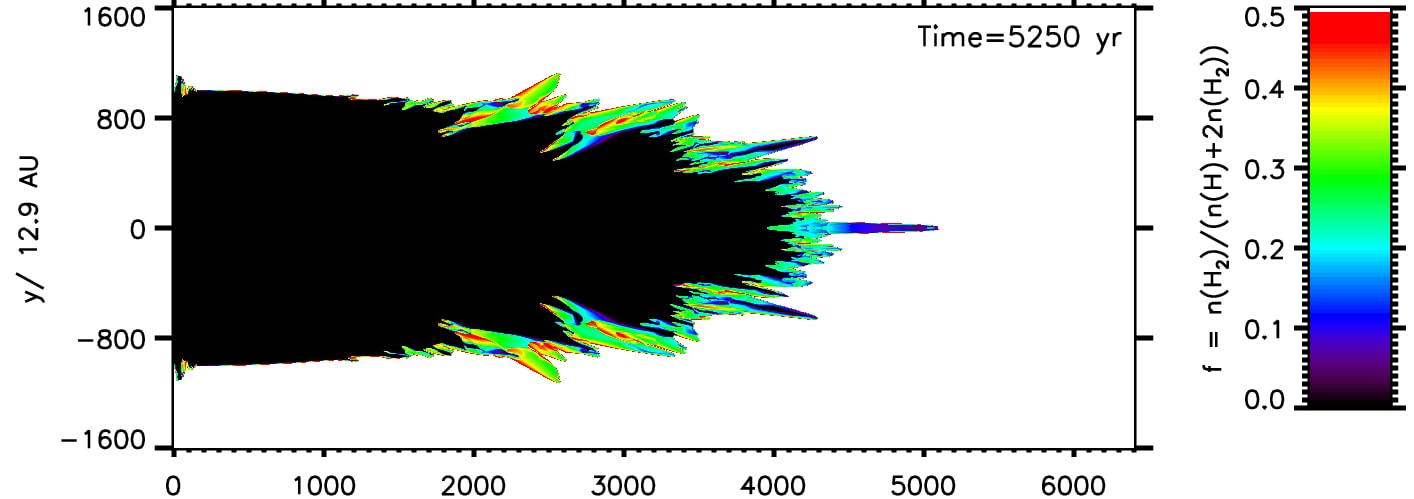} } \\
\subfloat[$\mathrm{V_w=140\, kms^{-1}}$]{
\label{subfig:notwhitelight}
\includegraphics[width=0.95\columnwidth]{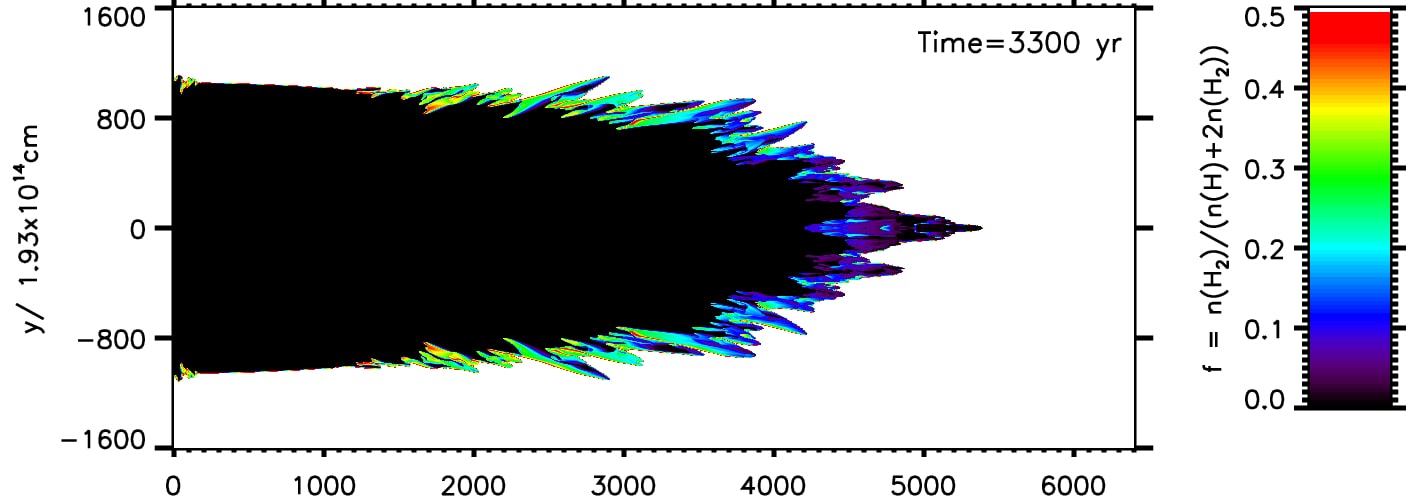} } \\
\subfloat[$\mathrm{V_w=200\, kms^{-1}}$]{
\label{subfig:nonkohler}
\includegraphics[width=0.95\columnwidth]{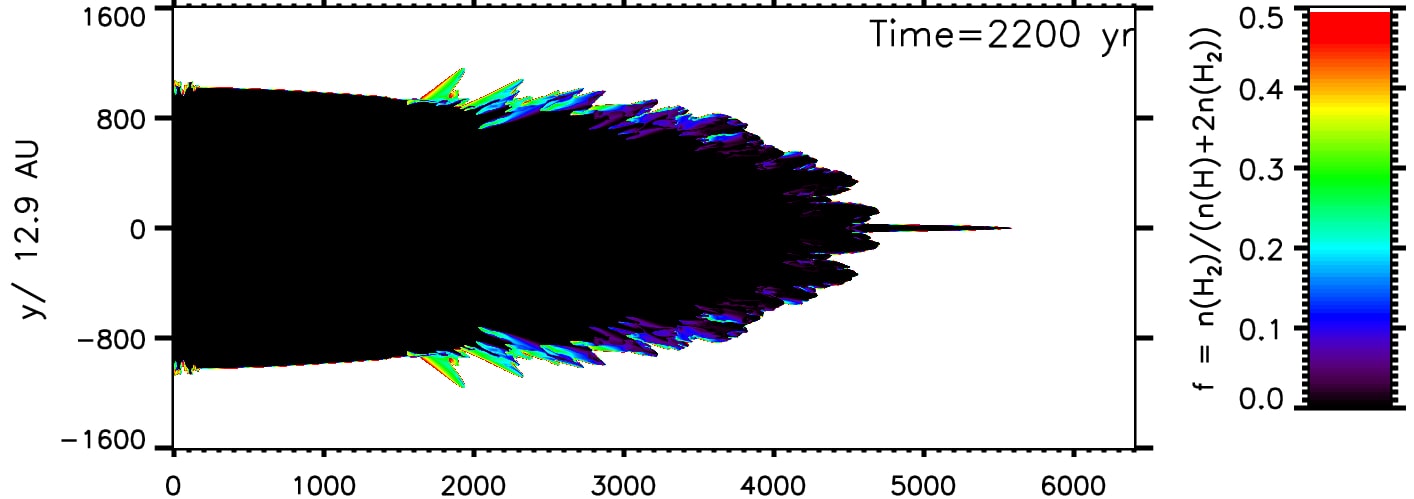}} 
  \caption{M3: $\rm{H}$ wind interacting with an $\rm{H}$2 ambient medium. Corresponding cross-sectional distributions of molecular fraction produced by 4.1 outflows.} 
\label{wind4mwaa}
\end{figure}

\begin{figure*}
\includegraphics[width=0.95\textwidth,height=0.48\textheight]{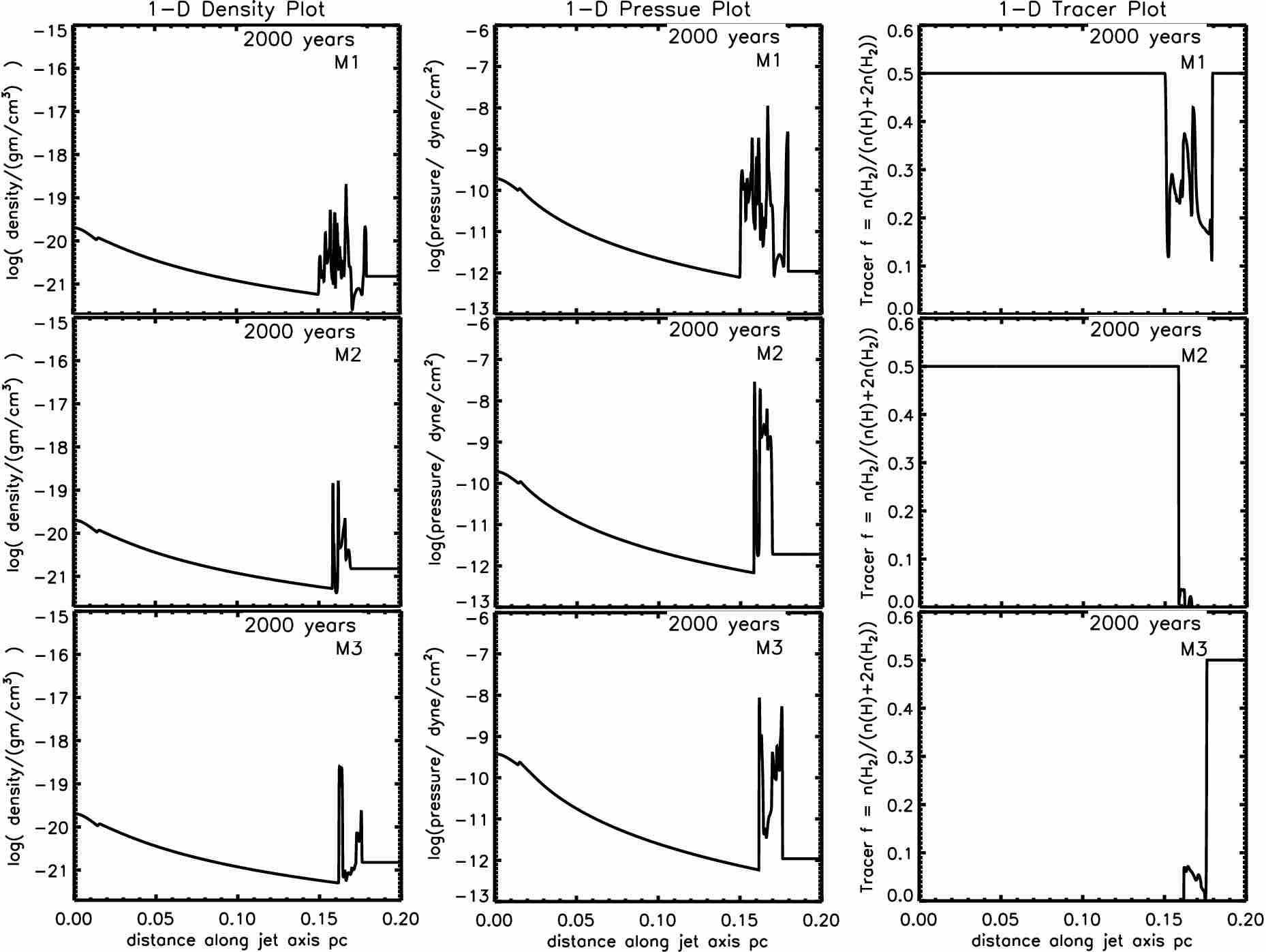}
\hspace{\fill}
\caption[Emission rates]
{One-dimensional plots of the physical parameters for all wind models after 2,000 years of evolution of a 140~km\,s$^{-1}$ wind.}  
\label{merged1D}
\end{figure*}

\begin{figure*}
\includegraphics[width=\textwidth,height=0.20\textheight]{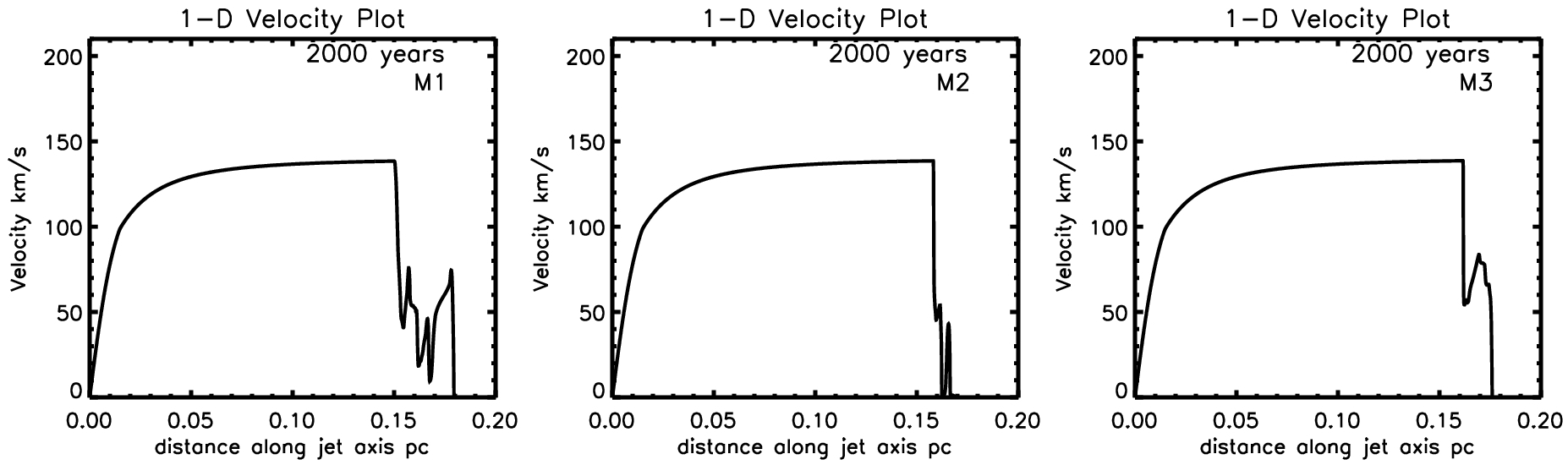}
\hspace{\fill}
\caption[Emission rates]
{One-dimensional velocity plots of the physical parameters for all wind models after 2,000 years of evolution of a 140~km\,s$^{-1}$ wind. The horizontal cut is taken at 200 zones above the symmetry axis in order to avoid on-axis numerical discrepancies. The axial component of the inner part of the wind is thus increasing.}  
\label{merged1D_vel}
\end{figure*}

\section{Results of Parameter Study}
\label{results}

\subsection{Composition: hydrogen dissociation}
All four extreme combinations of atomic and molecular winds into  atomic and molecular ambient media have been studied. First, however, we investigate the three combinations in which the molecular fraction changes as hydrogen partly dissociates within the shell. The atomic case is not interesting in this respect since there is insufficient time for molecules to reassociate on these scales. 

The molecular tracers are examined in Figs.~\ref{molfrac_2.1_80}, \ref{molfrac_2.1_140} and  \ref{molfrac_2.1_200} for the three speeds with a 2:1 elliptical wind. The cases are easily identfied with black representing fully molecular and white corresponding to atomic. It is evident that the shell breaks up into fingers in all these cases. The fingers are prominent toward the front section (with no imposed corrugations, the cylindrical coordinates are better aligned to the sphere approximation toward the symmetry plane).
The top-left panels exhibit the molecular wind-ambient interactions.

We first  take the case in which the axial speed is 80~km~s$^{-1}$, half this speed in the mid-plane, to illustrate the typical flows.
Panel (a) displays the molecular wind into a molecular ambient (denoted MWMA). Clearly, moderate dissociation occurs over the entire shell although the front half of the simulation
shows extended dissociation within a highly turbulent shell, as expected with the higher  jump in speed.  

Most significantly, Figs.~\ref{molfrac_2.1_80}~-~\ref{molfrac_2.1_200} demonstrate  that  the shock front is not a smooth surface but a rough surface, broken down into a large total area of weak oblique shocks with only a small fraction of strong shock apices. This holds for all three wind speeds.  It is also evident that the growing protrusions extend from the interface into the ambient medium. They are generally pockets of wind material channeled into the advancing fingers. Panels (b) and (c) illustrate this for the  MWAA and AWMA cases, respectively, since the make-up of the gas can be ascertained. Until the late stages of the flow, however, the leading shock which enters the atomic ambient medium is of course not visible in the MWAA run.

Panel (c) demonstrates that the molecular ambient gas becomes confined to partly-dissociated thin striations embedded in the atomic wind. In contrast, Panel (b) shows that the molecular wind can protrude quite far into the ambient medium, suggesting that these dense molecular fingers are promoted by the Rayleigh-Taylor instability of the decelerating shell.

At the higher speed shown in Fig.~\ref{molfrac_2.1_200}, we see the dissociation is more complete. The fingers are almost completely washed out at the highest speed by the bath of hot atomic gas although some thick fingers do survive in the MWMA case, a few molecular trunks remain in the wind in the AWMA case, and a few small (blue) wisps are ahead of the main shell in the MWAA case.

A further surprising feature is that the fingers are rarely directed in the radial direction, toward the star. The shells are roughly elliptical with the 2:1 shape as would be expected from the scaling when the shell is driven by a heavy impacting wind. However, at late times as the shell decelerates and is not being driven by a heavy wind, the shape begins to change and the ellipticity decreases. Hence, the wind and ambient fingers are subject to deflection by oblique shocks and will be directed according to the global shape as well as the local conditions.   

The displayed change in total mass in  molecular hydrogen within the 0.1\,pc cylindrical region is instructive. This is shown in terms of the mass fraction in the lower-right panels of Figs.~\ref{molfrac_2.1_80}, \ref{molfrac_2.1_140} and  \ref{molfrac_2.1_200}, giving an indication of the nature of the environment surrounding pPN. For example, at the lower speed, the H$_2$ evolution of MWAA and AWMA appears quite symmetric: the mass enters the environment at a uniform rate and there is a small amount of total dissociation. In contrast, at the high speeds, the molecules added in the wind are dissociated especially at late times when the reverse shock is strong. In summary, the mass fraction in molecules is reduced by 20\% across this range of speeds for all model compositions.

\subsection{Pressure, density and temperature}

The pressure, density  and temperature  distributions for the 2:1 140~km~s$^{-1}$winds are shown in Figs.~\ref{pressure21},   \ref{density21} and \ref{temperature21}. The 2D distributions highlight different sections of the shock structures. The pressure highlights the combined shocked layers, the temperature the immediate shock fronts while the density emphasises the cool compressed gas downstream.  Note that the pressure range shown in the colour bars in Fig.~\ref{pressure21} are identical. Hence, there is a lower pressure in the molecular wind/ambient run, as would be expected with the five degrees of freedom.
 
 These figures confirm that the shock front is  a rough surface, broken down into a large total area of weak shock arcs with only a small fraction of strong shock apices.
The temperature plots demonstrate that the shell is dominated by relatively cool gas globally. However, when inspecting the same images in detail, as shown in Fig.\,\ref{temperature21zoom}, we see that the individual fingers are topped by hot zones: strong shocks are confined to the tips. This result should impact on the H$_2$ excitation that would be measured when observations are made even at quite high resolution. 

It is also seen in Fig.\,\ref{temperature21zoom} that the purely atomic simulation shows different structure to the other three combinations. The AWAA case  displays a thicker  surface which remains turbulent or rougher close to the plane. The temperature distributions displayed in Fig.~\ref{temperature21} provide the interpretation: the shell  of the atomic AWAA run remains warm, just below 10,000 K whereas these layers of the molecular-atomic runs are closer to 1,000 K and a few hundred Kelvin for the MWMA run. Therefore, the atomic layer remains partly inflated and less susceptible to dynamic instabilities. The instabilities responsible for the corrugation of the front shock are not sensitive to the width of the shell. As shown by the resolution study in  Fig.~\ref{resolutions}, the physical width of the shell is independent of numerical resolution.  

\subsection{Structure of collimated winds}

The dependence on  the wind speed is well illustrated with the  4:1 ellipsoidal wind simulations. The molecular simulation is illustrated in Fig.\,\ref{wind3mwma} for three speeds. As expected, the highest pressures occur along the axis and in the fastest winds. The shell structure is best shown in Figs.\ref{wind4mwma} -- \ref{wind4mwaa} for three different axial  wind speeds and for all three composition combinations.

The sensitivity of the molecular fraction to the  wind speed and the distribution around the shell are prominent here. At low speeds, the dissociation remains at a few percentage along the shell. Large atomic bow-shaped shells result when the axial wind speed is high, even when both media are input as molecular. The bow is again broken up into fingers in cross-section. However, when the ambient medium is atomic (see Fig.~\ref{wind4mwaa}), the fingers become long thin molecular streaks which penetrate into the ambient medium and arc toward the axis. These striations, best observed by zooming in on the panels, are partly dissociated. Moreover, the fingers develop into longer streaks at the higher wind speeds.
Generally speaking, the properties are an exaggerated form of the 2:1 case: material does not flow tangentially to the shell but mixes locally.

\subsection{Nebula shaping: local or global flow?}
 
Significant to further simulations is that, in all cases, there is no flow along the shell toward the apex/axis  of the ellipse. This lack of global flow applies to the entire range of wind 
speeds and compositions and implies that flow details can be usefully investigated through simulations with a constricted opening angle. Furthermore, without a coherent  oblique 
shock front to deflect the wind into a high speed shear interface, it is likely that the sound speed in the shell will determine the maximum length scale for 
the growth of disturbances.  Here, there are two main timescales effecting the evolution of the shell morphology: the thermal gas response time at the sound speed, $\mathrm{t_s=\lambda/c_s}$, and the dynamical timescale, $\mathrm{t_d = R_{shell}/U}$.  Hence the maximum wavelength condition is
\begin{equation}
     \lambda \, \textless \, c_s \frac{R_{shell}}{U}, 
\end{equation}
which implies that the warmer atomic gas in the shell is likely to lead to wider and thicker coherent shell structures, as found above. The sound speed in the molecule dominated shell is typically between 0.2 and 0.8 km\,s$^{-1}$ which implies that the fingers may only be a few hundred AU wide. 

The growth rate of disturbances to the shell interface with the ambient medium will be controlled by the Rayleigh-Taylor instability and the Vishniac thin-shell instability 
\citep{1994ApJ...428..186V}. However, most shell simulations have had inadequate resolution to follow these instabilities  in radiative flows. The cylindrical symmetry does make this feasible and, as shown in several studies \citep{1991ApJ...367..619F,1998ApJ...497..267D,2000ApJ...528..989K},  protruding fingers and streaks are present even at relatively low resolution. Indeed, a number of remarkable observations of radial streaks could be interpreted  by the Rayleigh-Taylor instability associated with a decelerating shell \citep{2013MNRAS.429.2366S,2009ApJ...700.1067M}.
 
 The RT growth rate into the non-linear regime is given by $\mathrm{1/t_{RT} = \surd{(gk)}}$ where we approximate  the deceleration as $\mathrm{g \sim U^2/R_s}$ and the wavenumber $\mathrm{k = 2\pi \lambda}$
 \citep{1972MNRAS.156...67B}. This yields a sufficiently short growth time  $\mathrm{t_{RT} = \surd(\lambda/(2\pi R_s))}$ to ensure that the non-linear regime would be reached.

To elucidate the interface structure, we present one-dimensional cuts through the simulations in Fig.~\ref{merged1D}. The molecular tracer in the lower left panels is instructive and indicates how fragments of partially molecular content become trapped in the evolving flow. In fact, inspection of the movies\,\footnote{ \url{http://astro.kent.ac.uk/~in32/ppn.html} } corresponding to  Fig.~\ref{merged1D_vel} shows that the speed of the advancing shock at this location  falls when the molecular fragment is generated.  There is a clear dynamical instability involved. 

While thermal instability may contribute to enhance the growth of denser clumps, in general there is little pressure equilibrium within the interface which is dominated by compressions, rarefactions and sub-hocks as cold dense layers become trapped within warm regions. Moreover, we do not expect thermal instabilities alone to trigger such structure since such instability has been shown to be confined to a very narrow range of shock parameters  \citep{2003MNRAS.339..133S}. 

A classical sawtooth velocity profile distinguishes the shock sandwiches in one-dimensional simulations \citep{1997A&A...318..595S}. These are indeed produced but can only be easily identified early on in the simulations; the shock speed  soon begins to oscillate and generate layers of stronger molecular cooling which propagate from the shock front toward the interface. Remarkable density and pressure variations are then sustained. This is particularly noticeable in the molecular  wind-ambient simulation: rather than hot atomic gas inflating a thick interface, the molecular case results in stronger local pressure and density variations which act to maintain an interaction layer almost twice as thick.

\begin{figure*}
   \subfloat[\label{genworkflow}][$\mathrm{V_w=80\, kms^{-1}}$.]{%
      \includegraphics[width=0.33\textwidth,height=0.25\textheight]{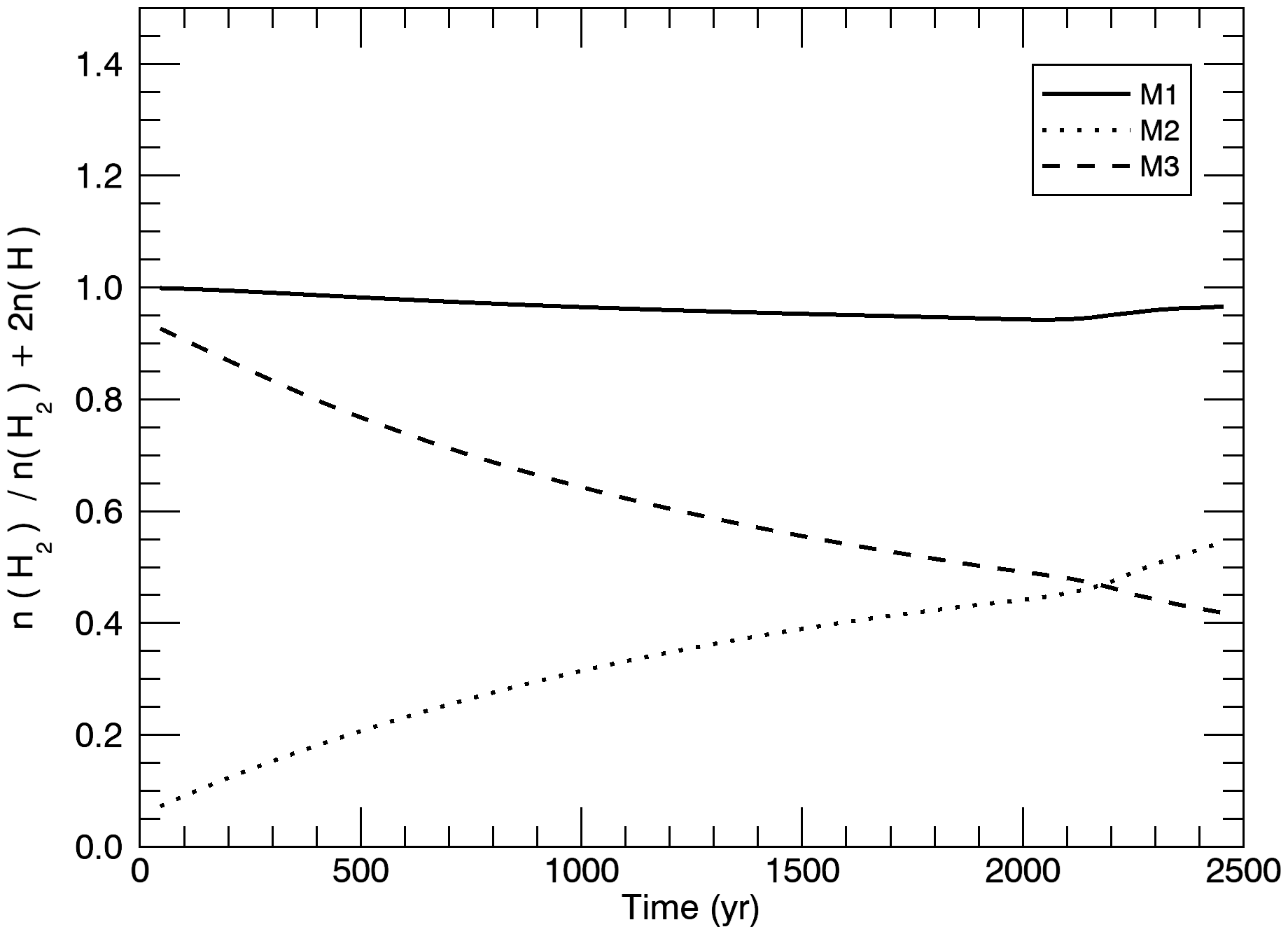}}
\hspace{\fill}
   \subfloat[\label{pyramidprocess}][$\mathrm{V_w=140\, kms^{-1}}$.]{%
      \includegraphics[width=0.33\textwidth,height=0.25\textheight]{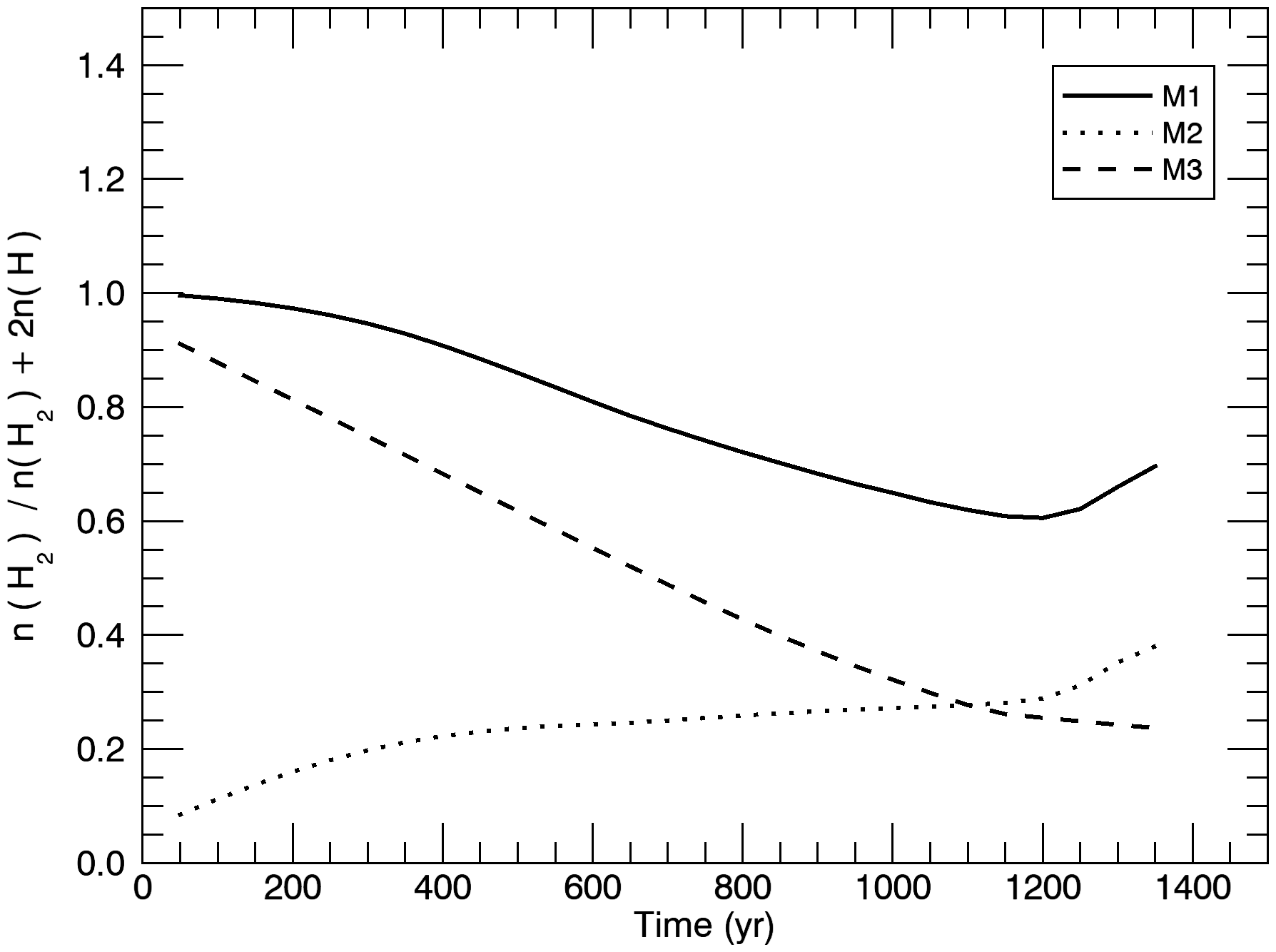}}
\hspace{\fill}
   \subfloat[\label{mt-simtask}][$\mathrm{V_w=200\, kms^{-1}}$.]{%
      \includegraphics[width=0.33\textwidth,height=0.25\textheight]{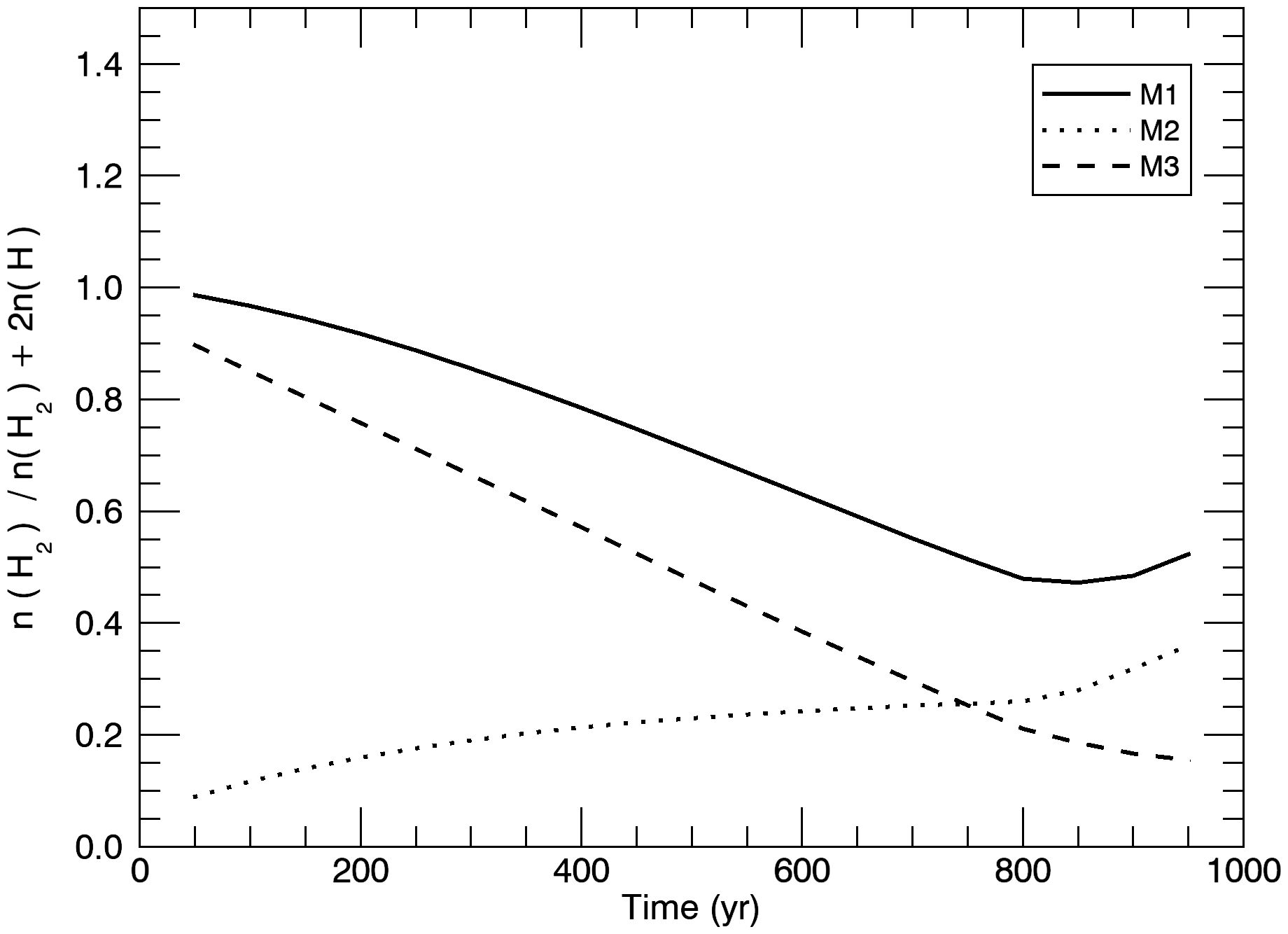}}\\
   \subfloat[\label{genworkflow}][$\mathrm{V_w=80\, kms^{-1}}$.]{%
      \includegraphics[width=0.33\textwidth,height=0.25\textheight]{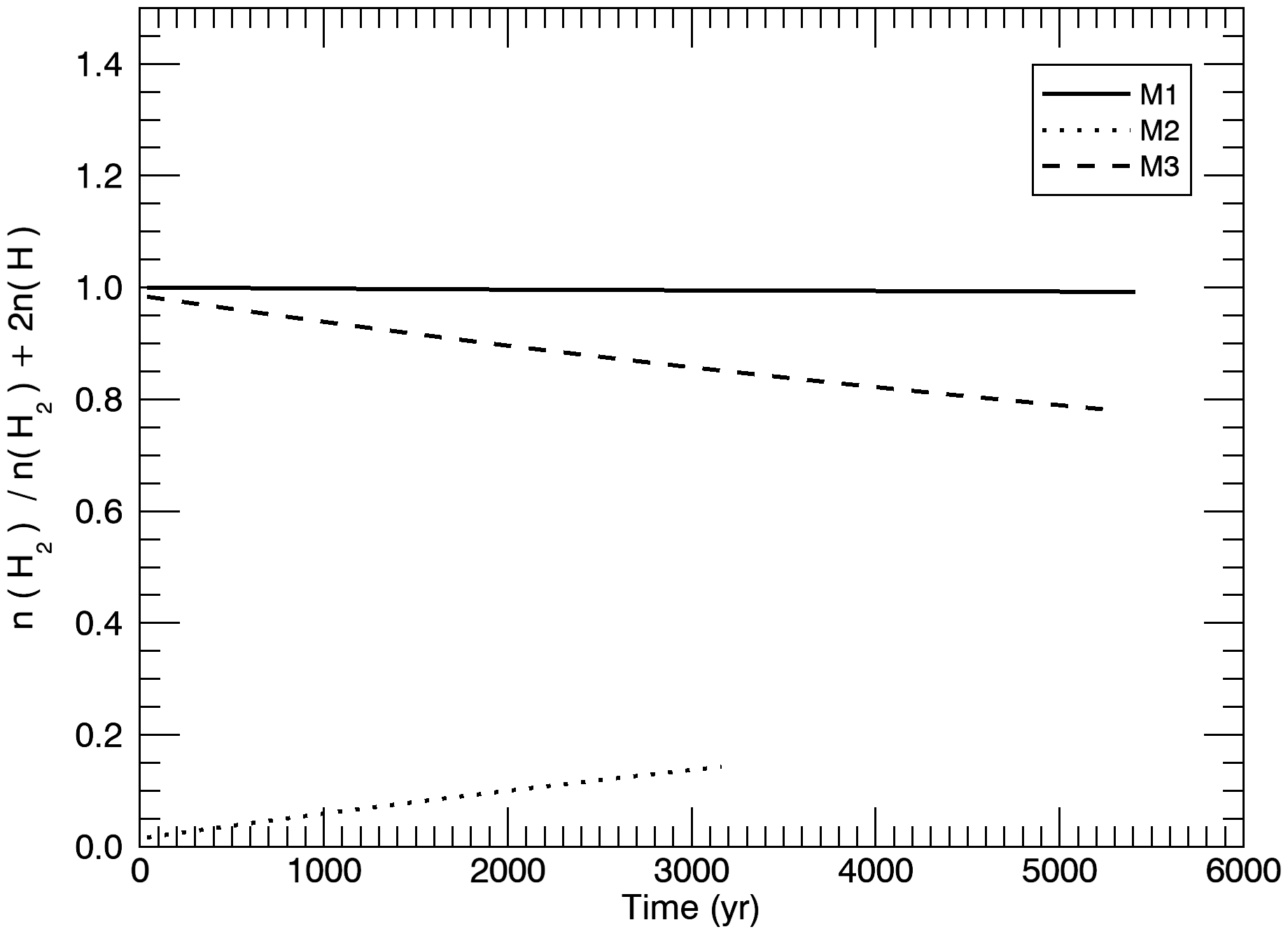}}
\hspace{\fill}
   \subfloat[\label{pyramidprocess}][$\mathrm{V_w=140\, kms^{-1}}$.]{%
      \includegraphics[width=0.33\textwidth,height=0.25\textheight]{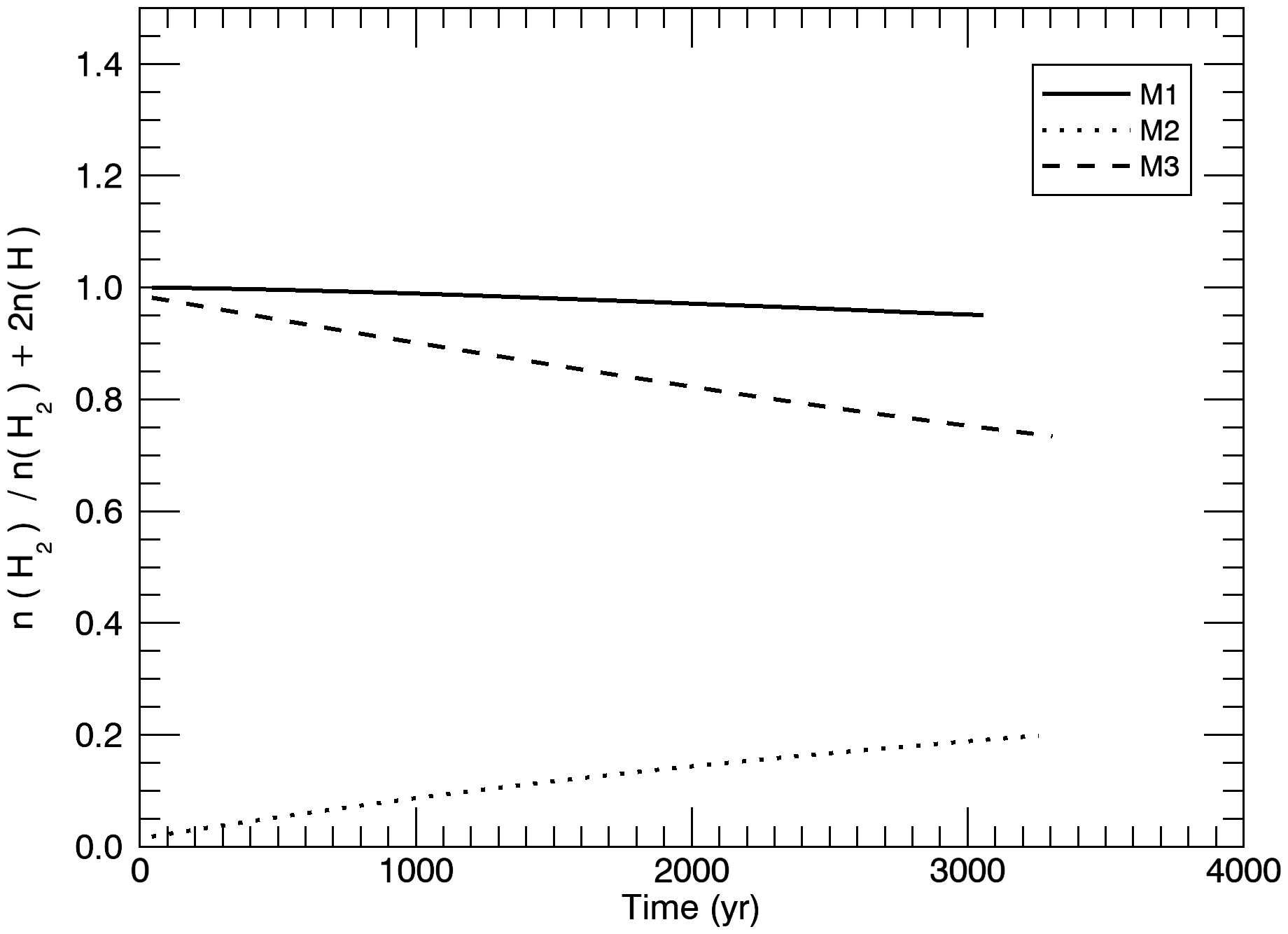}}
\hspace{\fill}
   \subfloat[\label{mt-simtask}][$\mathrm{V_w=200\, kms^{-1}}$.]{%
      \includegraphics[width=0.33\textwidth,height=0.25\textheight]{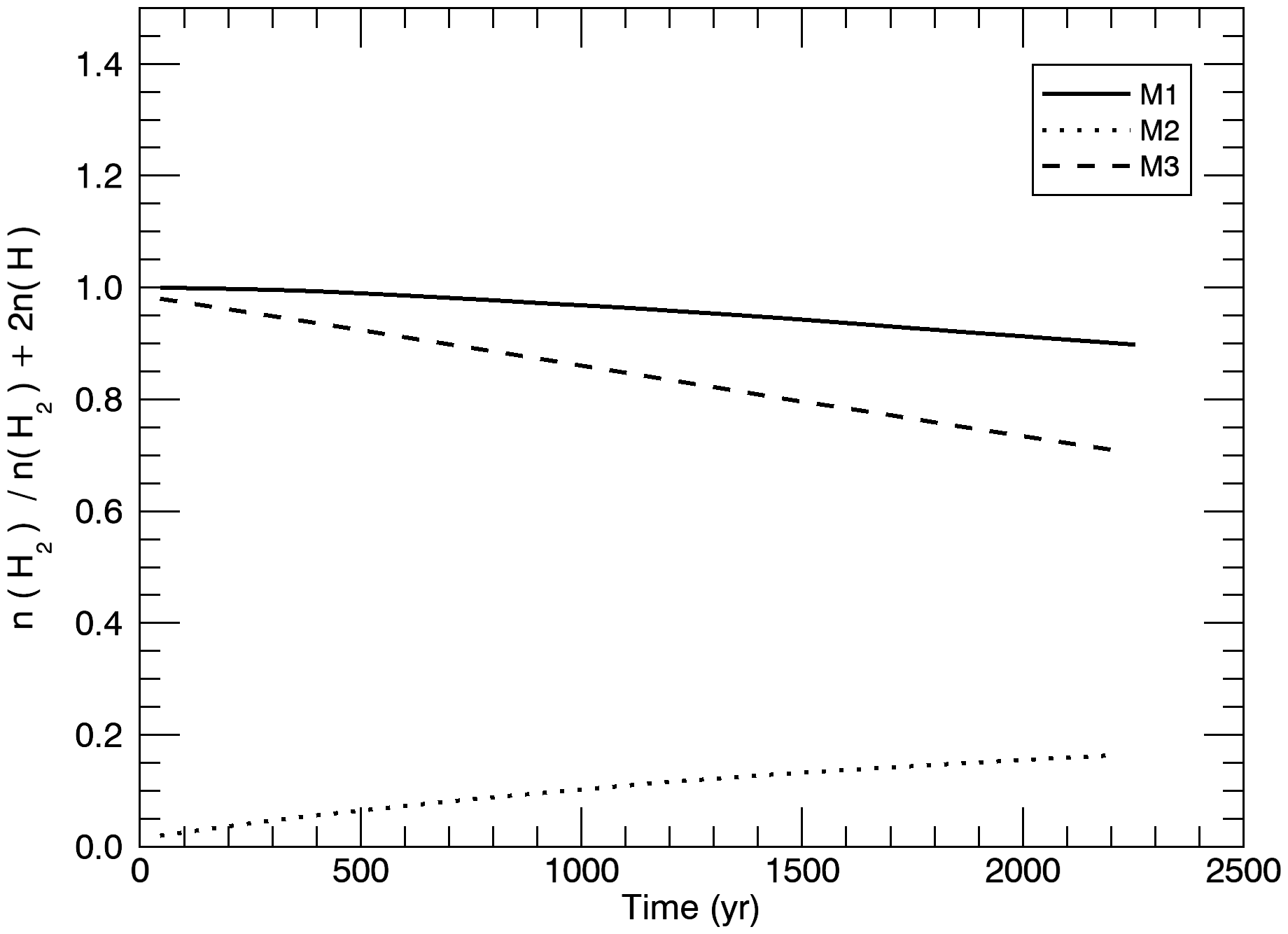}}\\\caption{\label{molfrac1.1}The evolution  of the total molecular fraction integrated over the grid as produced by a spherical wind (upper panels) and a 4:1 wind (lower panels) with M1, M2 and M3 outflows, at the three indicated wind speeds from left to right.}
\end{figure*}

Finally, we determine the integrated molecular fraction in Fig.~\ref{molfrac1.1} where the particle number fraction is presented. As the shock interface sweeps across the grid, the wind contents replaces the ambient medium with a significant fraction of the molecular medium being dissociated. Toward the end of the M1 (MWMA) run, the rate at which new molecular wind enters the grid 
is faster than the dissociation rate and the decrease in molecules is reversed. This analysis also clearly demonstrates that the sum of fractions for Models M2 (MWAA) and M3 (AWMA) at any given time will approximately yield the fraction for model M1.

\begin{figure}
\vspace{-12pt}
\subfloat[Subfigure 1 list of figures text][M1: early time, 700 yr.]{
\includegraphics[width=0.5\columnwidth,height=0.45\textheight]{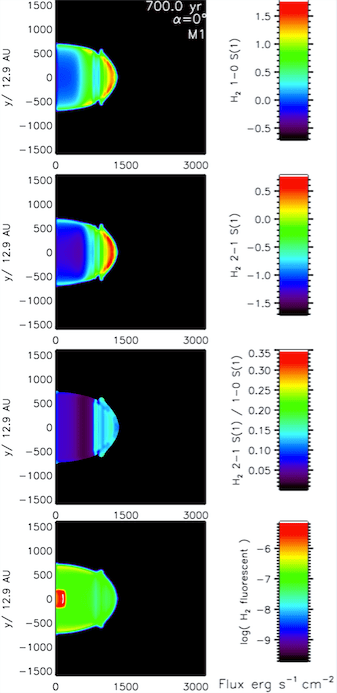}
\label{fig:subfig1e}}
\subfloat[Subfigure 2 list of figures text][M1: late time, 2,300 yr.]{
\includegraphics[width=0.5\columnwidth,height=0.45\textheight]{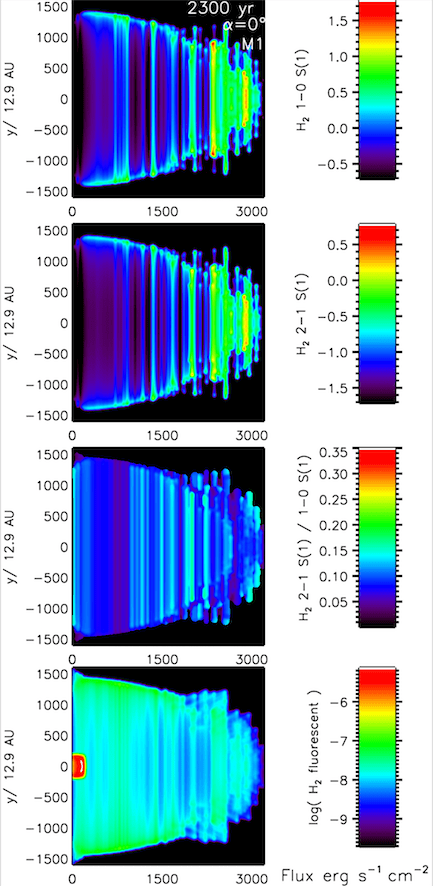}
\label{fig:subfig2e}}
\caption{H$_2$ line emission maps for a molecular wind impacting a molecular medium. Model M1: with  a 2:1 $V_{w}\sim 140 \,km\, s^{-1}$ wind in the plane of the sky at the indicated times.}
\label{M1a-H2-21-4panel}
\qquad
\end{figure}
 


\begin{figure}
\subfloat[M1: $\rm{H_2}$ wind interacting with an $\rm{H_2}$ ambient medium.]{
        \label{subfig:correct}
        \includegraphics[width=1.0\columnwidth,height=0.3\textheight,keepaspectratio]{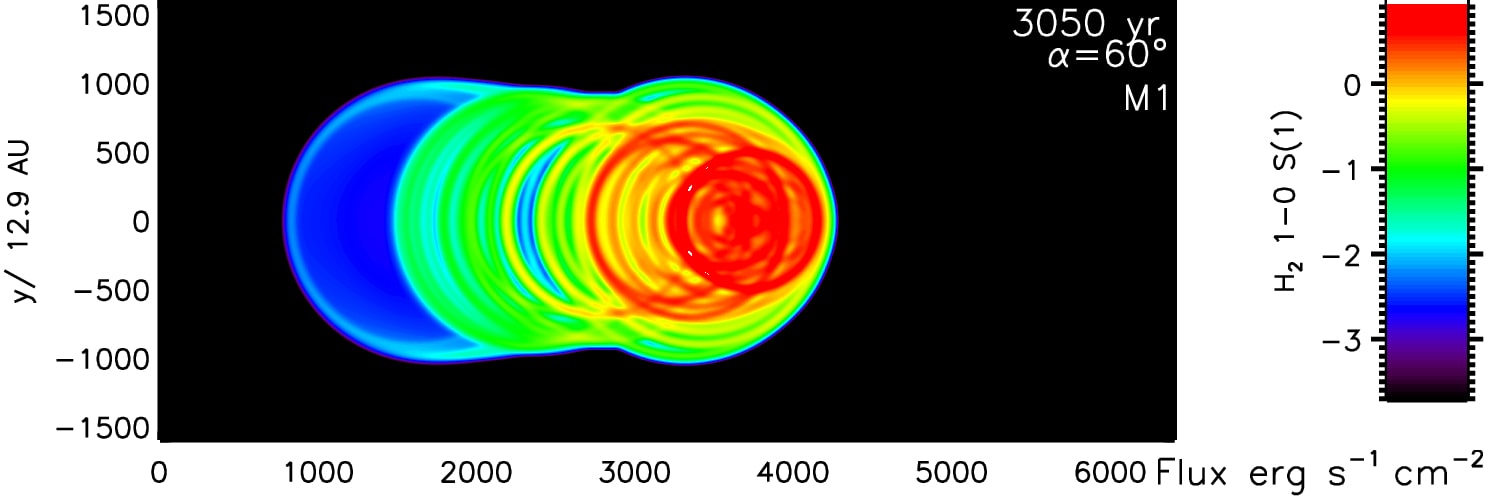} } \\
\subfloat[M2: $\rm{H_2}$ wind interacting with an $\rm{H}$ ambient medium.]{
        \label{subfig:notwhitelight}
        \includegraphics[width=1.0\columnwidth,height=0.3\textheight,keepaspectratio]{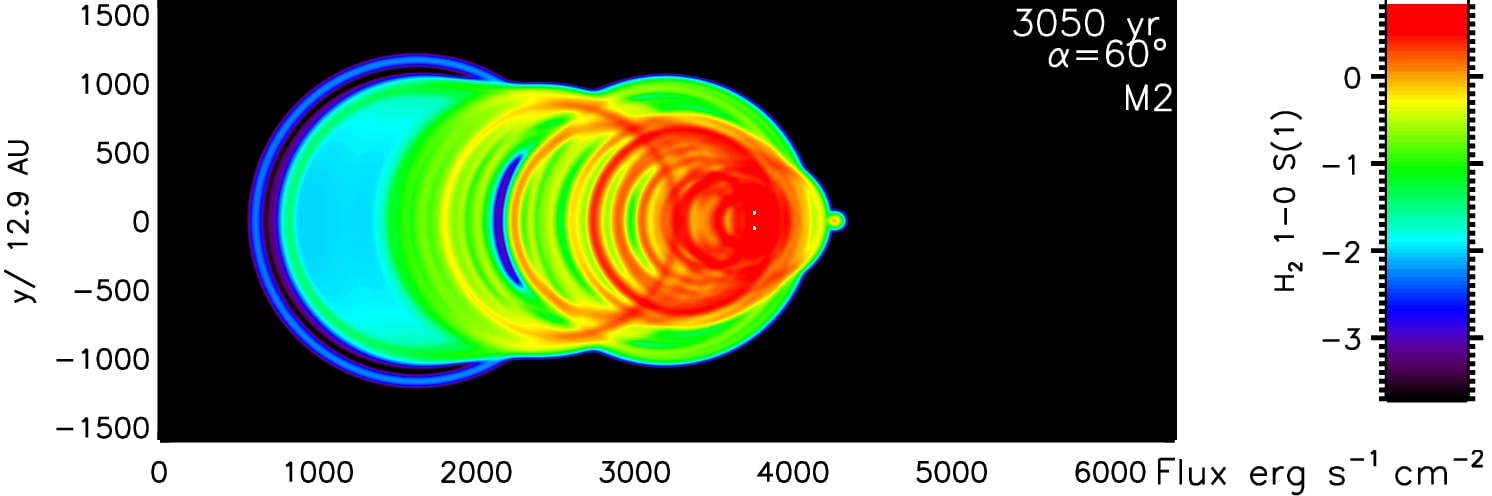} }\\
\subfloat[M3: $\rm{H}$ wind interacting with an $\rm{H_2}$ ambient medium.]{
        \label{subfig:nonkohler}
        \includegraphics[width=1.0\columnwidth,height=0.3\textheight,keepaspectratio]{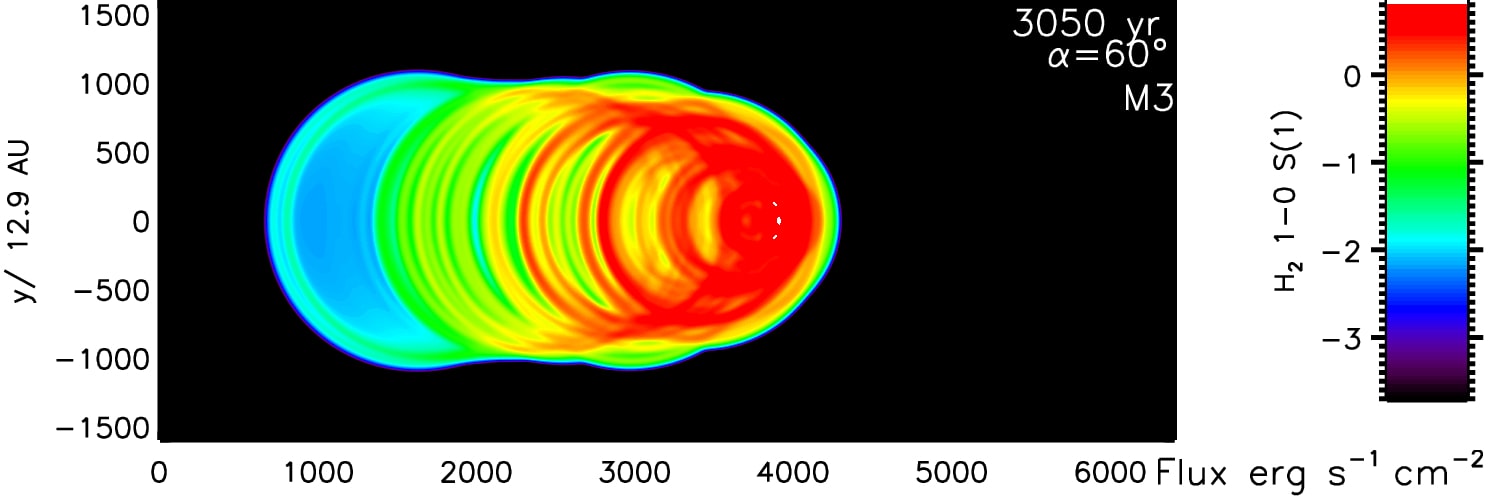}} 
        \caption{Emission as a function of wind-ambient type. Simulated H$_2$ emission from the $\mathrm{1\to0\,S(1)}$ line of 4:1 ellipticl winds at $60^{\circ}$ to the plane of the sky and axial speed of 140~km~s$^{-1}$. The origin of the wind is at zone (1,600,0).}
\label{H2-1-0-60deg-140}
\end{figure}


\begin{figure}
\subfloat[M1: $\rm{H_2}$ wind interacting with an $\rm{H_2}$ ambient medium.]{
        \label{subfig:correct}
        \includegraphics[width=1.0\columnwidth,height=0.3\textheight,keepaspectratio]{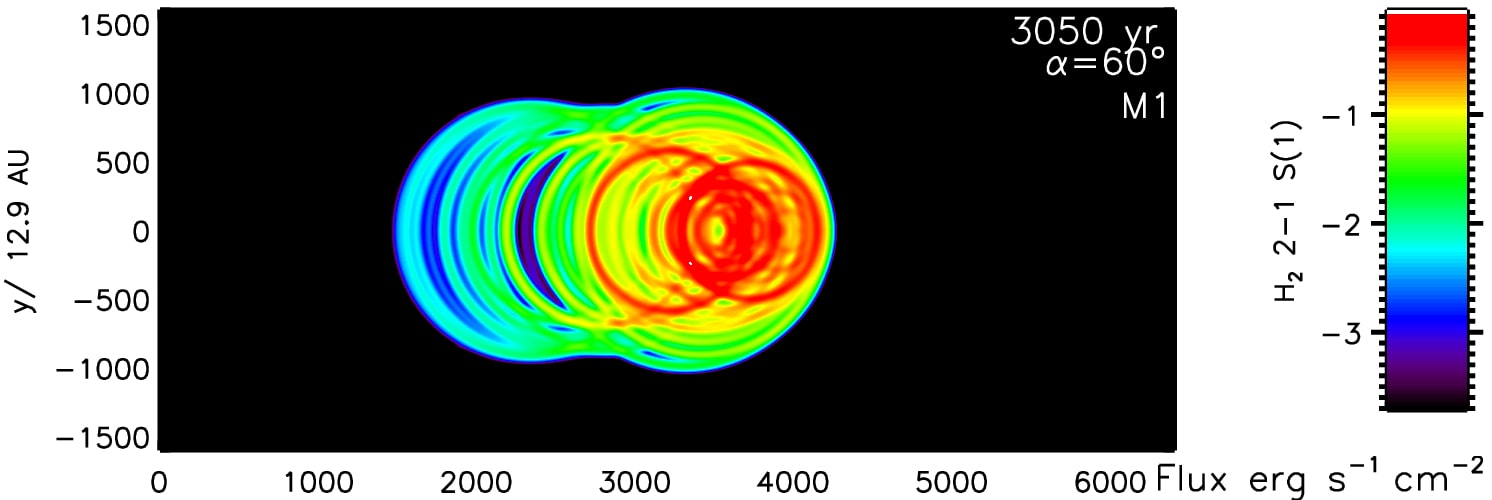} } \\
\subfloat[M2: $\rm{H_2}$ wind interacting with an $\rm{H}$ ambient medium]{
        \label{subfig:notwhitelight}
        \includegraphics[width=1.0\columnwidth,height=0.3\textheight,keepaspectratio]{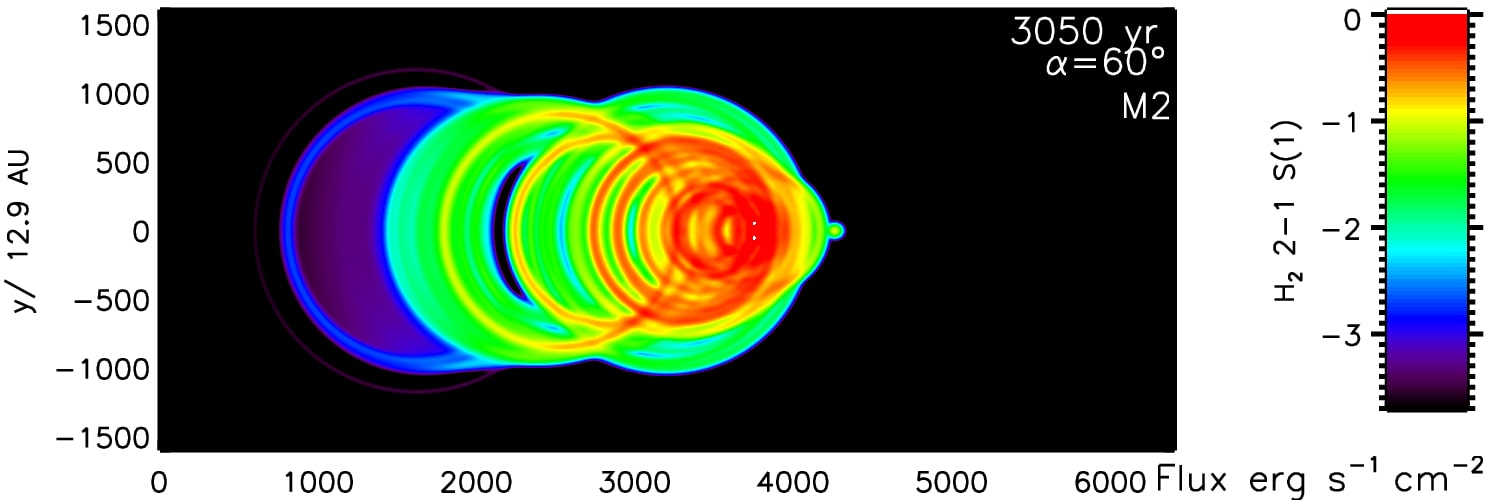} }\\
\subfloat[M3: $\rm{H}$ wind interacting with an $\rm{H_2}$ ambient medium]{
        \label{subfig:nonkohler}
        \includegraphics[width=1.0\columnwidth,height=0.3\textheight,keepaspectratio]{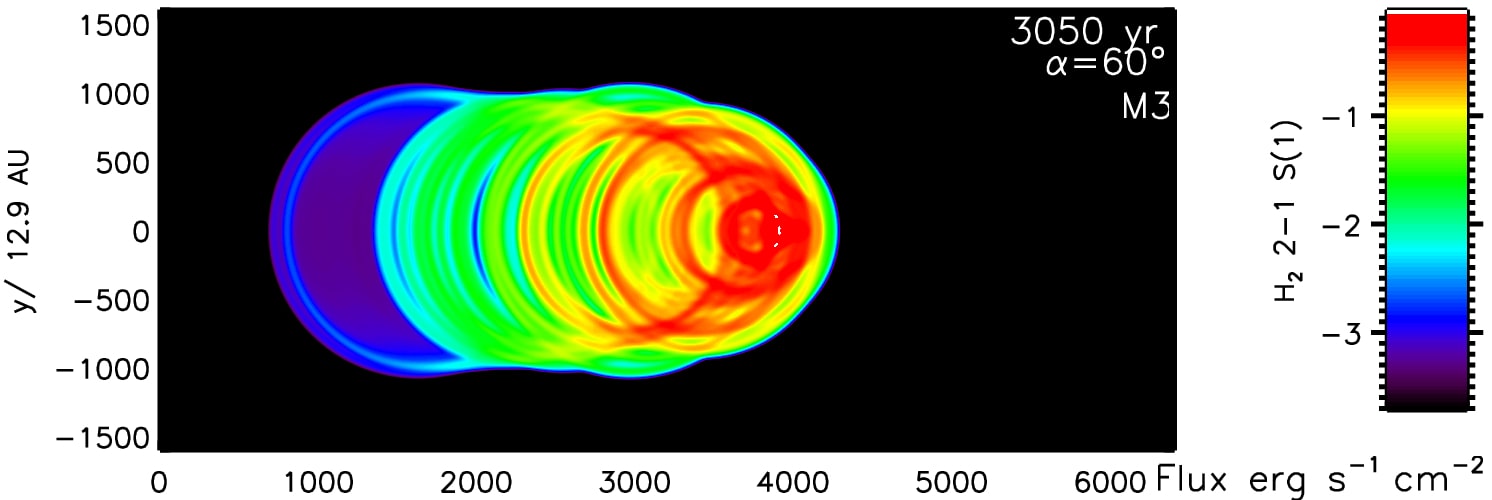}} 
        \caption{Emission as a function of wind-ambient type. Simulated H$_2$ emission from the $\mathrm{2\to1\,S(1)}$ line of 4:1 ellipticl winds at $60^{\circ}$ to the plane of the sky and axial speed of 140~km~s$^{-1}$.  The origin of the wind is at zone (1,600,0).}
\label{H2-2-1-60deg-140}
\end{figure}

\begin{figure}
\subfloat[M1: $\rm{H_2}$ wind interacting with an $\rm{H_2}$ ambient medium.]{
        \label{subfig:correct}
        \includegraphics[width=1.0\columnwidth,height=0.3\textheight,keepaspectratio]{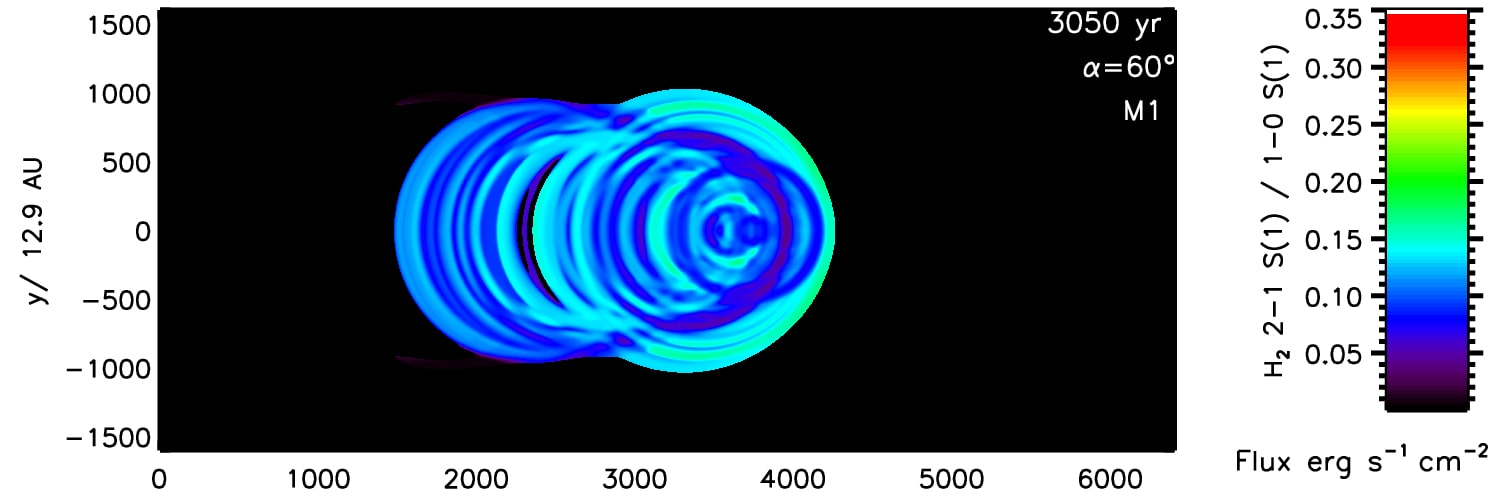} } \\
\subfloat[M2: $\rm{H_2}$ wind interacting with an $\rm{H}$ ambient medium.]{
        \label{subfig:notwhitelight}
        \includegraphics[width=1.0\columnwidth,height=0.3\textheight,keepaspectratio]{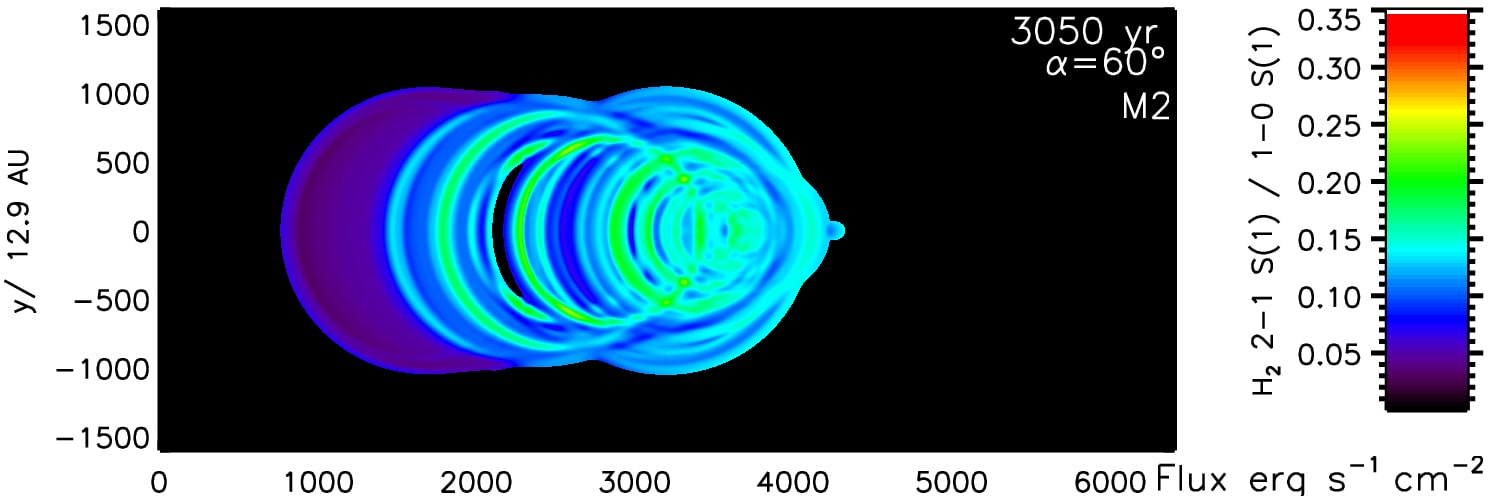} }\\
\subfloat[M3: $\rm{H}$ wind interacting with an $\rm{H_2}$ ambient medium.]{
        \label{subfig:nonkohler}
        \includegraphics[width=1.0\columnwidth,height=0.3\textheight,keepaspectratio]{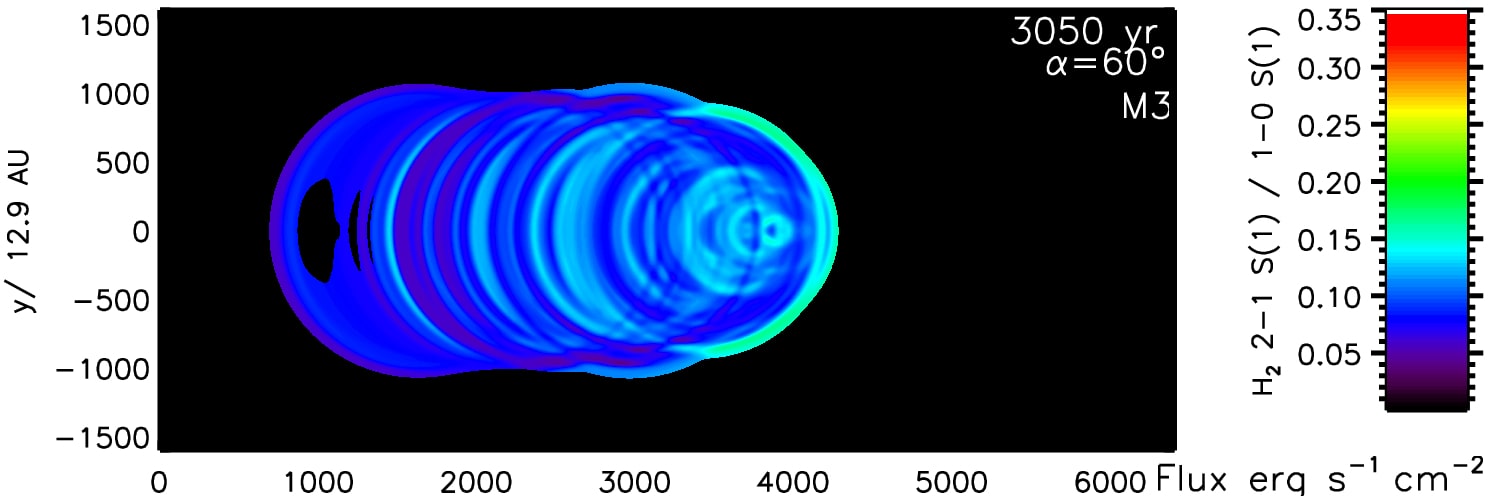}} 
        \caption{Vibrational excitation as a function of wind-ambient type. Simulated H$_2$ ratio mapping from the $\mathrm{2\to1\,S(1)/1\to0\,S(1)}$ line ratio of 4:1 ellipticl winds at $60^{\circ}$ to the plane of the sky and axial speed of 140~km~s$^{-1}$.  The origin of the wind is at zone (1,600,0).}
\label{H2-ratio-60deg-140}
\end{figure}

\begin{figure}
\subfloat[M1: $\rm{H_2}$ wind interacting with an $\rm{H_2}$ ambient medium.]{
        \label{subfig:correct}
        \includegraphics[width=1.0\columnwidth,height=0.3\textheight,keepaspectratio]{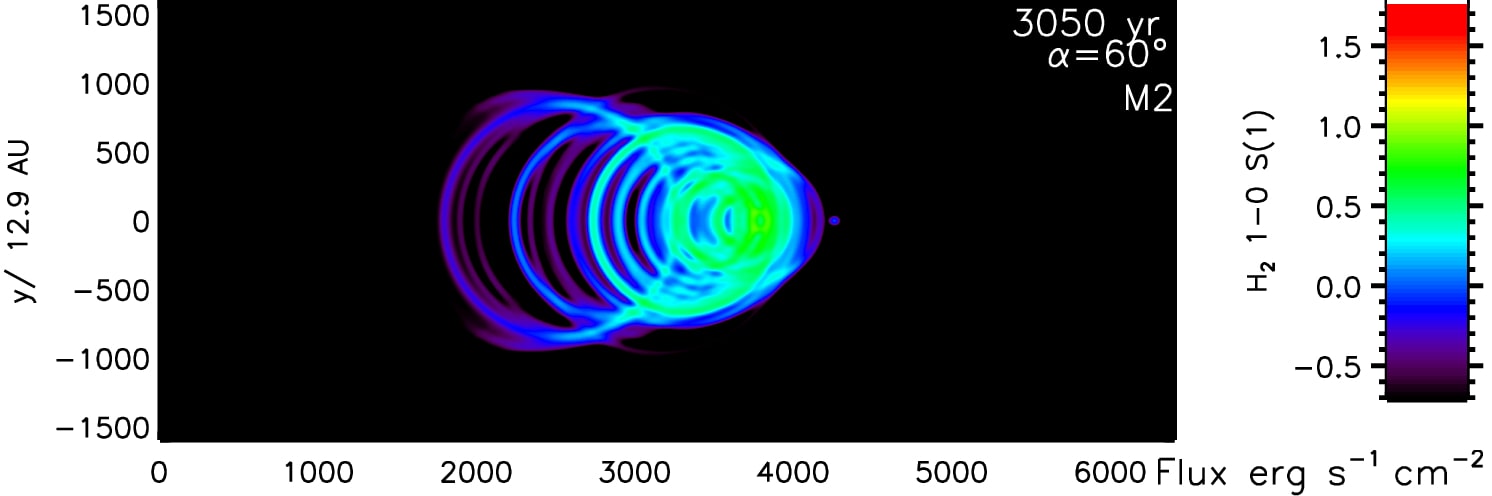} } \\
\subfloat[M2: $\rm{H_2}$ wind interacting with an $\rm{H}$ ambient medium.]{
        \label{subfig:notwhitelight}
        \includegraphics[width=1.0\columnwidth,height=0.3\textheight,keepaspectratio]{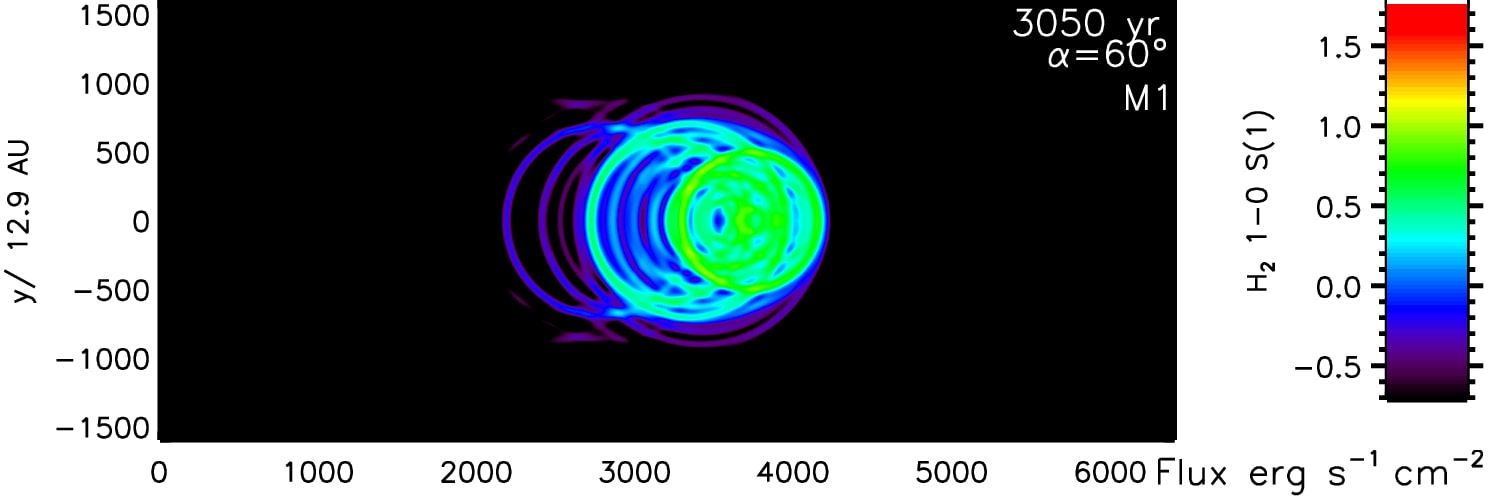} }\\
\subfloat[M3: $\rm{H}$ wind interacting with an $\rm{H_2}$ ambient medium.]{
        \label{subfig:nonkohler}
        \includegraphics[width=1.0\columnwidth,height=0.3\textheight,keepaspectratio]{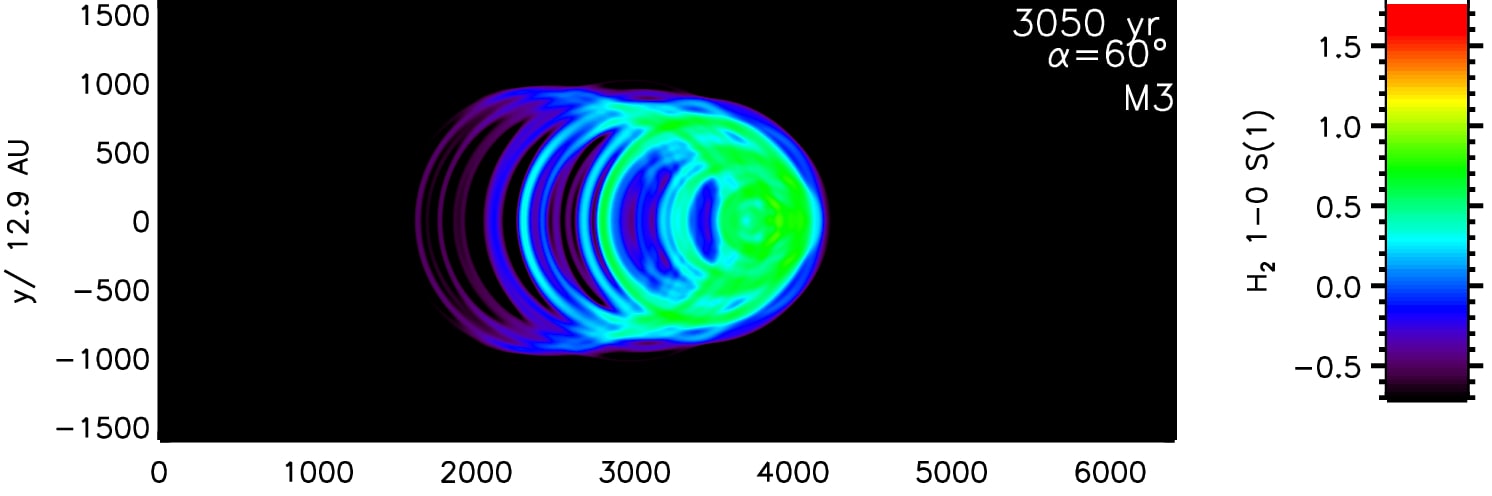}} 
        \caption{Emission as a function of wind-ambient type. Simulated H$_2$ emission of smaller range in flux from the $\mathrm{1\to0\,S(1)}$ line of 4:1 ellipticl winds at $60^{\circ}$ to the plane of the sky and axial speed of 140~km~s$^{-1}$.  The origin of the wind is at zone (1,600,0).}
\label{Linemap4:1a}
\qquad
\end{figure}

\begin{figure}
\subfloat[M1: $\rm{H_2}$ wind interacting with an $\rm{H_2}$ ambient medium.]{
        \label{subfig:correct}
        \includegraphics[width=1.0\columnwidth,height=0.3\textheight,keepaspectratio]{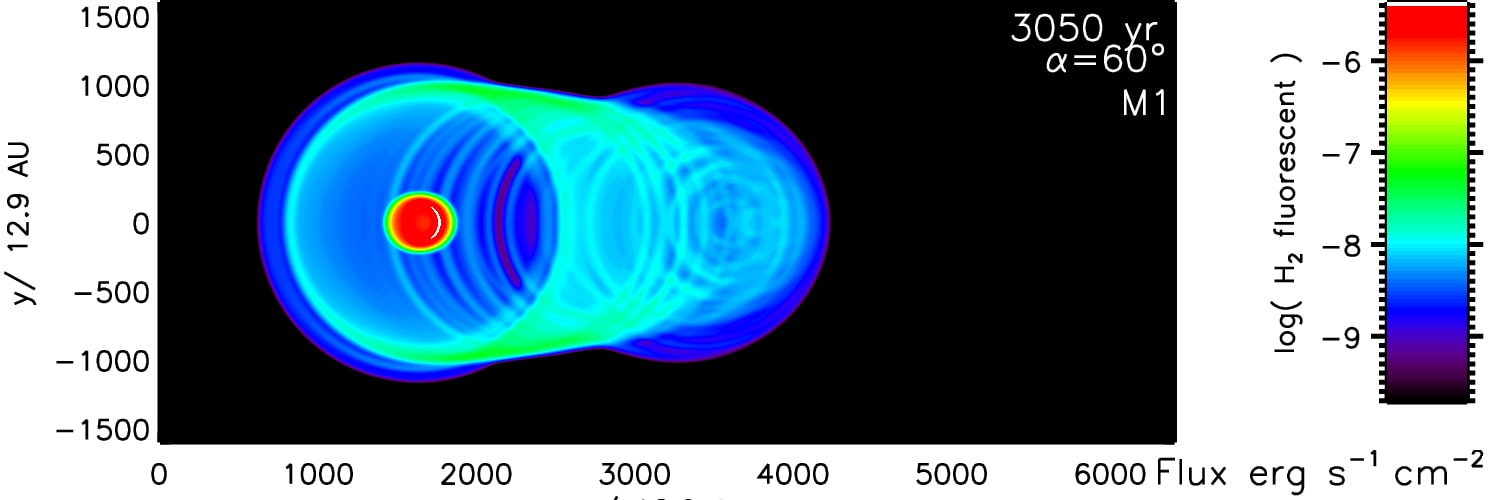} } \\
\subfloat[M2: $\rm{H_2}$ wind interacting with an $\rm{H}$ ambient medium.]{
        \label{subfig:notwhitelight}
        \includegraphics[width=1.0\columnwidth,height=0.3\textheight,keepaspectratio]{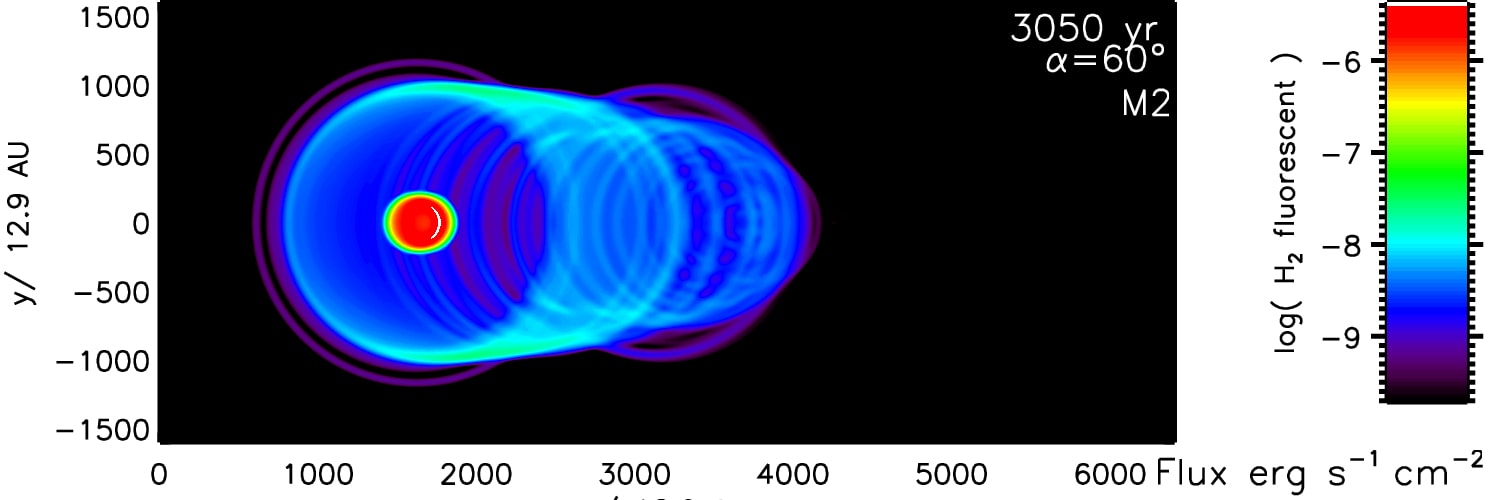} }\\
\subfloat[M3: $\rm{H}$ wind interacting with an $\rm{H_2}$ ambient medium.]{
        \label{subfig:nonkohler}
        \includegraphics[width=1.0\columnwidth,height=0.3\textheight,keepaspectratio]{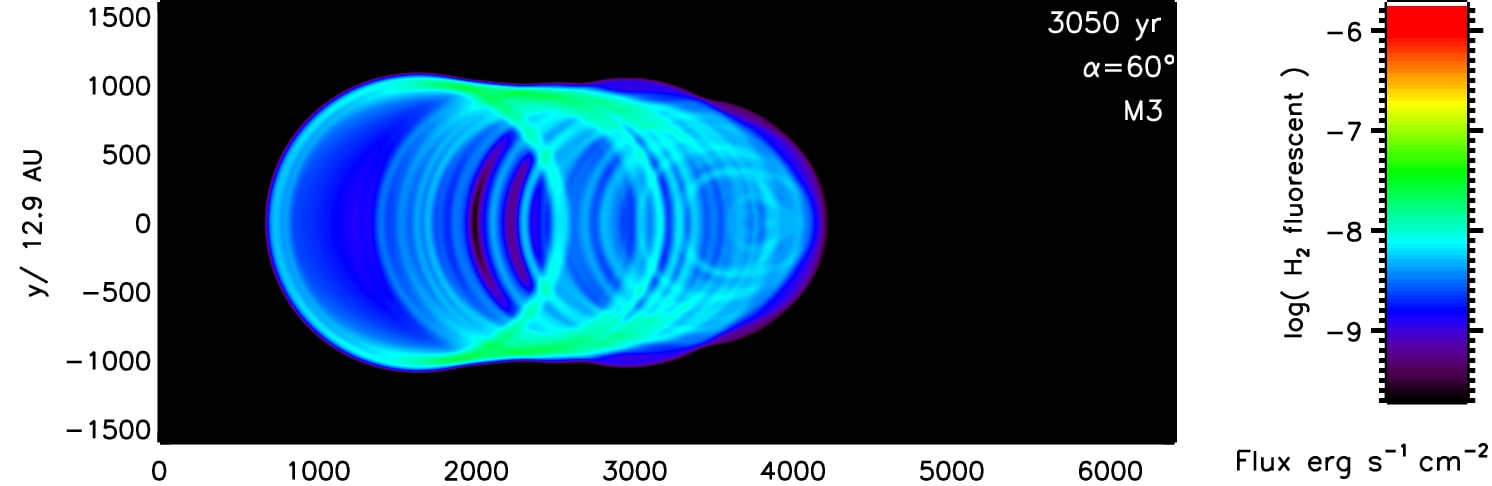}} 
        \caption{A possible measure of H$_2$ fluorescence, displayed here is the distribution of $\rho$(H$_2$)/$r^2$ where $r$ is the distance from the wind origin (assumed to be the  UV radiation source for the fluorescence) measured in arbitrary units.}
\label{H2-fluor-60deg-140}
\end{figure}

\section{Emission Maps}

The simulated images are displayed on an infrared array designed so that all the emission from the simulated 3D cylinder  can be projected on to the same array whatever the orientation to the line of sight. Thus, the projected wind origin remains fixed. Due to the axial symmetry and the formation of the warm streaks, the emission tends to form discrete rings in projection which would
probably appear as individual bullets or globules in full 3D simulations. Nevertheless, it is instructive to generate these images and determine where the emission is located and how excited the gas is.

Figure~\ref{M1a-H2-21-4panel} demonstrates the H$_2$ emission from a flow in the sky plane. Emission is generated across the entire cylinder although the flux dynamic range is over 1,000.

At 60$^\circ$ to the line of sight, Fig.~\ref{H2-1-0-60deg-140}   for the 
 $\mathrm{1\to0\,S(1)}$ line and Fig.~\ref{H2-2-1-60deg-140} for the 
 $\mathrm{2\to1\,S(1)}$ line, we illustrate that the simulated emission does show some significant variation across the three displayed models. 
 Of note is that the fully molecular interaction appears to show less emission from the flanks although there is no spatial shift of emission
 close to the apex. The cause for the lower flank emission when fully molecular can be discerned from the H$_2$ ratio image of Fig.~\ref{H2-ratio-60deg-140} which indicates a slightly higher excitation in the far flanks when an atomic component is present.
 
 The peak emission is set back from the leading edge in a similar way for all cases. The  suggestion is that the source geometry determines the distribution more than the particular reverse and advanced shock configuration.  

Taking a smaller range in flux, appropriate for most observations,  it is apparent that the majority of the molecular flux is located at a bow-shaped shock followed by circular cylindrical shells 
as illustrated in Fig.~\ref{Linemap4:1a}. It can be concluded that collimated pPN are produced with the present set up and butterfly-type bipolar nebulae are not generated.

It is interesting to also determine the fluorescent emission that would result from an optically thin gas. Here, in Fig.~\ref{H2-fluor-60deg-140},  we assume an emissivity proportional to the H$_2$ density and an inverse square law from a star at the origin (1,600 ,0). Now, the images are very different since a molecular wind generates strong emission from the denser inner parts of the wind while an atomic wind creates a cavity.
 
\begin{figure*}
   \subfloat[\label{genworkflow}][M1: $\rm{H_2}$ wind into $\rm{H_2}$ amb.]{%
      \includegraphics[width=0.33\textwidth,height=0.25\textheight]{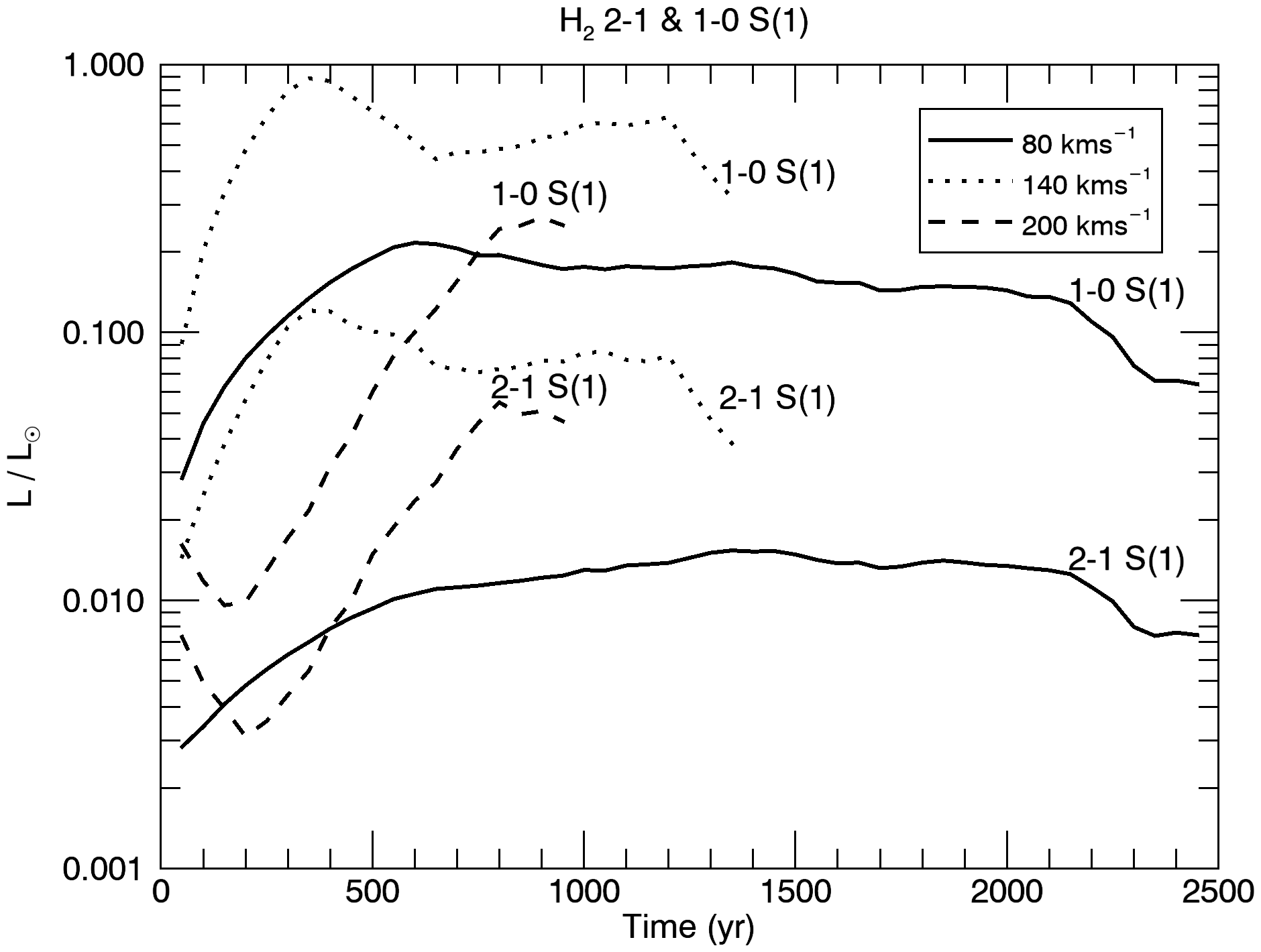}}
\hspace{\fill}
   \subfloat[\label{pyramidprocess}][M2: $\rm{H_2}$ wind into $\rm{H}$ amb.]{%
      \includegraphics[width=0.33\textwidth,height=0.25\textheight]{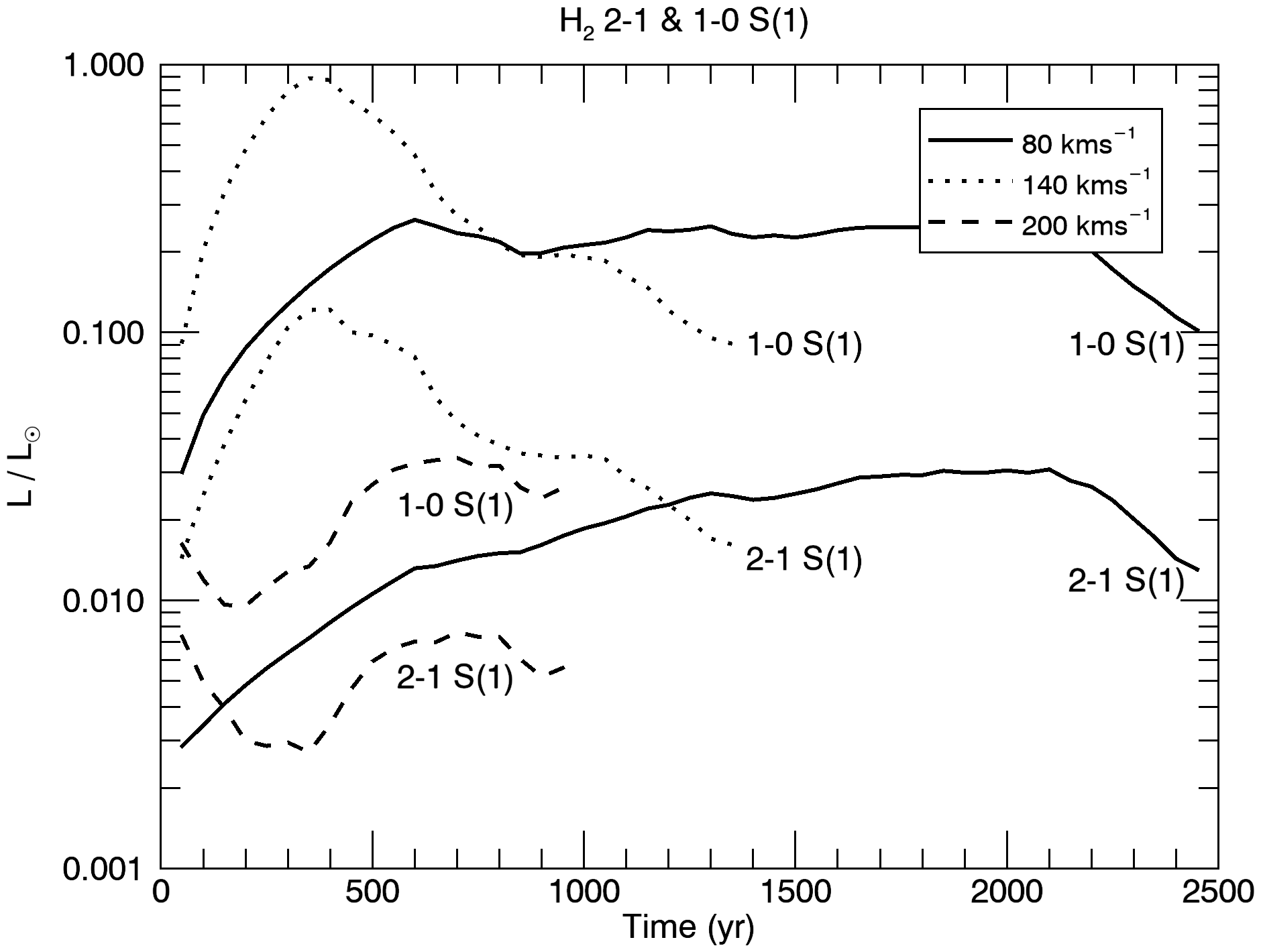}}
\hspace{\fill}
   \subfloat[\label{mt-simtask}][M3: $\rm{H}$ wind into $\rm{H_2}$ amb.]{%
      \includegraphics[width=0.33\textwidth,height=0.25\textheight]{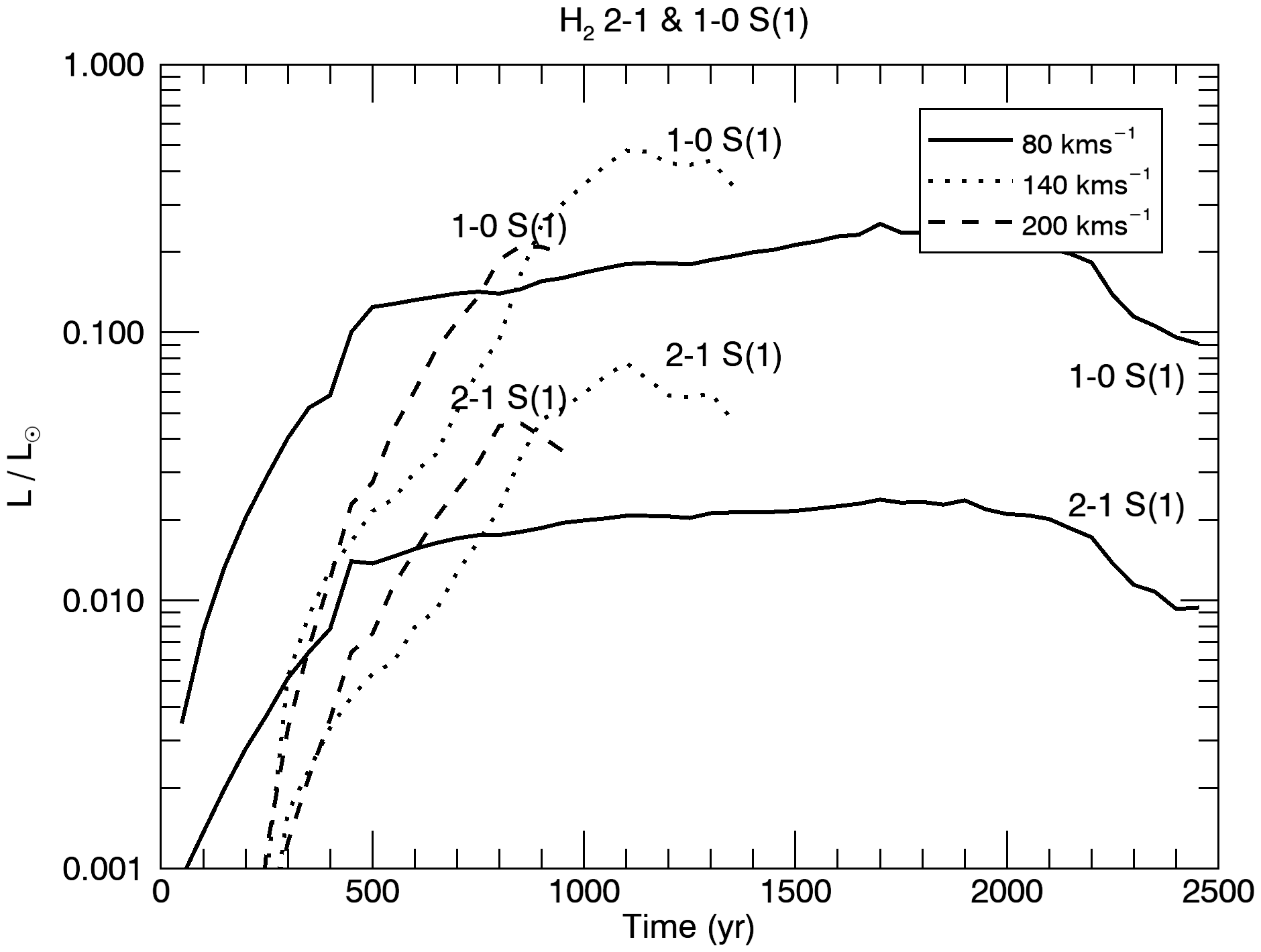}}\\
\caption{\label{1.1molemis} The evolution of the integrated emission in the indicated K-band vibrational transitions produced by 1:1 M1, M2 and M3 outflows for the three axial wind velocities.}
\end{figure*}


\begin{figure*}
   \subfloat[\label{genworkflow}][M1: $\rm{H_2}$ wind into $\rm{H_2}$ amb.]{%
      \includegraphics[width=0.33\textwidth,height=0.25\textheight]{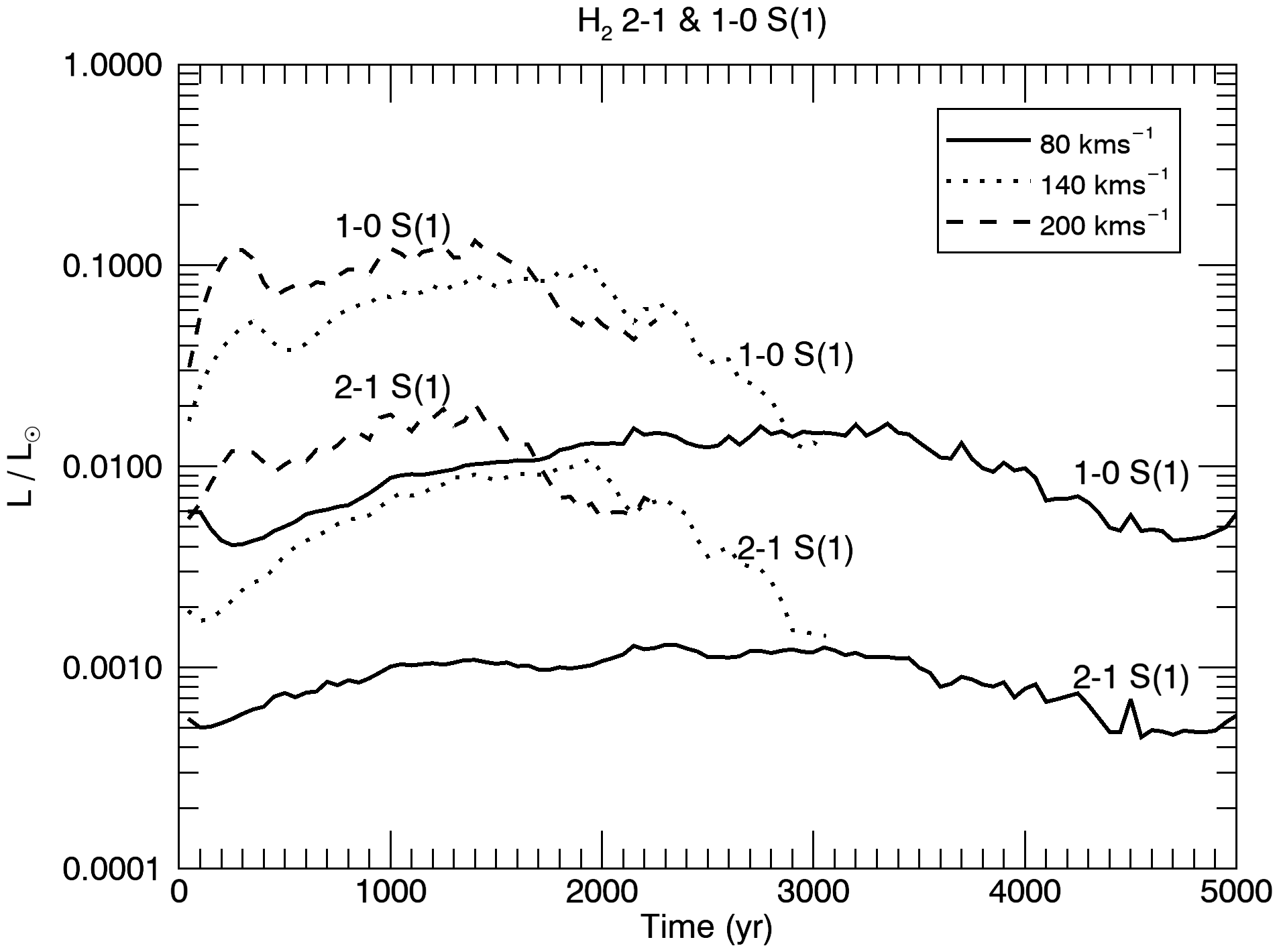}}
\hspace{\fill}
   \subfloat[\label{pyramidprocess}][M2: $\rm{H_2}$ wind into $\rm{H}$ amb.]{%
      \includegraphics[width=0.33\textwidth,height=0.25\textheight]{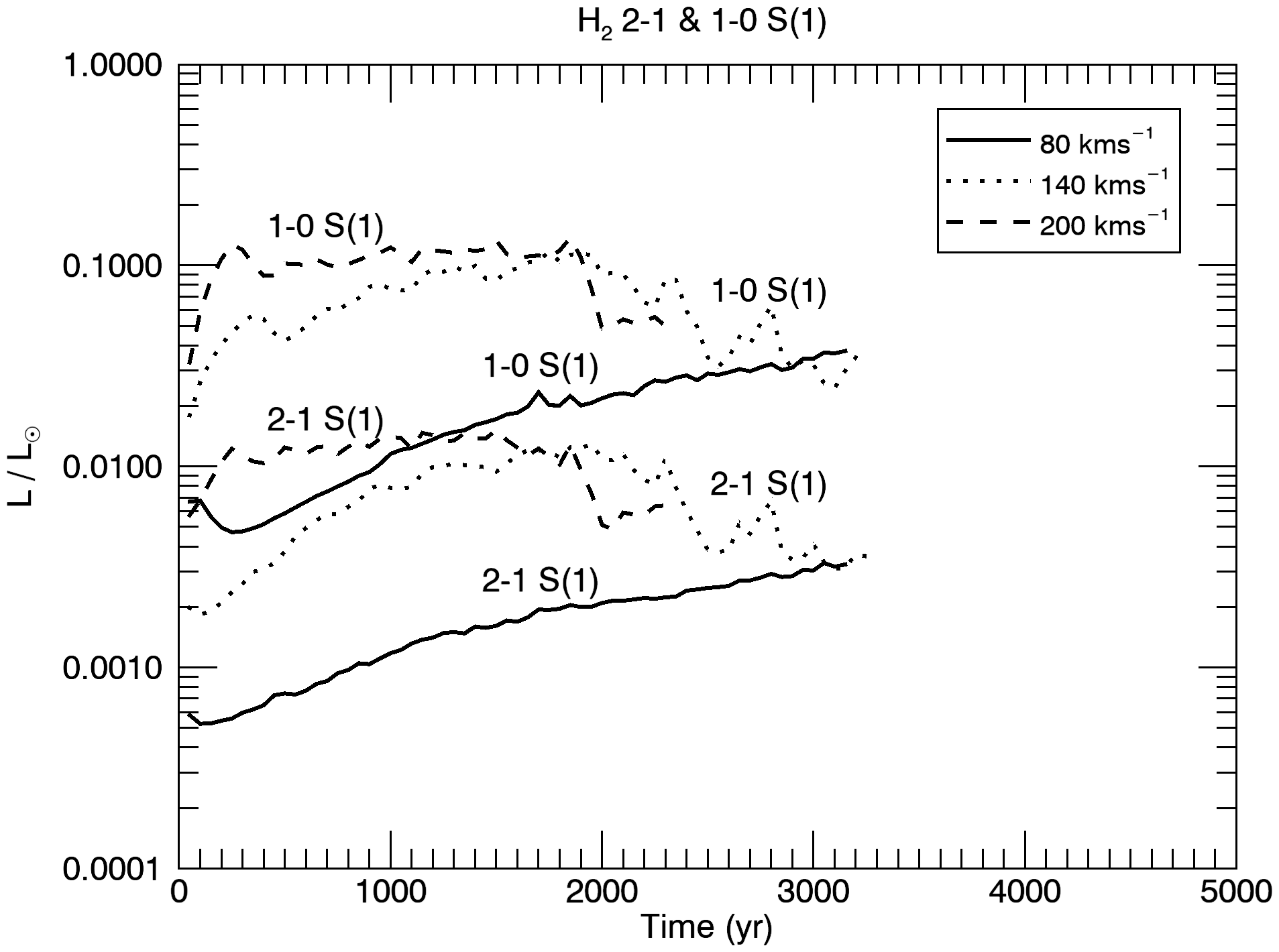}}
\hspace{\fill}
   \subfloat[\label{mt-simtask}][M3: $\rm{H}$ wind into $\rm{H_2}$ amb.]{%
      \includegraphics[width=0.33\textwidth,height=0.25\textheight]{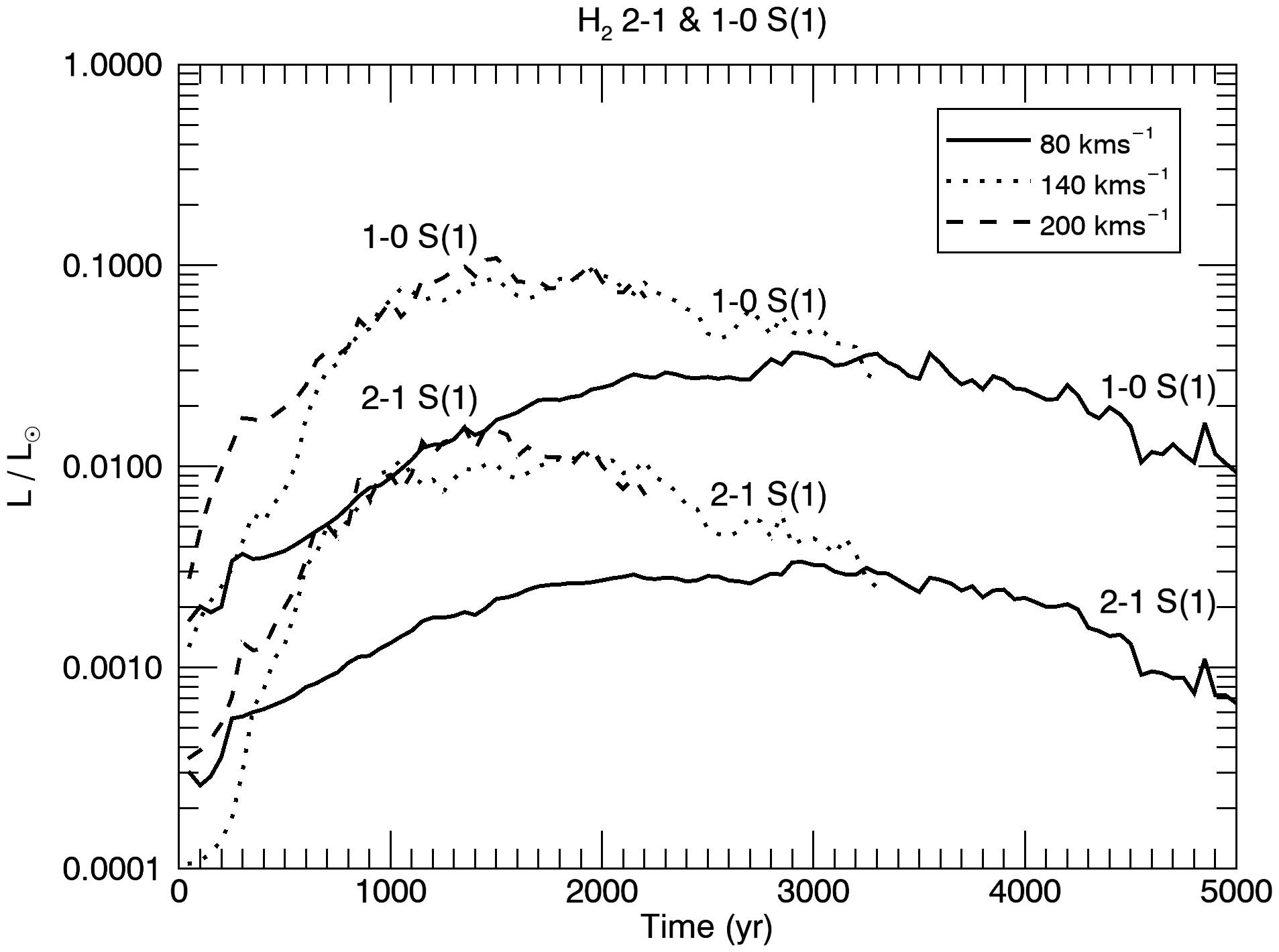}}\\
\caption{\label{4.1molemis}The evolution of the integrated emission in the indicated K-band vibrational transitions produced by 4:1 M1, M2 and M3 outflows for the three axial wind velocities.}
\end{figure*}

\begin{figure*}
   \subfloat[\label{genworkflow}][$\mathrm{V_w=80\, kms^{-1}}$.]{%
      \includegraphics[width=0.33\textwidth,height=0.25\textheight]{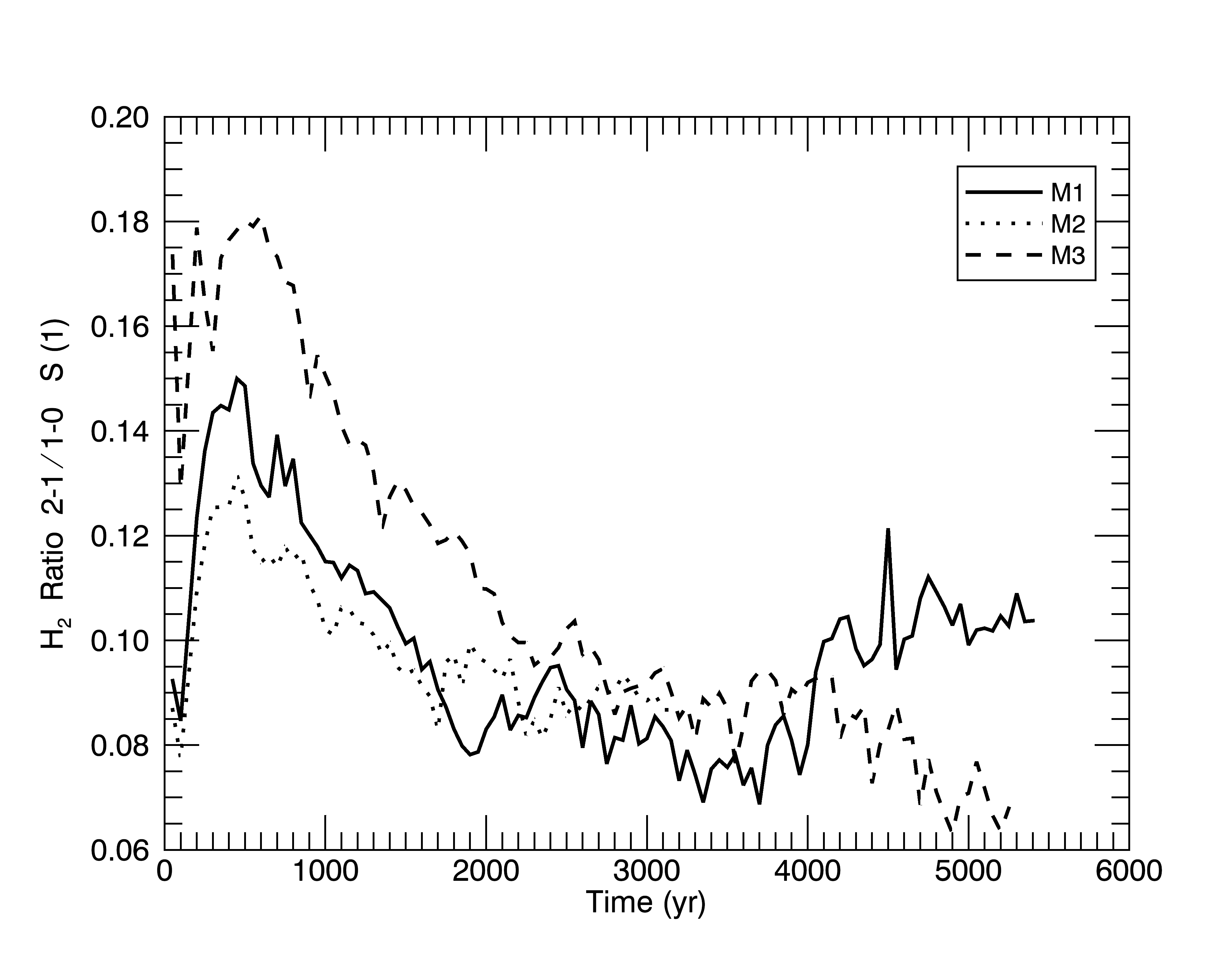}}
\hspace{\fill}
   \subfloat[\label{pyramidprocess}][$\mathrm{V_w=140\, kms^{-1}}$.]{%
      \includegraphics[width=0.33\textwidth,height=0.25\textheight]{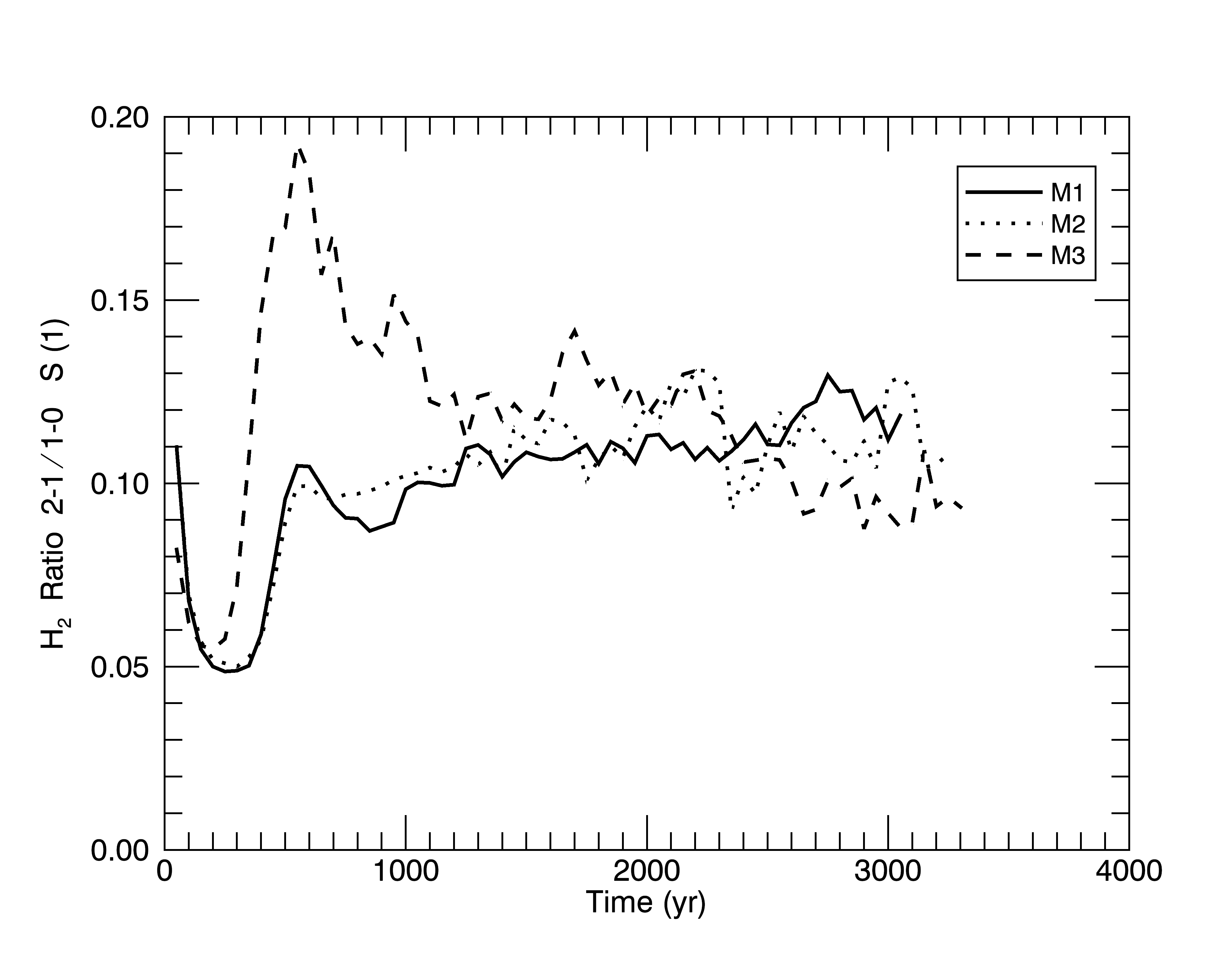}}
\hspace{\fill}
   \subfloat[\label{mt-simtask}][$\mathrm{V_w=200\, kms^{-1}}$.]{%
      \includegraphics[width=0.33\textwidth,height=0.25\textheight]{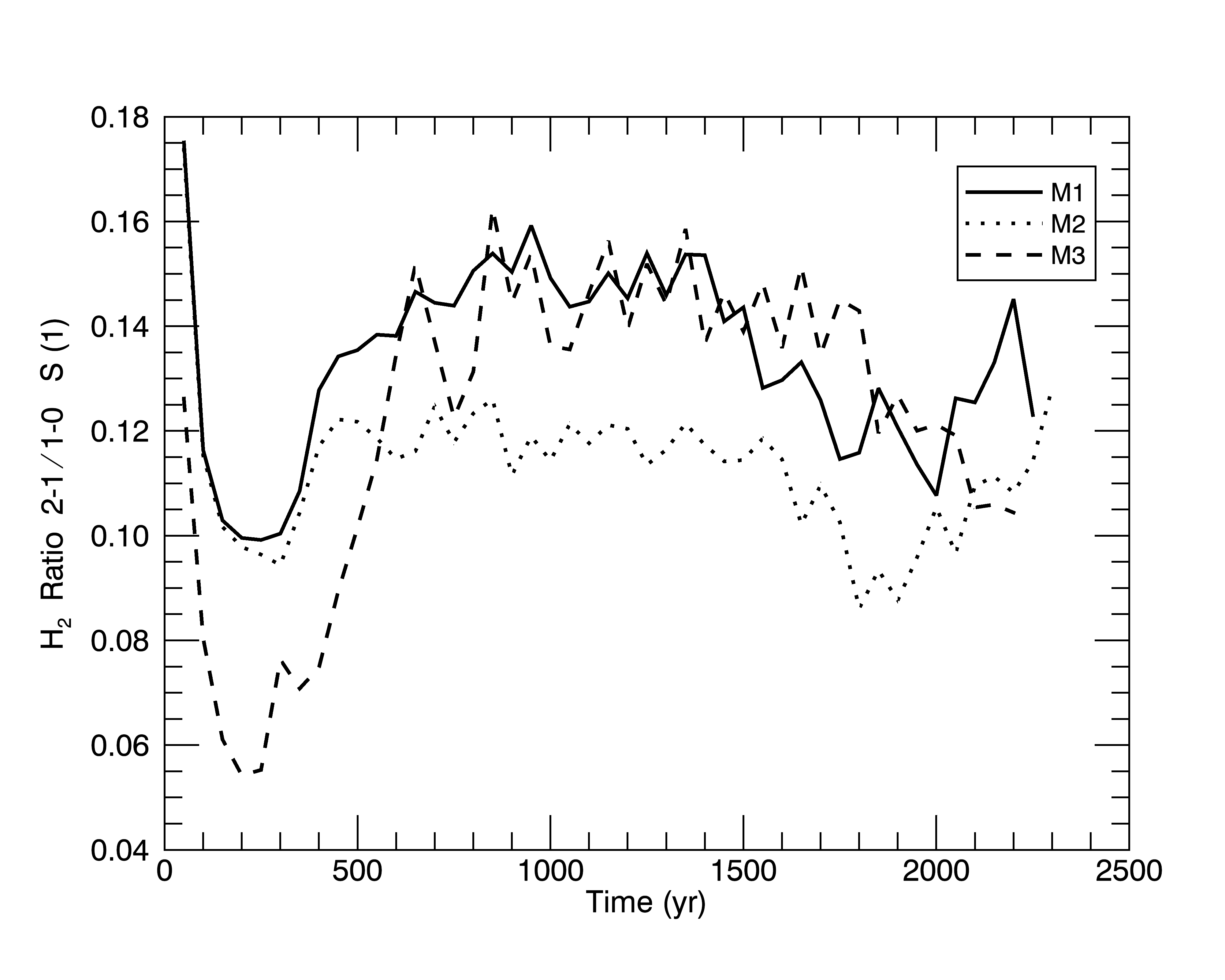}}\\
\caption{\label{4.1ratio}The ratio of the emissions from vibration transitions  originating from the second and first levels produced by 4:1 M1, M2 and M3 outflows
at 80~km\,s$^{-1}$ (left), 140~km\,s$^{-1}$ (middle) and  200~km\,s$^{-1}$  (right  wind velocities.}
\end{figure*}

A common diagnostic for measuring the excitation is the $\mathrm{2\to1/1\to0 \,S(1)}$ line ratio. One could expect that the excitation would increase toward the apex of the expanding prolate ellipsoid, as found for bow shocks \citep{2003MNRAS.339..524S}. However, as shown in Fig~\ref{H2-ratio-60deg-140} for the molecular run, the ratio remains between 0.1 and 0.2, typical of moderate shock excitation from gas 1,800\,K to 2,600\,K. This small range occurs despite the wide range in shell speeds in the 4:1 simulation and can be attributed to the fragmentation of the shock surface into a large number of streaks which encompass weak oblique shocks at all locations.   

The flux values and ratios are consistent with a collisional excitation scenario. A fraction of the molecules are abruptly dissociated after the shock front, radiative cooling follows as gas loses heat by radiation, forming  dense and cool post-shock relaxation layers \citep{2003MNRAS.339..133S}. Line map units are in $\mathrm{erg\,s^{-1}\,cm^{-2}}$. These results can be compared to $K$-band integral field spectroscopy that reveals bipolar $\mathrm{(\textgreater \, 150 \, km \,s^{-1}) \, H_2}$ outflows such as  associated with $\text{IRAS} \,16342-3814$ \citet{2012MNRAS.421..346G} and \citet{2015MNRAS.447.1080G}. 

A molecular wind can be differentiated from a molecular environment. At early evolution times, the advancing shock is strong.
To begin with Model M3, a dense wind pushes easily into the lower density ambient medium. This fast shock dissociates the H$_2$ in the ambient medium, so making the cap deficient in H$_2$,
 In contrast, the cap will be initially  bright in model M2. The remaining hot molecular gas generates very high excitation emission in M3 and the 2--1/1--0 ratio exceeds 0.4 over most of the cap.  In contrast, the strong molecular wind generates stronger infrared emission in Model M2 with lower excitation.

At later times, the advancing molecular shock has slowed down in Model M3, generating a typical large bow structure in which the excitation declines towards the flanks. On the other hand, this slow moving shell becomes the wall for strong emission and excitation of the molecular wind in Model M2. The reverse shock appears to generate a moderate excitation of $\sim$ 0.1 -- 0.15 over almost the entire shell.

The ability to observe pPN in the near-infrared lines is assessed by finding the total pPN emission in the two commonly observed near-infrared H$_2$ lines as a function of time.
These are displayed in Figs.~\ref{1.1molemis} and  \ref{4.1molemis} for all three wind speeds and for spherical and 4:1  elliptical winds. Particularly high line fluxes  are determined for spherical winds with 1--0\,S(1) luminosities  approaching one solar luminosity for the intermediate wind speed of 140~km\,s$^{-1}$. At this speed, most shock emission is channelled into the vibrational lines.
In addition, it should be noted that the implementation of the ellipticity in the velocity distribution does strongly reduce the energy flux away from the axis and this explains the general trend with ellipticity. However, the emission is also high for elliptical winds but at higher axial wind speeds, which may be expected given the distribution of interface speeds expected across the entire flow.
 
 Overall, H$_2$ line fluxes remain high once established, throughout the pPN wind phase. The peak H$_2$ luminosity typically occurs after 500\,--\,1,000 years which is enhanced  when a molecular wind is involved (M1 \& M2).  However, the flux remains high and slightly increases in the cases involving lower speed winds, corresponding to solid lines. 
 
The overall excitation is best analysed by plotting the 2--1/1--0 ratio as shown in Fig.~\ref {4.1ratio} for the 4:1 winds. There is no significant dependence on model type after the initial blast: the excitation is independent of whether ambient streams become trapped and compressed by the wind or vice versa. 
However, there is a significantly higher vibrational excitation  associated with the higher speed wind (right panel), a relationship which holds for all ellipticities studied. 
The ratio depends on both  density and temperature since the H$_2$ will not possess LTE distributions at the pPN densities but the range displayed between 0.08 and 0.14 corresponds to
temperatures of$\sim$ 1,700\,K -- 2,500\,K.

\section{Spectroscopic Properties}

\begin{figure*}
   \subfloat[\label{genworkflow}][M1: (MWMA).]{%
      \includegraphics[width=0.3\textwidth]{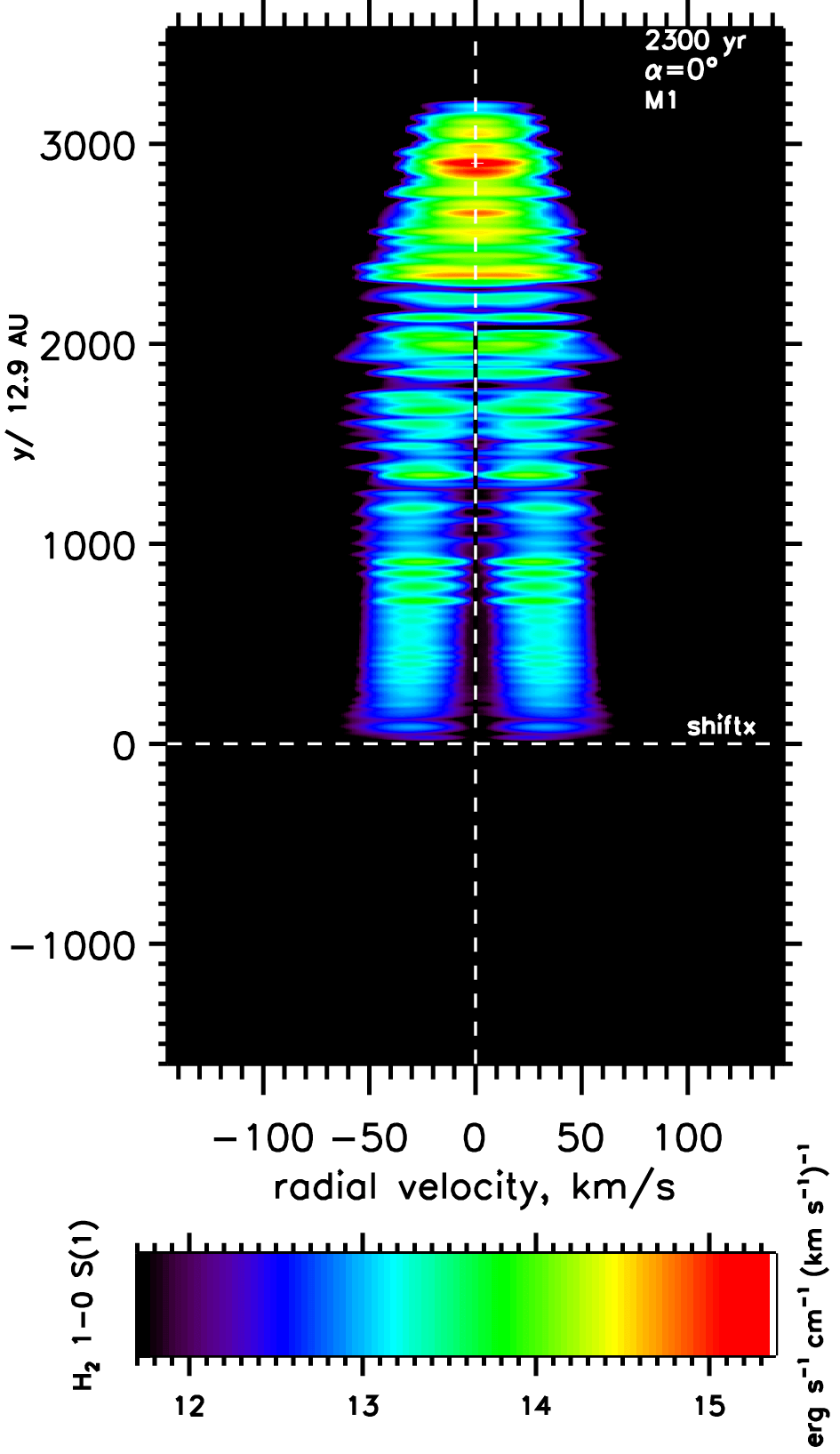}}
\hspace{\fill}
   \subfloat[\label{pyramidprocess}][M2: (MWAA).]{%
      \includegraphics[width=0.3\textwidth]{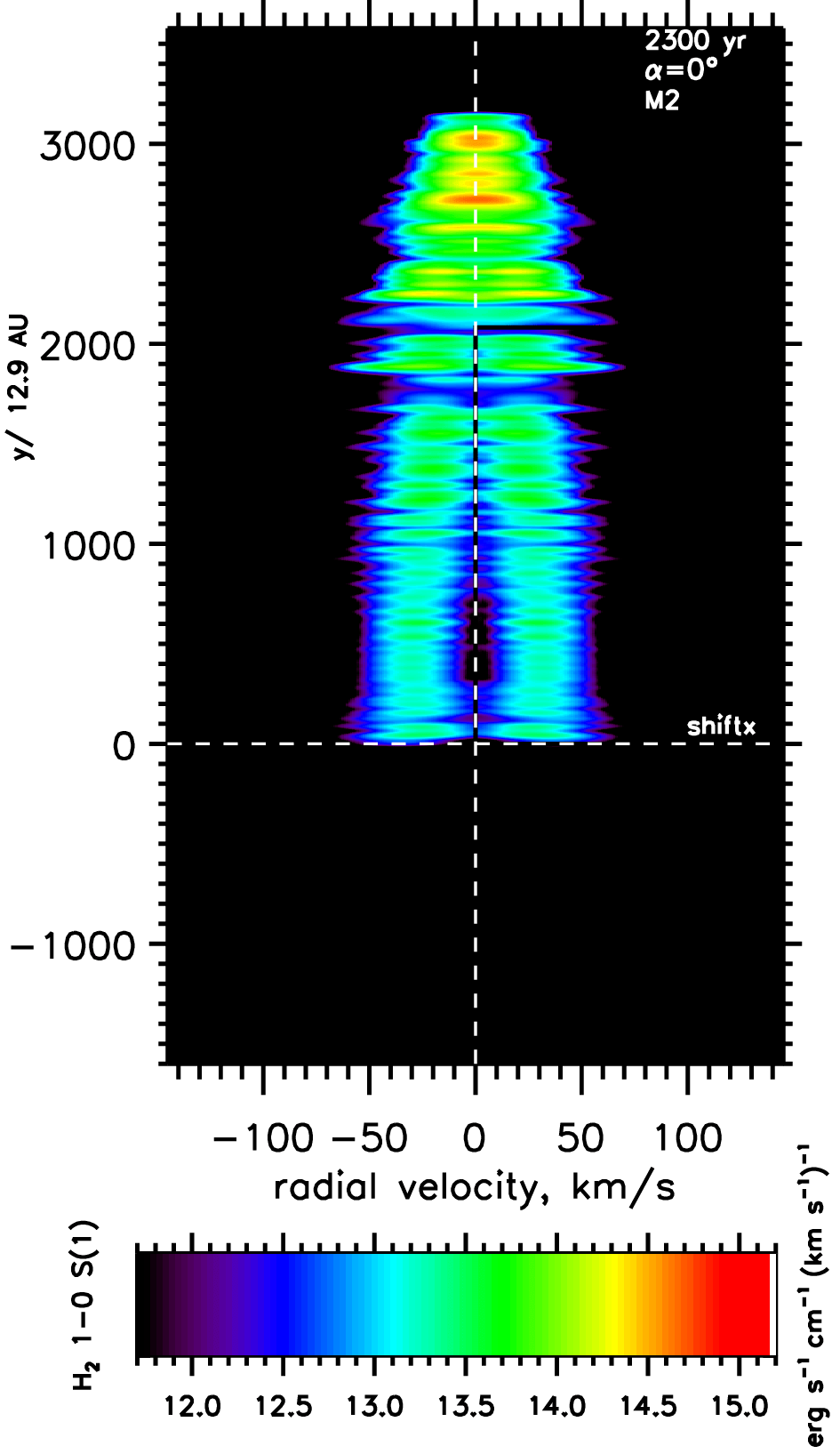}}
\hspace{\fill}
   \subfloat[\label{mt-simtask}][M3: (AWMA).]{%
      \includegraphics[width=0.3\textwidth]{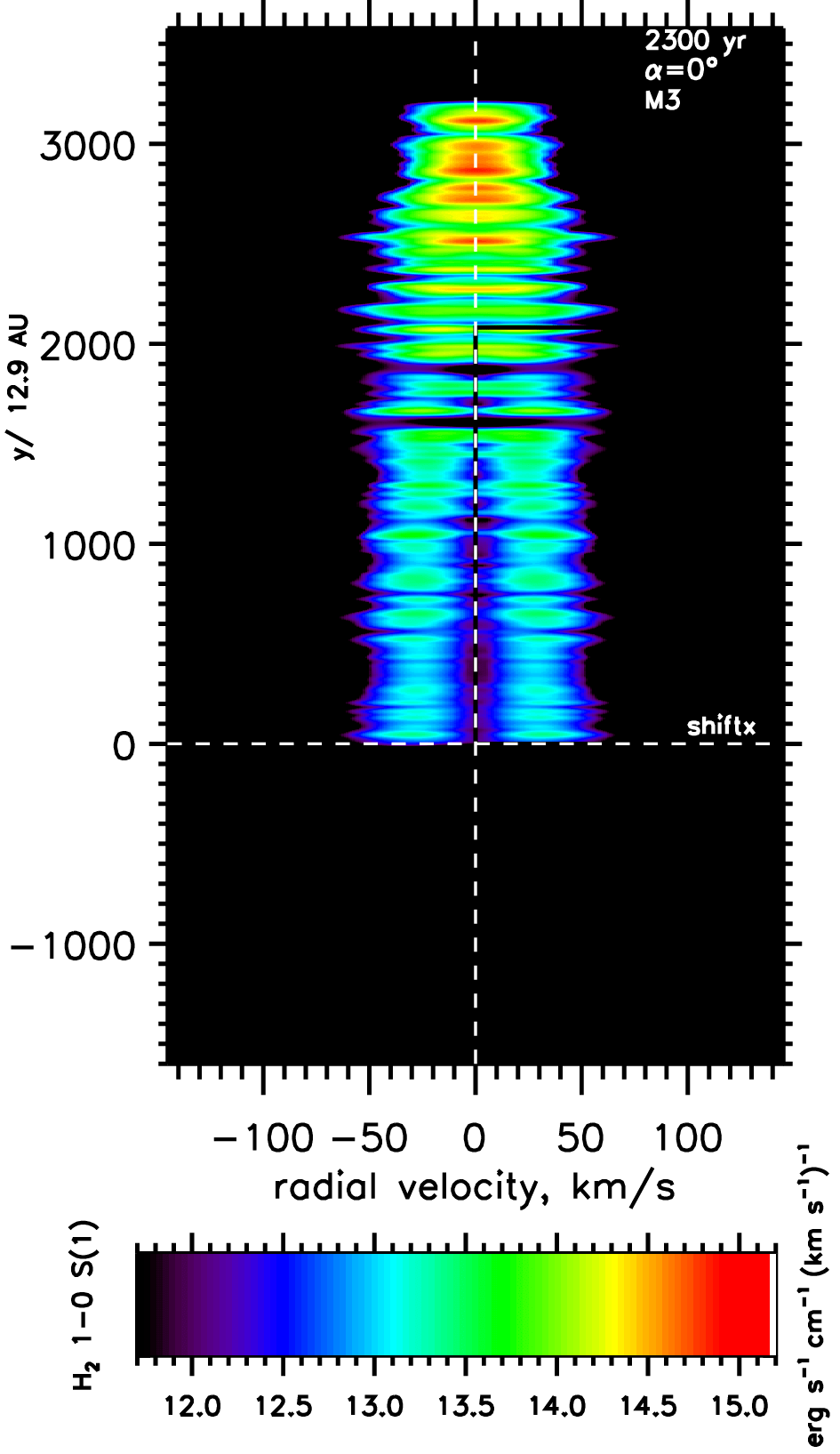}}\\
\caption{\label{2.1molpv0}Position-Velocity diagrams for the H$_2$ 1--0\,S(1) emission from the 2:1 elliptical wind with the long axis in the plane of the sky 
$\mathrm{\alpha = 0^\circ}$  at a late stage of 140\,km\,s$^{-1}$ wind expansion. The three composition models are as indicated. }
\end{figure*}

\begin{figure}
      \includegraphics[width=0.49\columnwidth]{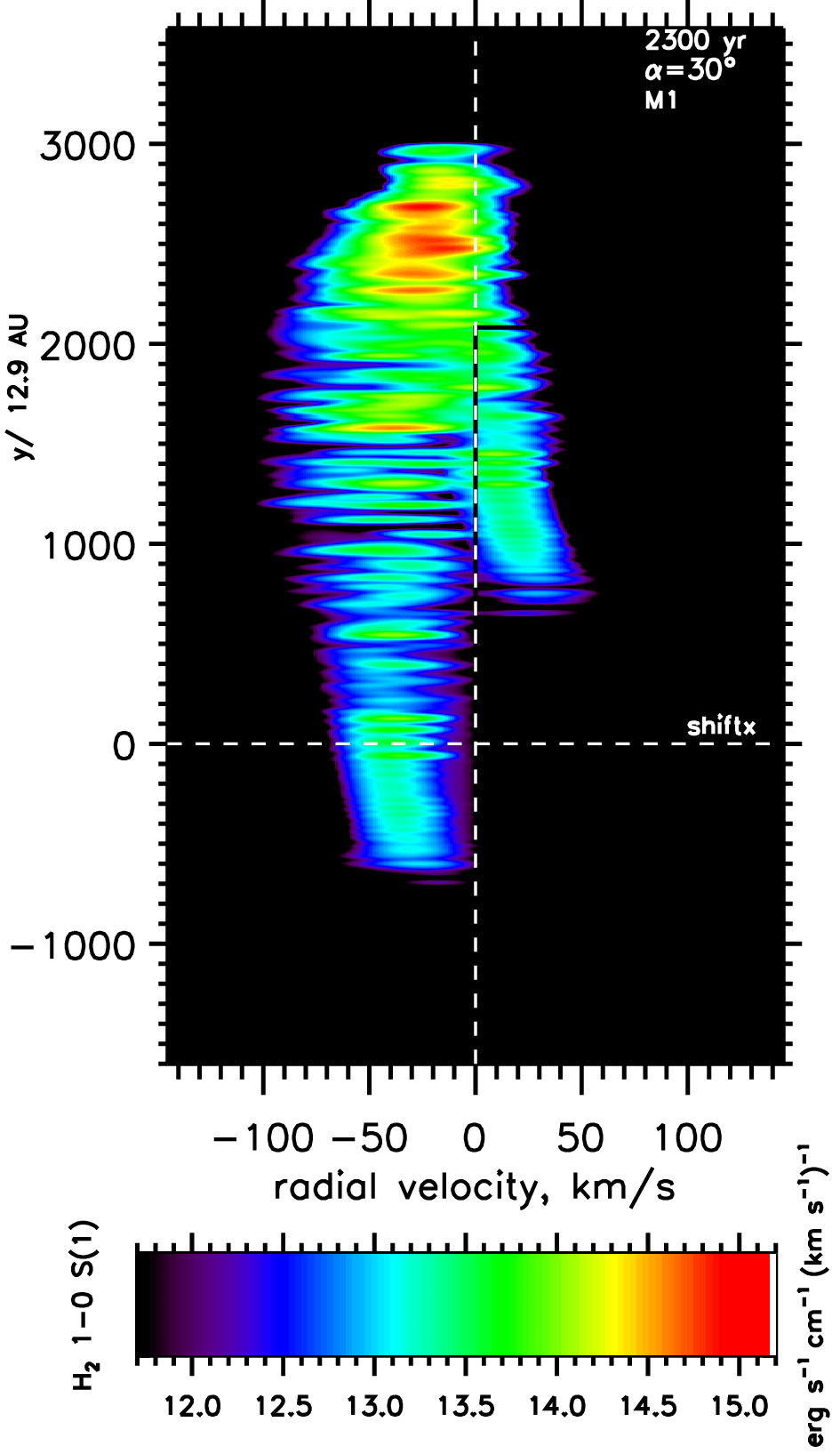}
       \includegraphics[width=0.49\columnwidth]{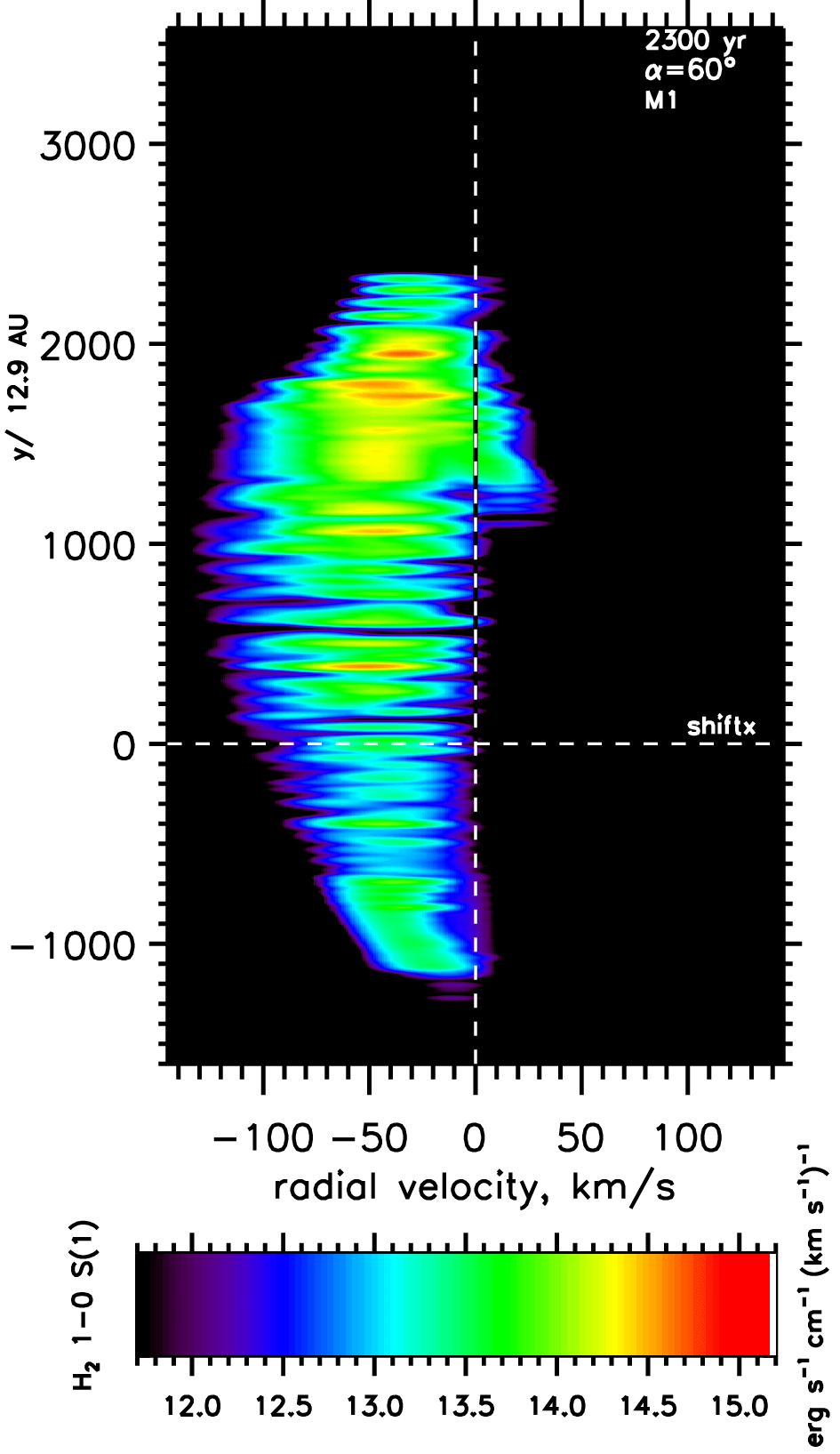}  
\caption{\label{2.1molpv60}Position-Velocity diagrams for the H$_2$ 1--0\,S(1) emission from the 2:1 elliptical wind with the long axis
$\mathrm{\alpha = 30^\circ}$ and  $\mathrm{\alpha = 60^\circ}$ out of  the plane of the sky, 
at a late stage of 140\,km\,s$^{-1}$ wind expansion. The M1 model for a molecular wind and ambient medium is displayed.}
\end{figure}

We have generated Position-Velocity diagrams for the H$_2$ emission, as displayed in Figs.~\ref{2.1molpv0}~--~\ref{2.1molpv60} for the 2:1 case.
We apply a Gaussian smoothing to both the radial velocity and position. However, no explicit thermal Doppler broadening is included.
All composition models are shown (from left to right as indicated)  at three orientations for the 2:1 wind speed ratio. The origin is set to (0,0) and the position range is set automatically so that the emission remains on the diagram however the axis is orientated. The emission is integrated over the transverse direction within a long slit positioned along the axis with a width of 200 pixels.
The radial velocity relative to the star is binned in 1~km\,s$^{-1}$ intervals.
   
The narrowest lines are generated in Model 3 (right panels) where the wind pushes the distorted shocks into the molecular ambient medium. Since in this work the ambient medium is stationary, the line width is limited, expected to be approximately given by twice the dissociation speed limit on theoretical considerations \citep{1994MNRAS.266..238S}. In shocks which are magnetic-field cushioned
with a low ion fraction allowing ambipolar diffusion to dominate (C-type), much narrower H$_2$ lines are produced. Here we are modelling hydrodynamic jump shocks in which the H$_2$ emission will mainly occur after the shocked molecular gas has cooled back down to about 2,000\,K and accelerated to a high fraction of the shock speed. Hence, we find lines of full width exceeding 80\,km\,s$^{-1}$ in some locations.

\begin{figure*}
   \subfloat[\label{genworkflow}][$\mathrm{\alpha=0^{\circ}}$.]{%
      \includegraphics[width=0.30\textwidth,height=0.2\textheight]{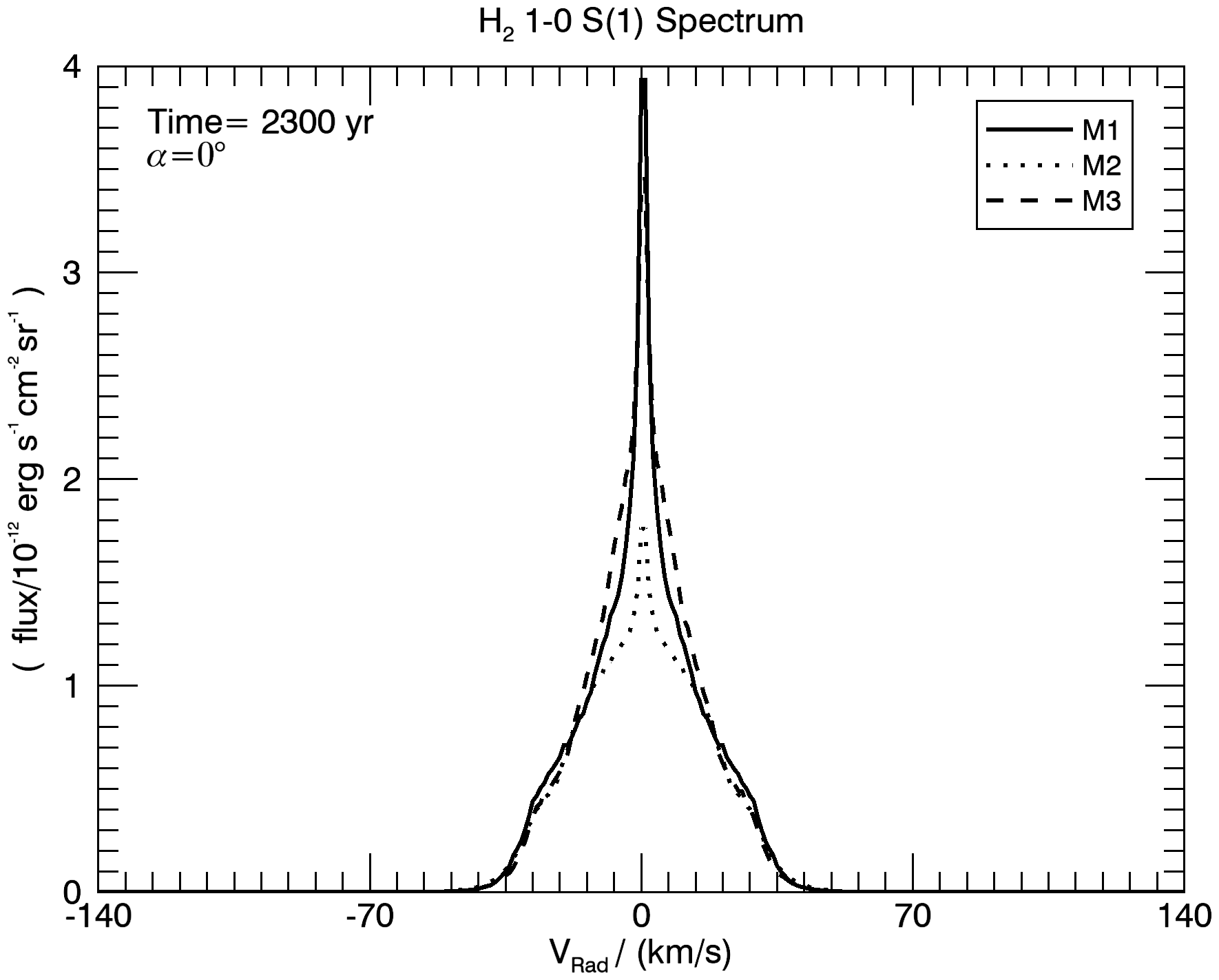}}
\hspace{\fill}
   \subfloat[\label{pyramidprocess}][$\mathrm{\alpha=30^{\circ}}$.]{%
      \includegraphics[width=0.30\textwidth,height=0.2\textheight]{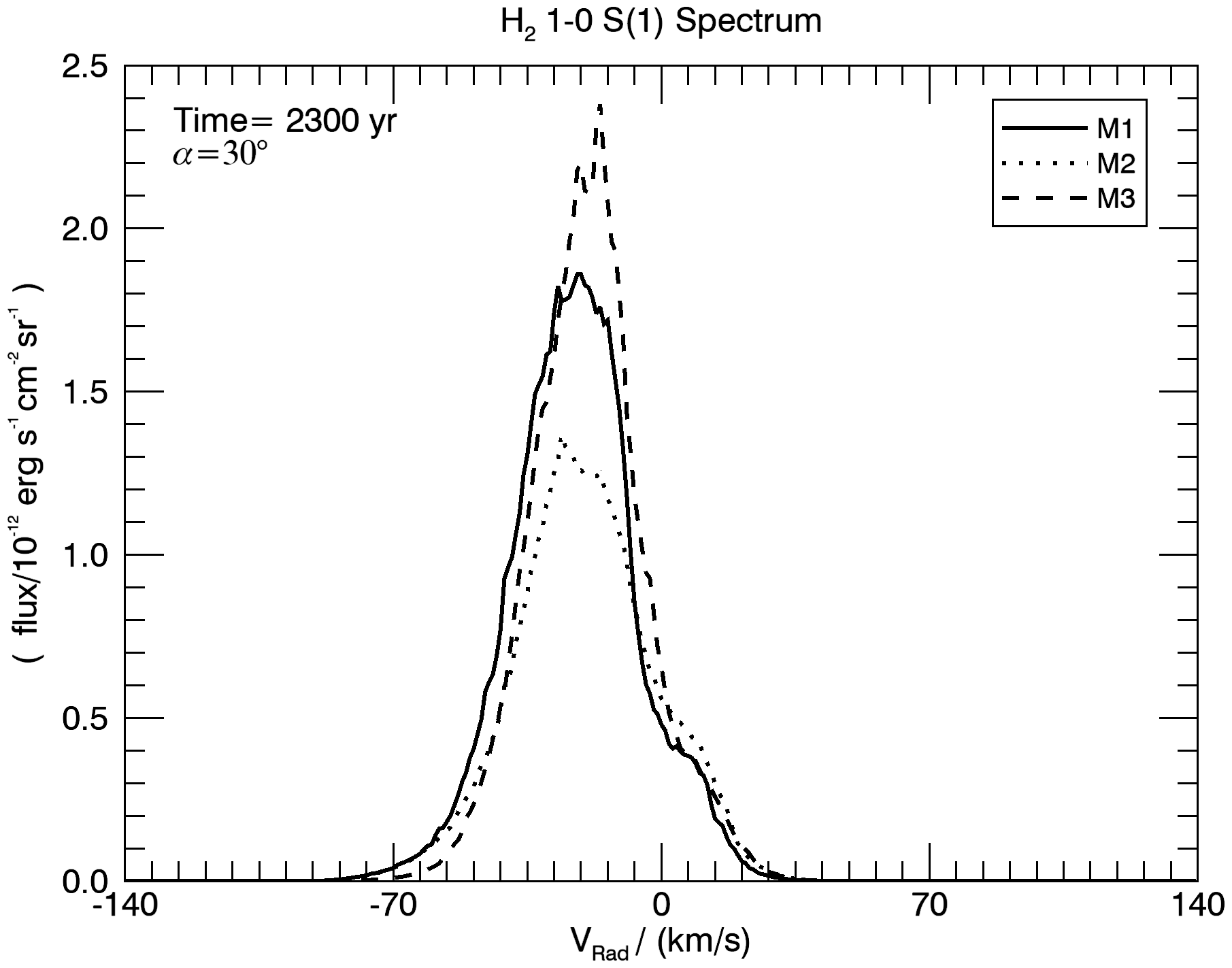}}
\hspace{\fill}
   \subfloat[\label{mt-simtask}][$\mathrm{\alpha=60^{\circ}}$.]{%
      \includegraphics[width=0.30\textwidth,height=0.2\textheight]{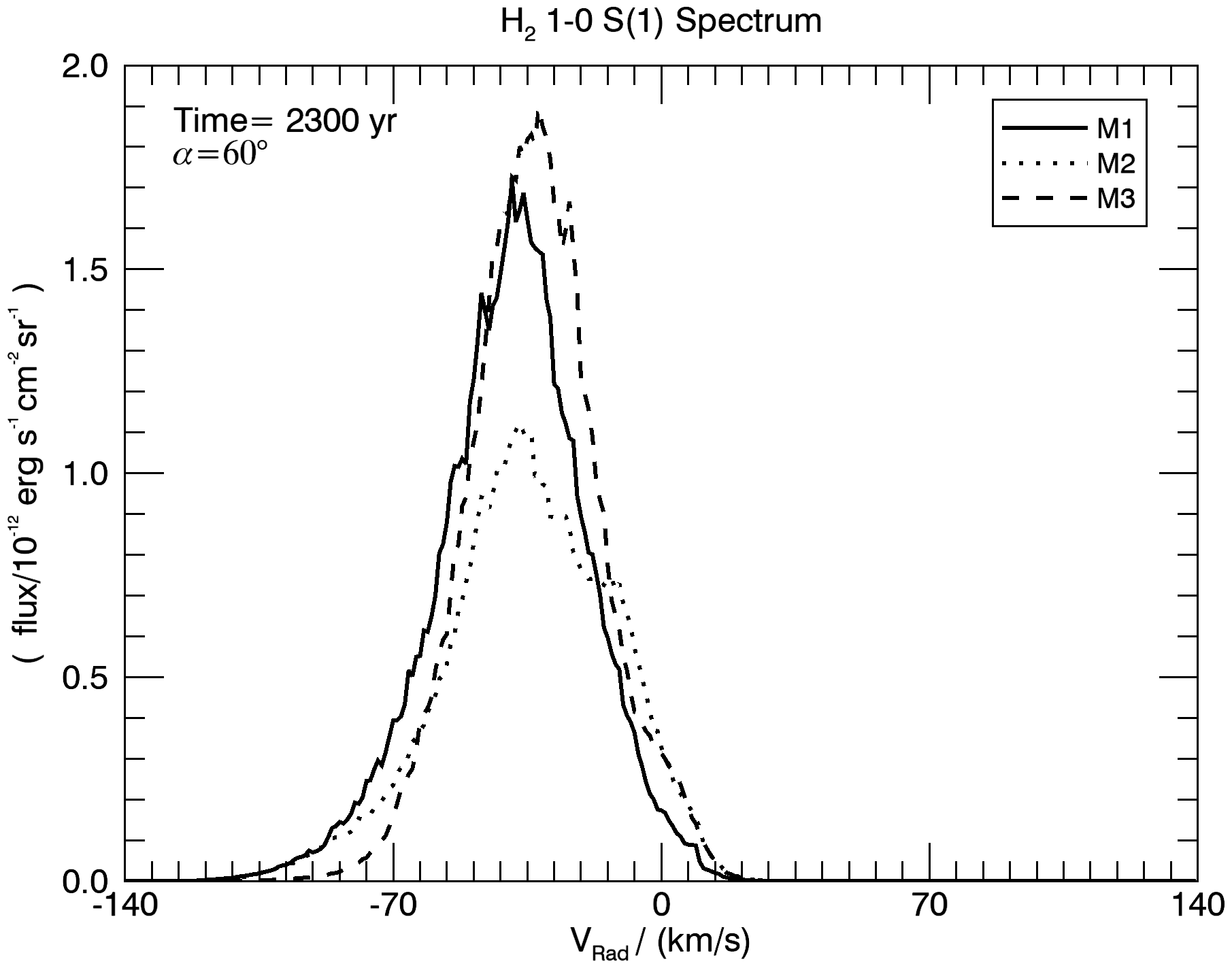}}\\
\caption{\label{lineprofile}Line profiles of the entire pPN in  H$_2$ 1--0\,S(1) emission from  2:1 elliptical $\mathrm{V_w=140\, km s^{-1}}$ wind with the symmetry axis at  $0, \,30 \,\& \,60^{\circ}$ to the plane of the sky.}
\end{figure*}

The broader lines are generated from the molecular wind model, M2. It is clear that radial speeds can exceed 100\,km\,s$^{-1}$ when the symmetry axis is close to the line of sight.
However, during the stage displayed where the pPN has evolved, the emission will still be associated with molecular gas that has been decelerated and compressed within  radiatively cooling zones. Therefore, the PV diagrams, as well as the line profiles shown in Fig.\,\ref{lineprofile}, tend to describe the shocked layer as a whole.

\begin{figure}
      \includegraphics[width=0.49\columnwidth]{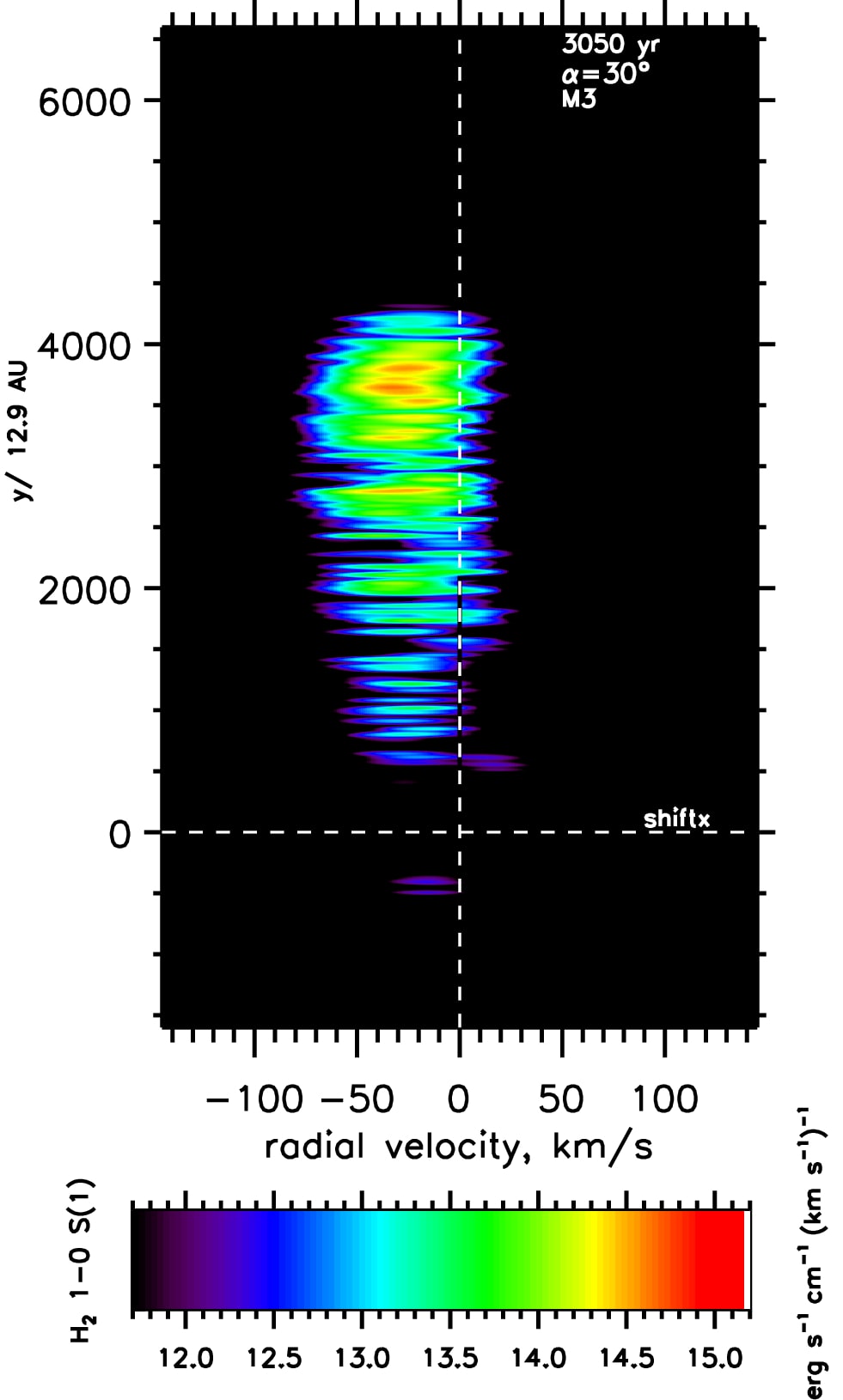}
      \includegraphics[width=0.49\columnwidth]{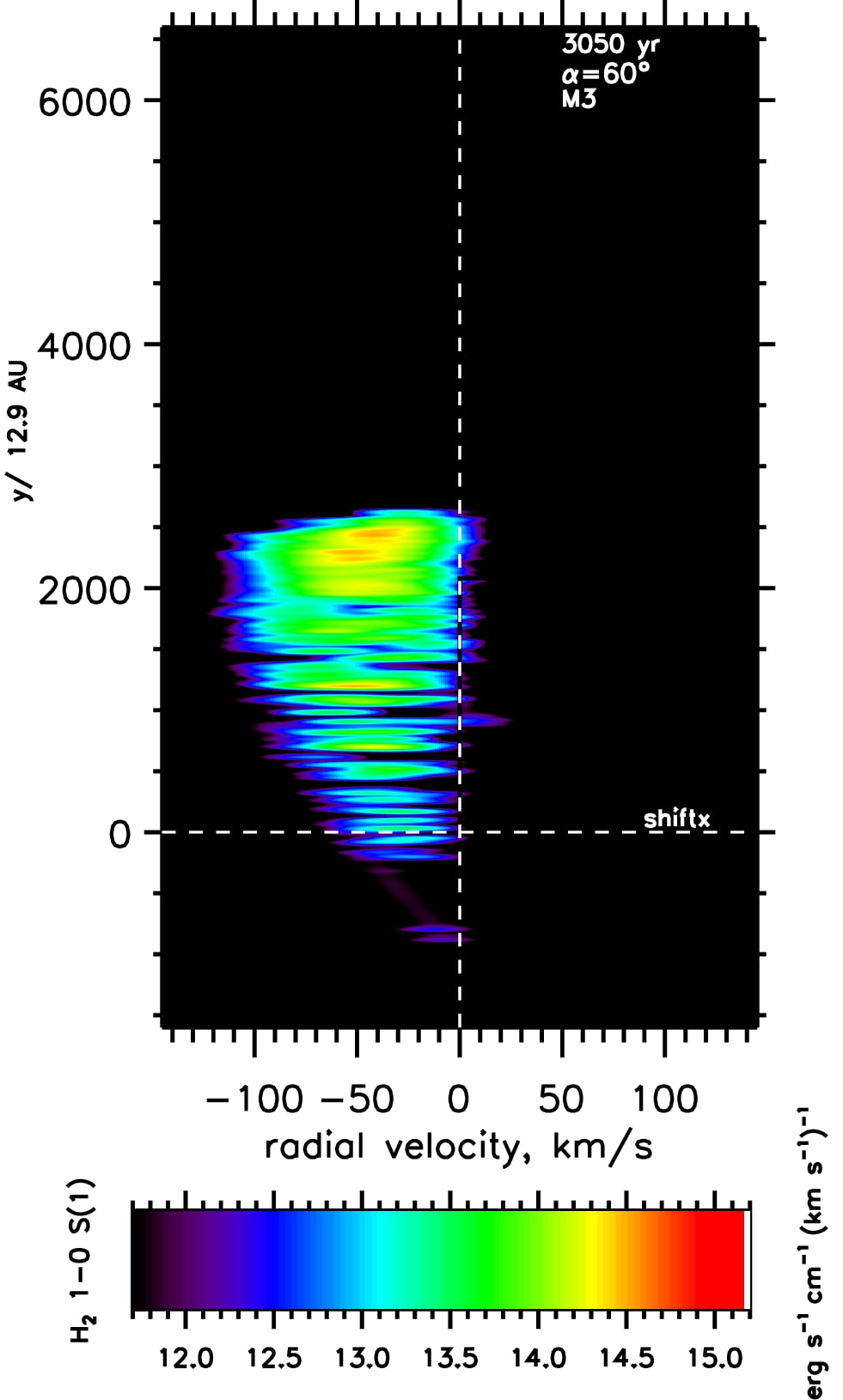}
\caption{\label{4.1molpv}Position-Velocity diagrams for the H$_2$ 1--0\,S(1) emission from the 4:1 elliptical 140\,km\,s$^{-1}$ wind with the long axis at
$\mathrm{\alpha = 30^\circ}$ (left panel) and  $\mathrm{\alpha = 60^\circ}$ (right panel)
out of  the plane of the sky.   Model M3 with a molecular wind is taken as the example here.}
\end{figure}

With an orientation close to the line of sight, the blue-shifted emission dominates. The emission is now wider and stronger toward the end of the lobe. This is as expected especially from the 4:1 wind displayed in Fig.~\ref{4.1molpv} since the H$_2$ excitation is much stronger towards the front which is driven by the 140\,km\,s$^{-1}$ wind while this is reduced to just 35\,km\,s$^{-1}$ in the transverse direction.

it is also interesting that the PV diagrams for these elliptical winds are morphologically different from the jet-driven bows in which both a molecular jet and molecular ambient medium can contribute \citep{2014MNRAS.443.2612S}. Whereas jet-driven flows generate an apparent acceleration signature termed a Hubble law, the elliptical winds yield a more constant radial speed which is still high at the source location. Imposed on this wide ridge is an S-type meander in the PV pattern with substantial redshifted H$_2$ at $2.12\,\mathrm{\mu m}$ even when the pPN axis is at 30$^\circ$ to the line of sight.

Finally, we can now also integrate over the entire pPN to obtain a single line profile. 
Fig.~\ref{lineprofile} displays a characteristic triangular shape when the flux is expressed logarithmically. Once again, the properties appear to be related to the geometry and orientation rather than the Model. The profile peak shifts with orientation by $\sim$ 35~km~s$^{-1}$ even with a wind speed four times larger. Line widths are $\sim$ 50--60~km~s$^{-1}$ at 10\% maximum.

\section{Discussion}

The phase signalled by protoplanetary nebula  bridges the AGB and planetary nebula phases. Previously ejected material and swept up ambient gas are subject to the impact of a wind which may have a speed of 60 -- 300~km\,s$^{-1}$, generating a shell expanding at a speed of several  tens of km\,s$^{-1}$. We have made an attempt here to simulate this phase in which molecular shock physics may dominate and the observed emission is shock-induced. 

Specifically, the inclusion of molecular chemistry and cooling is vital for  correct understanding because of the strong cooling and compression within the shock layers which, rather than generating a thin shell, distorts and fragments into molecular streams and fingers. These fingers protrude from the molecular medium across the interface. Thus, classical cometary globules which are exposed to the wind and roughly face the central star or wind-generated bullets advancing ahead of the interface can result. We find that the direction of these streams are not constrained to lie on the radial line to the central star but appear to divert according to the local conditions. 

Molecule dissociation is also significant. Overall, we note that in the early stages of wind expansion, the molecules in the ambient medium are destroyed due to a strong forward shock. For the later stages as the pPN radius expands, the situation becomes reversed and it is the molecules in the wind that get dissociated. This shock structure is time dependent and leads to a variety of emission stages that can be followed with emission maps generated by molecular cooling routines.

Planetary nebula which are elliptical could be feasibly generated through a variety of mechanisms. As summarised by \citet{2012MNRAS.424.2055H}, these include (1) an  early jet phase followed by a spherical wind, (2) a wind that is shock-focussed or otherwise redirected by an early AGB equatorial wind or shell and (3) an early spherical AGB envelope or wind with a subsequent collimation axial flow. Here, we simply assume an elliptical velocity distribution, a hybrid jet-wind mechanism. The resulting molecular simulations provide a set of templates against which we can test complex evolutionary scenarios which may be relevant to protostellar outflows as well as pPN.

These wind interactions possess some resemblance  in their set up to those occurring in diverging jet or conical atomic outflow \citep{2003ApJ...586..319L} which lead to bow-shock dominated flows, and to collimated molecular winds \citep{2009ApJ...696.1630L}using the same molecular physics as in the original version of the present ZEUS code  \citep{1997A&A...318..595S}. However, the outcomes are very different since the jet interface is then tangential to the flow and exposed to the ambient medium. Hence, the ratio of the ambient thermal pressure to the wind pressure becomes of crucial importance to the wind/jet expansion.

The comparison to observations must take into account the limitations imposed on the simulated physics and dynamics in order to incorporate the chemistry with moderate accuracy. Firstly, we assume cylindrical symmetry which influences how perturbations grow from an inner spherical surface  and how observations of the fingers appear in projection.  In addition, magnetohydrodynamics would limit the compression, but possibly introduce ambipolar diffusion effects before ionisation levels are raised sufficiently by the emerging white dwarf. Finally, we have taken the environment to be uniform and homogeneous in this first work. Such an assumption would appear to eliminate the possibility of generating bipolar nebula (with a constricted or hour-glass structure),
although this was a priori not necessarily the case, given that a deflected wind could, in theory,  generate candle-flame shapes, if the wind flow were deflected by a `mid-plane' oblique shock as shown by \citep{1986MNRAS.223...57S}. 

Molecular hydrogen emission is associated with both  bipolar and elliptical nebula \citep{2013MNRAS.429..973M} which may well follow entirely distinct evolutions.
There is indeed gathering  evidence  that the PN phenomenon is heterogeneous; PN are  formed from multiple evolutionary scenarios
\citep{2010PASA...27..129F}. The ejection models may well depend on the progenitor mass, binarity and detailed stellar physics. 

 Collision shock waves are the dominant source of energy for the H$_2$ excitation in the least evolved pPN \citep{2003MNRAS.344..262D}.
 Fluorescence, on the other hand, becomes more important at intermediate evolutionary stages (i.e. in `young' PN when the UV from the exposure to the hot central star first becomes strong), particularly in the inner core regions and along the inner edges of the expanding post-asymptotic giant branch  envelope.   This is consistent with the finding that, during the  developed PN  stages, molecular hydrogen remains strongly  associated with bipolar nebula rather than elliptical nebula \citep{1999ApJS..124..195H}. 

The distribution of H$_2$ emission in pPN suggests that a fast wind is impinging on material in the cavity walls and tips \citep{2011MNRAS.411.1453G}. Examples of prolate ellipsoidal nebula are found .\citep{2008ApJ...688..327H}  even though the bipolar PN are particularly associated with H$_2$, possibly because of a previously ejected slow molecular wind that possesses an oblate ellipsoidal structure.

 The simulations here produce morphological structures similar to cometary knots found in high resolution images of objects such as the Helix nebula \citep{2011MNRAS.416..790A}. The density of the knots present in pure molecular model M1 is found to be significantly denser $\sim 10^6$ compared to surrounding ambient and wind media. Knots and filamentary condensations are featured in $\mathrm H\alpha$ and $\mathrm{1\to0\,S(1)} \, \mathrm H_2$ images, marking the transition region between warm ionised and neutral $\mathrm H$.

 Knots have also been detected in H$_2$ emission from low-ionization structures  in two planetary nebula \citep{2017MNRAS.465.1289A}. The H$_2$ 2-1/1-0\, S(1) line ratio ranges from 0.1 to 0.14, suggesting the presence of shock interactions, consistent with the predictions here over wide regions of pPN.
The strongest line, 1--0\,S(1), was also detected in several low-ionization knots located at the periphery of the elliptical planetary nebula NGC\,7662. 
Only four knots of these exhibit detectable 2--1\,S(1)  and therefore possess high H$_2$ excitation with  line ratios between 0.3 and 0.5.
It was  suggested that the emission is induced by ionizing ultraviolet from the central star although we have found here that isolated hot-shock zones can possess similar values at early stages.
In particular, Model M3 with an ambient molecular medium contains regions of very high excitation even at a later evolutionary stage.
 
Position-velocity diagrams are being increasingly extracted from long-slit and integral-field spectroscopy. A distinct elliptical morphology was found in IRAS\,16594-4656 \citep{2008ApJ...688..327H}. A comparison with the PV diagrams generated here indicates an axis very close to the plane of the sky. However, the predicted strong increase of brightness at the leading edge is not detected. While the general expanding-shell structures are very similar to the PV diagrams generated here, the observed velocity width is only $\sim$ 20~km~s$^{-1}$, about half that predicted in Fig.~\ref{2.1molpv0}.

Position-velocity diagrams for Hen\,3-401 display two converging arcs with the radial velocity increasing uniformly with axial distance   \citep{2008ApJ...688..327H}.
Such a structure is found here at 30$^\circ$ out of the sky plane as shown by Fig.~\ref{2.1molpv60} if one assumes the redshifted lobe has point-symmetry with the displayed blue-shifted one. Again, however, our models predict a prominent region of high flux at the leading edge of the fast flow along the axis, which is not observed.

Position-velocity diagrams were presented from long-slit spectroscopy along the central axis of the PN M2--9 by \citet{2005AJ....130..853S}. These display two components which correspond to the near and far sides of the expanding elliptical structure. However, we cannot distinguish between models presented here since the emission  is shell-dominated in all three cases.

\section{Conclusions}

We have investigated the consequences of introducing a supersonic wind with a prolate ellipsoidal velocity distribution into a uniform circumstellar environment. We consider all combinations of fully molecular and fully atomic wind and ambient media. Conditions may be relevant to that of pPN. Early in the runs, the forward shock is dominant and molecules in the ambient medium are partly dissociated. As the wind slows down as it expands, the backward shock begins to dominate and molecules in the wind are dissociated.

We found that an ellipsoidal shell interface forms but fragments into finger-like protrusions which poke from the molecular medium into the atomic medium. When both media are atomic, the fingers are not apparent. Instead, the atomic shell thickens into a turbulent crust.

To create links to observations, we produced $\mathrm H_2$ emission maps for the  $\mathrm{1\to0\,S(1)}$ line. By employing 2-D axisymmetric winds, we gain in 
accuracy and resolution of the temperature and molecular fraction. This, however, means that the simulated structures are limited to ring-like formations. 
Gaussian smoothing is applied to facilitate a comparison to the observations data with a standard deviation of $\mathrm{\sigma=20}$ zones, and flux spread over $\mathrm{4\sigma}$ in order to reduce the calculation time for the convolution.  

Further spectroscopic studies (position-velocity diagrams) enable us to distinguish between the geometry and orientation. We uncover elliptical and Hubble-like PV diagrams from the same simulation with differing orientations. However, velocity widths are roughly double what is currently observed. This suggests that observed pPN are already quite evolved and the advancing shock has further decelerated. Alternatively, magnetohydrodynamic C-shock physics should be considered. The H$_2$ emission from C-shocks is produced well before the gas is shock-accelerated, as opposed to the J-shocks considered here where the jump occurs before the emission. The result would be lower velocity widths  and narrower line profiles if the main source of emission originates from a slow-moving ambient medium.

The model set up here generates lobe-type nebula with a leading bow-shaped region. We have also indicated, however, that fluorescent H$_2$ distributions would demonstrate 
a classical bipolar configuration with arcs stemming from the equatorial plane. 

There are some significant differences in trends between the models. For  fully molecular Model 1 (MWMA), both shocks contribute to the total emission produced which therefore holds quite constant.   The increasing reverse shock strength leads to dissociation and raises the overall line ratio, whereas the winds' ellipticity lowers the $\mathrm{1\to0 \,S(1)}$ and $\mathrm{2\to1 \,S(1)}$ lines and raises the overall line ratio in the low=speed flows, but the $\mathrm H_2$ excitation remains constant in the high-speed elliptical wind since the major location of vibationally-excited gas gradually moves across the surface with time. 

Model 2 (MWAA) with the molecular wind displays more time-dependent emission. obtaining lower flux at the bow shock but more compact overall flux density compared to M1 (MWMA) suggesting that M2 is overall more dissociative. 
PV diagrams show that the shock front becomes more blunt in the spherical wind case, but stretches out with increasing ellipticity becoming similar to M1 \& M3. 
Model 3 (AWMA), with a molecular ambient medium, displays the most time-dependent integrated emission. The emission falls off quite rapidly in the latter stages. This is related to the diminishing speed of the forward shock front. At the same time,  the vibrational excitation remains relatively high consistent with the vibrational 
temperature of the H$_2$ being fixed by the collection of shock strengths within the disrupted shell.

Vibrational excitation remains quite constant across the lobes and also in time due to the presence of  a full turbulent spectrum of shock speeds rather than a single shock surface. 
This may help explain the enignmatic absence of high-excitation H$_2$ where expansion speeds are thought to be highest. 

Our second paper will study atomic tracers  from these pPN by $1.64\,\mathrm{\mu m} \,\mathrm {[FeII]}$ and $\mathrm{[SiII]\,6716\lambda}$  forbidden lines, the $\mathrm{[OI]\,6300\lambda}$ airglow line, and $\mathrm [H\alpha]\,6563\lambda$ emission. The wind and ambient media  will have the same initial setups, but this time we will be focusing more on Models 2, 3 and 4. Our final aim is to then reproduce the most realistic pPN environment by introducing Generalized Interacting Stellar Winds (GISW) model with the addition of an earlier AGB slow wind. Further shaping can be done with precessing jet and pulsations with three-dimensional hydrodynamical simulations. The shock structure can also be altered with the addition of magnetic fields introducing a  `C-shock' environment with lower shock velocities and $\mathrm H_2$ excitation. 

\section{Acknowledgements}
\label{acks}
We thank the CAPS members at UKC, especially Mark Price for maintaining the SEPnet funded Forge supercomputer. Special thanks to Justin Donohoe for helping out with IDL code algorithms.    

\bibliography{ppn_mol_astroph}

\setcounter{equation}{0}
\renewcommand\theequation{A.\arabic{equation}}

\section*{Appendix A}
\label{appa}
The cooling functions used in numerical code Eq. ~\ref{eq:8} -- \ref{eq:9} are listed below.

\begin{itemize}
\item $\Lambda_{1}$: cooling from the dissociation of molecular hydrogen with reaction rate coefficients.
\end{itemize}
\begin{equation}
\Lambda_{1}=7.18\times10^{-12}(n^2_{H}(k_{D,H_2})+n_{H}n_{H_2}k_{D,H}) \,{\rm \bigl[erg\,s^{-1}\,cm^{-3}\bigr]}
\end{equation}
Where the units of collisional de-excitation rates are in ${\rm {cm^{3}\,s^{-1}}}$
\begin{flalign}
& k_{D,H}=1.2 \times 10^{-9} \exp(-52400/T) \times [0.0933 \exp(17950/T)]^{\beta} \nonumber & \\
& k_{D,H_2}=1.3 \times 10^{-9} \exp(-53300/T) \times [0.0908 \exp(16200/T)]^{\beta} 
\end{flalign}
With coefficient $\beta$ being the fraction of thermally released energy.
\begin{equation}
\beta= \left\{ 1.0+n\left[2f\left(\frac{1}{n_2}-\frac{1}{n_1}\right)+\frac{1}{n_1}\right]\right\}^{-1}.
\end{equation}
The following two approximations represent critical densities for dissociation by collisions of molecular hydrogen with atomic hydrogen, $n1$, and with itself, $n2$.
\begin{align}
& n1 = dex(4.0 - 0.416x - 0.327x^{2}) {\rm \bigl[cm^{-3}\bigr]}, \nonumber \\
& n2 = dex(4.845 - 1.3x + 1.62x^{2}) \,\,\, {\rm \bigl[cm^{-3}\bigr]}.
\end{align}

\begin{itemize}
\item $\Lambda_{2}$: collisional cooling, 
\end{itemize}
associated with vibrational and rotational modes in molecular hydrogen, based on equations (7)--(12) in \citep{1983ApJ...270..578L}.
\begin{equation}
\Lambda_{2}=n_{H_2} \left[  \frac{L_v^H}{1 + (L_v^H)/(L_v^L)} + \frac{L_r^H}{1 + (L_r^H)/(L_r^H)} \right],
\end{equation}
where the vibrational cooling coefficients of high and low density are
\begin{flalign}
& L_v^H = 1.10 \times 10^{-18}exp(-6744/T) \,\,\,\,\,\,\,\,{\rm \bigl[erg\,s^{-1}\bigr]},  \nonumber \\
& L_v^L = 8.18 \times 10^{-13}\exp(-6840/T) \nonumber \\ & \,\,\,\,\,\,\,\,\,\,\,\,\, \times[n_{H_2}k_{H}(0,1)+n_{H_2}k_{H_2}(0,1)] {\rm \bigl[erg\,s^{-1}\bigr]}
\end{flalign}
Collisional excitation rates are represented by the terms $k_H (0,1)$ and $k_{H_2} (0,1)$ with $\Delta v=1-0$. The exponential term $\exp(-6840/T)$ converts these to de-excitation rates:
\begin{equation}
k_{H}(0,1)=
\begin{cases}
1.4 \times 10^{-13} \exp[(T/125)-(T/577)^2], & T \textless T_{v}   \\
1.0 \times 10^{-12}T^{1/2}\exp(-1000/T)   , & T \textgreater T_{v}.  
\end{cases}
\end{equation}
where the limit $T_{v}=1635\,\mathrm{K}$, and
\begin{equation}
k_{H_2}(0,1)=1.45 \times 10^{-12}T^{1/2}\exp[-28728/(T+1190)].
\end{equation}
The rotational cooling rate coefficient at high density is
\begin{equation}
L_{rH}=
\begin{cases}
dex[-19.24+0.474x-1.247x^2], & T \textless T_{r} \\
3.90 \times 10^{-19}\exp(-6118/T) , & T \textgreater T_{r},  
\end{cases}
\end{equation}
with the limit $T_{r}=1087\,\mathrm{K}$, and $x=log(T/10 000\,\mathrm{K})$. For low density, the coefficient is 
\begin{equation}
\frac{L_{rL}}{Q(n)}=
\begin{cases}
dex[-22.90-0.553x-1.148x^2], & T \textless T_{l}   \\
1.38 \times 10^{-22}\exp(-9243/T) , & T \textgreater T_{l},  
\end{cases}
\end{equation}
with the limit $T_l=4031\,\mathrm{K}$, and the coefficient
\begin{equation}
Q(n)=\left[ (n_{H_2})^{0.77}+1.2 (n_H)^{0.77} \right]
\end{equation}

\label{lastpage}

\end{document}